\documentclass[aps,showpacs,nofootinbib,superscriptaddress,preprintnumbers,epsf,psf,12pt]{elsarticle}

\journal{Nuclear Physics B}

\usepackage{style_sw}

\def\li{.17\columnwidth}
\def\ra{0.05\columnwidth}
\def\si{\hspace{.9cm}}
\def\sci{\hspace{1.7cm}}

\newcommand{\qb}{{\bar q}}
\newcommand{\sla}{\slash \!\!\!}

\renewcommand{\xb}{{\underline x}}  
\newcommand{\eb}{{\underline e}}
\newcommand{\Lb}{{\underline L}}

\newcommand{\lqqb}{{\underline \ell}_{{\bar q}q}}

\newcommand{\Pu}{\underline{e}_{\gamma}}
\newcommand{\eu}{\underline{e}^*_{\rho}}
\begin{document}

\begin{frontmatter}






\title{The dipole representation of vector meson electroproduction beyond leading twist}
%

\author[LPT]{A. Besse}

\author[NCBJ]{L. Szymanowski}

\author[LPT,UPMC]{S. Wallon}

\address[LPT]{LPT, Universit{\'e} Paris-Sud, CNRS, 91405, Orsay, France}

\address[NCBJ]{National Center for Nuclear Research (NCBJ), Warsaw, Poland}

\address[UPMC]{UPMC Univ. Paris 06, facult\'e de physique, 4 place Jussieu, 75252 Paris Cedex 05, France}

\begin{abstract}
We link the recent computation beyond leading twist of the impact factor of the transition $\gamma^*_{T}\to \rho_{T}$ performed in the light-cone collinear approach, to the dipole picture 
 by expressing the hard part of the process through its Fourier transform in coordinate space. We show that in the Wandzura-Wilczek approximation the impact factor up to twist 3 factorises in the wave function of the photon combined with the distribution amplitudes of the $\rho-$meson and the colour dipole scattering amplitude with the $t-$channel gluons. We show also that beyond the Wandzura-Wilczek approximation, the hard contribution of the amplitude still exhibits the signature of the interaction of a single colour dipole with the $t-$channel gluons. 
 This result allows a phenomenological approach of the helicity amplitudes of the leptoproduction of vector meson, by combining our results to a dipole/target scattering amplitude model.
%
\end{abstract}
%

%

\end{frontmatter}


\tableofcontents
\noindent


\setcounter{footnote}{0}

\section{Introduction}
\label{Sec_introduction}

In the limit of asymptotical energy, the understanding of hadronic scattering processes
is a longstanding question, which can be addressed in perturbative QCD when a hard scale, generically denoted as $Q^2$,  justifies the applicability of perturbation theory. In this so-called perturbative Regge limit, in which $s \gg -t \sim Q^2 \gg \Lambda_{QCD}^2$, the scattering amplitude is dominated by the exchange of reggeized gluons (named gluonic reggeons) in the $t-$channel. This regime is governed 
by the  BFKL equation, derived in the leading order approximation (LLA) in Refs.~\cite{Fadin:1975cb, Kuraev:1976ge, Kuraev:1977fs, Balitsky:1978ic}. The scattering  
amplitude is a convolution in transverse momentum of impact factors - describing the coupling of the targets with reggeons - with the reggeon-reggeon scattering amplitude, in the $k_T$-factorisation framework.

In the same kinematics, a dipole representation
\cite{Nikolaev:1990ja} was proposed, and its dynamics was
studied at LLA in Refs.~\cite{Nikolaev:1994vf} and \cite{Mueller:1993rr, Mueller:1994jq}. In
this
picture, the degrees of freedom are coloured dipoles. The scattering of two colourless
objects
(heavy meson for example, or virtual photon, called generically onium) is then described
completely
in terms of these dipoles, and does not involve 
the notion of reggeization. It was checked at the level of the kernel
\cite{Chen:1995pa} and then at the level of full amplitude for onium-onium scattering \cite{Navelet:1997tx} that both BFKL and dipole models are equivalent, at large $N_c$.

At even higher energies, based on the growing of partonic densities which should be limited due to unitarity arguments, many theoretical investigations have been made in order to resum perturbative sub-series which contribute to the scattering amplitude, which could be responsible for the required non-linearities leading to recombination effects between partons. This includes the {\it Generalized Leading Log Approximation}, which takes into account any {\it fixed} number $n$ of
$t$-channel exchanged reggeons \cite{Bartels:1978fc, Bartels:1980pe, Jaroszewicz:1980mq, Kwiecinski:1980wb}, and the
{\it Extended Generalized Leading Log Approximation} (EGLLA) \cite{Bartels:1991bh, Bartels:1992ym, 
Bartels:1993ih, Bartels:1994jj, Bartels:1999aw}, in which the number of reggeon in
$t-$channel is not conserved. In  EGLLA, the simplest new building block  is the triple $\pom$omeron vertex 
\cite{Bartels:1993ih, Bartels:1992ym, Bartels:1994jj, Bartels:1995kf}.

In the Wilson line formalism,  non-linear equations have been derived \cite{Balitsky:1995ub, Balitsky:1998kc, Balitsky:1998ya, Balitsky:2001re},
based on the concept of factorisation of the scattering
amplitude in rapidity space and on the extension  of the operator product
expansion technique to high-energy Regge limit.
Its simplest version, the Balitsky-Kovchegov (BK) 
equation, has also been derived independently by Kovchegov \cite{Kovchegov:1999yj, Kovchegov:1999ua} in the dipole model. The triple 
$\pom$omeron vertex is the building block responsible for non-linearities in the BK equation. Its complete expression beyond large $N_c$, known in EGLLA \cite{Bartels:1994jj}, was rederived based on the Wilson line formalism in 
\cite{Chirilli:2010mw} in a very compact way.
Similarly the Wilson line approach, the Color Glass Condensate (CGC) \cite{JalilianMarian:1997gr, JalilianMarian:1997dw, JalilianMarian:1997jx, JalilianMarian:1998cb, Kovner:2000pt, Iancu:2000hn, Iancu:2001ad,
Ferreiro:2001qy, Weigert:2000gi} leads to an equivalent set of non-linear equations.

 The dipole model itself has been the basis of many studies in order to unitarize the theory \cite{Mueller:1994jq, Mueller:1994gb, Kovchegov:1997dm}. As for other models, it can describe both the $\pom$omeron and the $\odd$dderon \cite{Lukaszuk:1973nt} degrees of freedom 
 \cite{Kovchegov:2003dm}. Besides its dynamics at large $s$, the dipole representation of the probe has been used to 
build  observables sensitive to saturation effects. In particular, the geometrical scaling is a natural consequence of saturation. Although strictly speaking, geometrical scaling does not implies saturation, the fact that geometrical scaling  has been seen at HERA for Deep Inelastic Scattering (DIS)
 \cite{Stasto:2000er}, for moderate $Q^2$ (virtuality of the photon probe) and very small $x$ is a good sign that QCD probes the saturation regime. 
Basically, geometrical scaling means that the $\gamma^* P$ total cross-section, which is a function of $Q^2$ and $x$, can be described as a function of a single variable, which is ratio of $Q$ and of a saturation scale $Q_{s}$.
In the dipole representation, a simple model was introduced by 
Golec-Biernat and W\"usthoff (GBW)  \cite{GolecBiernat:1998js,GolecBiernat:1999qd} for describing the total $\gamma^* p$ cross-section  \cite{GolecBiernat:1998js} as well as diffractive events \cite{GolecBiernat:1999qd}.
The basic  idea is to describe the $\gamma^*-p$ interaction at small $x$ as the scattering of a $q \bar{q}$ pair (a dipole), formed
long before the scattering off the proton (in a non-symmetric frame where the nucleon is at rest).
This initial dipole is characterized by a transverse size $r$ and by a relative fraction $\alpha$ of longitudinal momentum carried by the quark and the antiquark.
 One should then parametrise the dipole-nucleon scattering cross-section.
The total cross-section can be expressed as (here $r = |\rb|$)
\be
\label{eqSigmaL_T}
\sigma_{T,L}(x,Q^2)\:=\:\int \!d\,^2  \rb \! \int_0^1 \!d\alpha \:  
| \Psi_{T,L}\,(\alpha,\rb) | ^2 \: \hat{\sigma}\,(x,\rb^2)\;,
\\
\ee
where $\Psi_{T,L}$ is the  photon wave function 
for the transverse $(T)$ and longitudinally polarized  $(L)$ photons, which can be expressed in terms of Bessel functions.
One then uses an effective description of the saturation dynamics,  implemented in a simple functional  form for
 the effective dipole cross section
$\hat{\sigma}(x,r)$ which encodes the interaction of the $q\bar{q}$ dipole with
a nucleon,
\bea 
\label{sigmahat}
\hat{\sigma}(x,r^2)\,=\,\sigma_0\,\left \{1\,-\,
\exp\left(-\frac{r^2}{4 R_0^2(x)}\right) 
\right\}\;,
\eea
where the $x$-dependent radius $R_0$ is given by
\bea\label{R_0}
R_0(x)&=&\frac{1}{GeV}\;\left(\frac{x}{x_0}\right)^{\lambda/2} \;.
\eea
One gets for $\hat{\sigma}(x,r^2)$ a usual linear perturbative description in the  regime $r \ll R_0\,.$ In that limit the dipole cross-section indeed scales like $r^2\,,$ a typical behaviour for colour-transparency. This is for example what is obtained when computing the dipole-dipole scattering through a two gluon exchange (form of which just comes out of an eikonal treatment \cite{Navelet:1997tx}), in a simple model where the nucleon probe is made of dipoles. The power-like behaviour of $R_0(x)$ is inspired by the hard Pomeron description of the HERA data in the linear regime, obtained within the Mueller dipole model when dressed by small-$x$ gluon emissions \`a la BFKL (dipoles in Mueller approach) the initial
$q \bar{q}$ dipole. 

On the other hand, when $r$ gets larger, of the order of $2 R_0\,,$ the cross-section saturates, with for small $Q^2$ a maximal value which equals $\sigma_0\,.$ The critical saturation line
is thus given by $Q = Q_s(x)$ with 
\beq
\label{defQs}
Q_s(x) = 1/R_0(x)\,.
\eq
To get the geometric scaling prediction, one should note that  the non-trivial functional dependency of the $\gamma^*$ wave function is 
completely expressed as a shape in the dimensionless variable $\bar{r} \equiv |\rb| \, Q\,.$ In this variable, the dipole cross-section (\ref{sigmahat})
has now the functional form
\beq
\label{sigma_hat_rescale}
\hat{\sigma}\left(\frac{Q^2_s(x)}{Q^2}\ \bar{r}^2 \right)
\eq
after using the saturation scale (\ref{defQs}).
Combining Eqs.~(\ref{sigma_hat_rescale}) and (\ref{eqSigmaL_T}), it is now clear that the functional form of the dipole
cross-section (\ref{sigmahat}) remains intact after the perturbative dressing due to the $\gamma^*$
coupling,
based on the fact that the squared wave functions are peaked at $r \sim 2/Q\,.$
One can therefore expect the
geometric scaling of the total cross-section at small $x$
\be
\sigma^{\gamma^*p\rightarrow X}_{tot}(x,Q^2)=
\sigma^{\gamma^*p\rightarrow X}_{tot}(\tau)\ ,\hspace{1cm}
\tau=Q^2/Q_s^2(x)\ .\label{test0}
\ee
This saturation model was further extended by including the effect of DGLAP evolution \cite{Bartels:2002cj}. This can be done by replacing in the dipole cross-section
$\hat{\sigma}(x,r^2)$ the elementary $r^2$ dipole-dipole
 cross-section of the colour transparency regime by a more elaborate cross-section
 which includes DGLAP evolution through the gluon density entering the scattering off the dipole. 
In Ref.~\cite{Marquet:2006jb}, a 
 compilation of all available data 
\cite{Adloff:2000qk, Breitweg:2000yn, Chekanov:2001qu,
Adams:1996gu, Arneodo:1996qe} for $F_2$ has been used, leading to an almost perfect agreement with geometric scaling. 


Another sign of the saturation effect was presented in Ref.~\cite{Iancu:2003ge} were a model, in the spirit of  GBW, was constructed  inspired by the BK equation, smoothly interpolating between the saturated regime and the linear regime. This model was able to describe 
the most recent very precise $F_2$ data \cite{Adloff:2000qk, Breitweg:2000yn, Chekanov:2001qu} in the moderate domain of $Q^2$, explicitly exhibiting a need for saturation effects
at small $x$ and small $Q^2$ in comparison with a pure linear evolution \`a la BFKL. 
Geometrical scaling can be extended at each impact parameter $b$, as  shown in Ref.~\cite{Munier:2003bf}. The idea of a $b-$dependent saturation scale is due to Mueller \cite{Mueller:1989st},
while a $b-$dependent $S-$matrix for dipole-proton scattering was studied for diffractive meson electroproduction in Ref.~\cite{Munier:2001nr}.
\\

Beside this framework whose main focus is devoted to QCD dynamics in the asymptotical limit, there has been much progress in the understanding of the partonic structure of mesons at leading and beyond leading twist \cite{Wallon:2011zx}.
This is mainly based on exclusive vector meson production at intermediate energies (HERMES, JLab) and asymptotical energies (H1, ZEUS). 
The natural theoretical framework is collinear factorisation, which describes the scattering amplitude as a convolution (in the space of longitudinal momentum fraction) of an hard part with a Distribution Amplitude (DA). This framework is under control at leading twist for many processes, but faces severe potential divergencies, in particular for processes involving or dominated by contributions beyond leading twist.
Recent data have shown that some $\gamma^* \to \rho$ transitions, which vanish
at twist 2, are copiously produced and thus deserve an understanding from first principles.
In Refs.~\cite{Anikin:2009hk,Anikin:2009bf}, a new framework was developed in order to deal with such transitions, and have been recently applied to describe H1 and ZEUS data \cite{Anikin:2011sa}. In this study, a non negligible contribution from the low vituality $t-$channel exchanged gluons was found. This could be a sign that saturation effects are important for both twist 2 and twist 3  $\gamma^* \to \rho$ transitions.

The natural question which we want to address in this paper is therefore whether it is possible to construct a consistent framework to deal with exclusive processes, for example exclusive vector meson electroproduction, beyond leading twist, and at asymptotical energies where saturation effects are expected. Our strategy is to combine the dipole framework with collinear factorisation beyond leading twist, that is to construct the 
$\gamma^* \to \rho$ transition in the coordinate space for the $t-$channel gluons, for finite $N_c$, including both 2-parton and 3-parton contributions.  This treatment is different in spirit from the pure 2-parton approach based on models for the light-cone $\rho-$meson wave function \cite{Nemchik:1996cw, Forshaw:2003ki, Kowalski:2006hc, Forshaw:2010py, Forshaw:2011yj}. Our result shows that the coupling of the proton 
to this transition only involves a dipole coupling, and does not require any quadripole structure, which in principle can arise in high energy scattering \cite{Kovchegov:1997dm}. The phenomenological implementation of our result,  which requires to combine it with a model for the proton - dipole scattering amplitude, is left for a forthcoming publication.       

The structure of the paper is the following. In Section \ref{Sec_2-parton}, we derive the dipole representation of the twist 3 $\gamma^* \to \rho $ impact factor
in the 2-parton approximation. The crucial ingredient to get this dipole representation is the use of the QCD equation of motion (EOM). In Section \ref{Sec_3-parton}, we extend this analysis to the 3-parton sector, again 
using the QCD EOM including now genuine twist 3 effects.
This leads to the main results of our analysis, shown in Eqs.~(\ref{PhiNf}, \ref{Phif}).
%

\section{2-parton impact factors up to twist 3 in the dipole picture}
\label{Sec_2-parton}
\subsection{Impact factor representation}
\label{SubSec_Impact-factor}

\begin{figure}
\psfrag{A}[cc][cc]{\raisebox{-.7cm}{$\kappa$}\rotatebox{-70}{$\underbrace{\rule{1.1cm}{0pt}}$}}
\begin{tabular}{cc} \epsfig{file=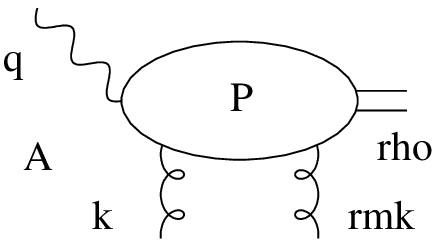,width=\widss} &\qquad \epsfig{file=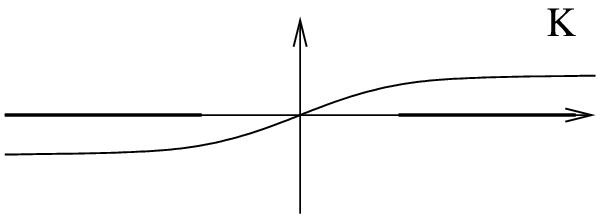,width=\widss}
\end{tabular}
\caption{a: $\gamma^* \to \rho$ impact factor. b: $\kappa$ integration contour entering the definition of the $\gamma^* \to \rho$ impact factor.}
\label{Fig:impactRhoANDkappaint}
\end{figure}

In the $k_{T}$ factorisation approach \cite{Catani:1990xk, Catani:1990eg, Collins:1991ty, Levin:1991ry}, the description of the scattering amplitude for the process
\begin{eqnarray}
\label{prgP}
\gamma^*(q)+N(p_2)\to \rho(p_1)+N
\end{eqnarray}
involves the $\gamma^* \to \rho $ impact factor $\Phi$  of the subprocess
\begin{eqnarray}
\label{ggsubpr}
   g(k_1,\varepsilon_{1})+\gamma^*(q)\to g(k_2, \varepsilon_{2})+\rho(p_1) \,.
\end{eqnarray}
This impact factor is
illustrated in Fig.~\ref{Fig:impactRhoANDkappaint}a,
in the  kinematical
region where virtualities of the photon, $Q^2$, and of $t-$channel gluons
$k_\perp^2$, are of the same order, $Q^2\sim k_\perp^2$, and much larger than
$\Lambda_{QCD}^2$. Neglecting masses, one considers that in the reaction (\ref{prgP}) the vectors  $p_1$ and $p_2$ are light-like. This sets a natural Sudakov basis for momenta, and we denote $2\,p_1\cdot p_2=s$.
\psfrag{P}[cc][cc]{$\Large\Phi$}
\psfrag{q}[cc][cc]{$q$}
\psfrag{k}[cc][cc]{$\, \, k_1$}
\psfrag{rmk}[cc][cc]{$k_2$}
\psfrag{rho}[cc][cc]{$\rho$}
\psfrag{K}[cc][cc]{$\kappa$}
The $\gamma^* \to \rho $ impact factor is defined as
 the integral of the  $S$-matrix element
 ${\cal S}^{\gamma^*_T\, g\to\rho_T\, g}_\mu$ with respect to the Sudakov component of the t-channel $k$
momentum along $p_2\,,$ which is directly related to
 $\kappa=(q+k_1)^2\,,$ the Mandelstam variable $s$ for the subprocess (\ref{ggsubpr}), as illustrated in Fig.~\ref{Fig:impactRhoANDkappaint}a.
Closing the $\kappa$ contour of integration below shows that the impact factor can  equivalently be written (see Fig.~\ref{Fig:impactRhoANDkappaint}b)
as the integral of the $\kappa$-channel discontinuity of the  $S$-matrix element\footnote{In this paper, underlined letters like $\underline{\ell}$ denote euclidean momenta, such that $\ell_\perp^2=-\underline{\ell}^2$. Overlined letters like $\bar{y}$ denote $1-y.$}
 ${\cal S}^{\gamma^*_T\, g\to\rho_T\, g}_\mu$
\begin{eqnarray}
\label{imfac}
\Phi^{\gamma^*\to\rho}(\kb,\,\rb-\kb)&=&e^{\gamma^*\mu}\, \frac{1}{2s}\int\limits^{+\infty}_{-\infty}\frac{d\kappa}{2\pi}
\,  {\cal S}^{\gamma^*\, g\to\rho\, g}_\mu(\kb,\,\rb-\kb)\nonumber\\
&=&
 e^{\gamma^*\mu}\, \frac{1}{2s}\int\limits^{+\infty}_0\frac{d\kappa}{2\pi}
\, \hbox{Disc}_\kappa \,  {\cal S}^{\gamma^*\, g\to\rho\, g}_\mu(\kb,\,\rb-\kb)
\,.
\end{eqnarray}
The two reggeized
gluons have so-called non-sense polarisations $\varepsilon_1=\varepsilon_2^*=p_2\sqrt{2/s}\,.$

\begin{figure}[htb]
\psfrag{l}[cc][cc]{$\ell$}
\psfrag{lm}[cc][cc]{}
\psfrag{q}[cc][cc]{$\gamma^*$}
\psfrag{H}[cc][cc]{$H$}
\psfrag{S}[cc][cc]{$\Phi$}
\psfrag{Hg}[cc][cc]{$H_\mu$}
\psfrag{Sg}[cc][cc]{$\Phi^\mu$}
\psfrag{k}[cc][cc]{}
\psfrag{rmk}[cc][cc]{}
\psfrag{rho}[cc][cc]{$\rho$}
\begin{tabular}{cccc}
\includegraphics[width=5.8cm]{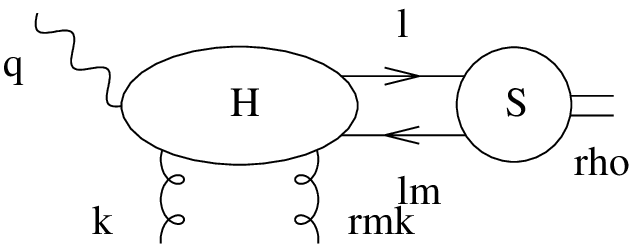}&\hspace{.2cm}
\raisebox{1.2cm}{+}&\hspace{.3cm}\includegraphics[width=5.8cm]{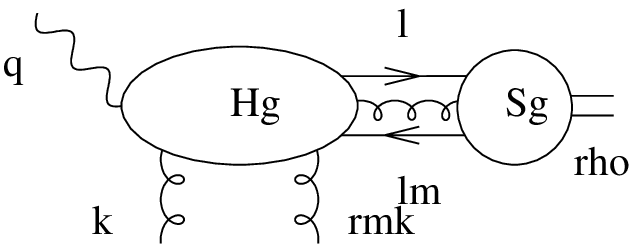}
&
\end{tabular}
\caption{2- and 3-parton correlators attached to the hard scattering amplitude of the   $\gamma^* \to \rho$ impact factor, where vertical lines are hard $t-$ channel gluons in the colour
singlet state.}
\label{Fig:NonFactorized}
\end{figure}
Up to twist 3, the impact factor receives contributions both from 2-parton and 3-parton diagrams, as illustrated in Fig.~\ref{Fig:NonFactorized}. In this section we only consider the 2-parton contributions, in the so-called Wandzura-Wilczek (WW) approximation. The 3-parton contribution will be discussed in Section \ref{Sec_3-parton}.

\subsection{Collinear factorisation beyond leading twist, in the transverse coordinate space}
\label{SubSec_Collinear-fact}

\begin{figure}[htb]
\begin{tabular}{ccc} \hspace{1cm}\epsfig{file=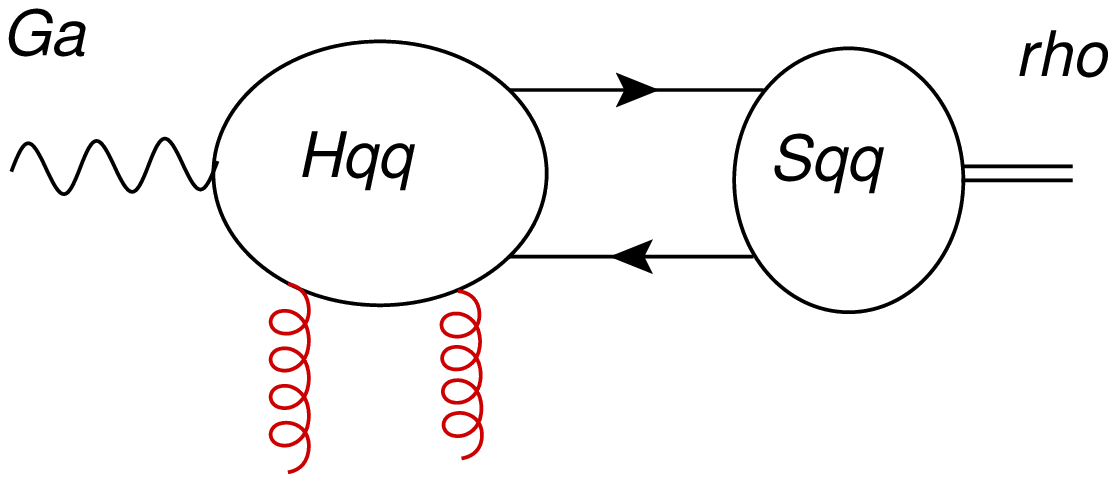,width=\widss} &\raisebox{1cm}{$\longmapsto$}& \epsfig{file=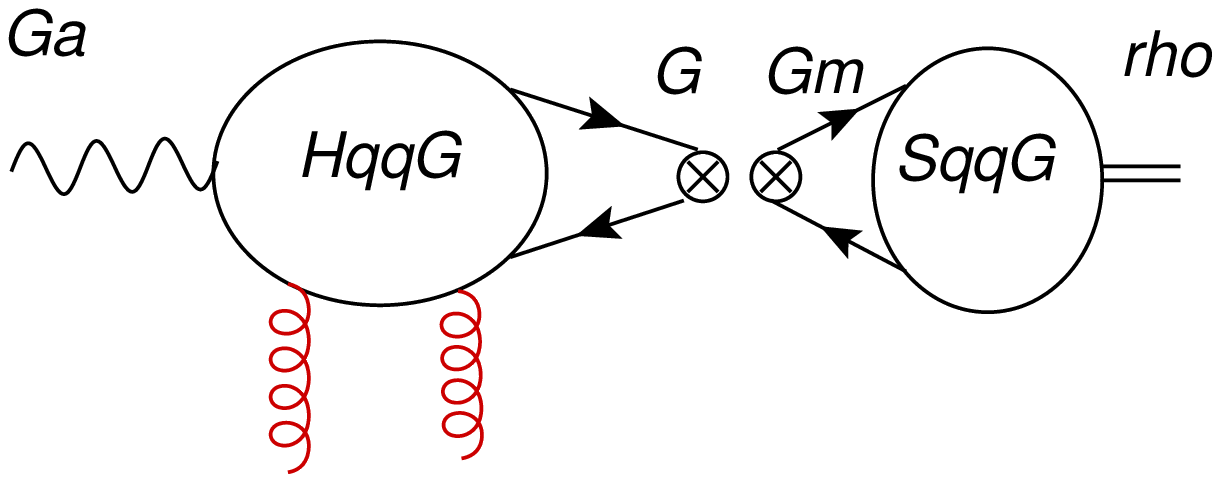,width=\widss}
\end{tabular}
\caption{Factorization with Fierz identity of the 2-parton $\gamma^*\to\rho$ impact factor.}
\label{Fig:Fierz2parton}
\end{figure}

The impact factor contains hard and soft parts, illustrated in Fig.~\ref{Fig:NonFactorized}, that we need to factorise out in Dirac, colour and momentum spaces.
Using the Fierz identity in the standard form \cite{ItZ}, 
\beq
\label{decompositionFierz1}
X = x_\alpha \, \Gamma^\alpha=\frac{1}4 \Gamma^\alpha \, {\rm Tr} \, (X \Gamma_\alpha) = \frac{1}4 \Gamma_\alpha \, {\rm Tr} \, (X \Gamma^\alpha)\,,
\eq 
with $\Gamma_\alpha \equiv (\Gamma^\alpha)^{-1},$ the contribution of the 2-parton Fock state 
 to the amplitude defining the impact factor $\gamma^* \to \rho$ reads\footnote{The minus sign arises due to the fermionic nature of quark and antiquark fields.}
\bea
\Phi^{\gamma^* \to \rho}&=&\int \frac{d^4 \ell}{(2\pi)^4} \,H(\ell) \, S(\ell)= -\frac{1}{4} \int \frac{d^4\, \ell}{(2\pi)^4} \, \text{tr}(H(\ell) \Gamma^{\alpha})\,S^{\Gamma_{\alpha}}(\ell) \,,
\label{A-Fierz}
\eea
as illustrated in Fig.~\ref{Fig:Fierz2parton}. Spinor indices, as well as colour indices by using Fierz identity in colour space, of the hard and the soft parts are now disconnected, which achieves the factorisation of the amplitude in Dirac and colour spaces.

\psfrag{Ga}[cc][cc]{$\gamma^*$}
\psfrag{Hqq}[cc][cc]{$H$}
\psfrag{Sqq}[cc][cc]{$S$}
\psfrag{G}[cc][cc]{}
\psfrag{Gm}[cc][cc]{}
\psfrag{HqqG}[cc][cc]{$H^{\Gamma^{\alpha}}$}
\psfrag{SqqG}[cc][cc]{$S^{\Gamma_{\alpha}}$}
\psfrag{rho}[cc][cc]{$\rho$}

Using Sudakov variables for the loop momentum $\ell = \alpha p +\beta n +\ell_{\perp}$ and for its Fourier conjugate coordinates $z=\alpha_{z}p+\beta_z n+z_{ \perp}$, the $\ell$ integration reads 
\bea
\Phi^{\gamma^* \to \rho}&=& -\frac{1}{4} \int d y \int \frac{d^2\ell_{\perp}}{(2\pi)^2} \, \text{tr}[H(y,\ell_{\perp})\Gamma^{\alpha}]\int \frac{d\alpha_{\ell}}{2\pi} \int \frac{d\beta_{\ell}}{2\pi} \int \frac{d\lambda}{2\pi} e^{i\lambda (\alpha_{\ell}-y)}\nonumber\\
&\times& \int d^4 z\, e^{-i \ell \cdot z}\langle \rho(p)|
\bar\psi(z) \,\Gamma_{\alpha}\, \psi(0)| 0 \rangle \nonumber\\
&=& -\frac{1}{4} \int d y \,\int \frac{d^2\ell_{\perp}}{(2\pi)^2} \, \text{tr}[H(y,\ell_{\perp})\Gamma^{\alpha}] \int \frac{d\lambda}{2\pi} e^{-i\lambda y} \int d^2 z_{\perp}\, e^{-i \ell_{\perp} \cdot z_{\perp}}\nonumber\\
&\times& \langle \rho(p)|
\bar\psi(\lambda n +z_{\perp}) \,\Gamma_{\alpha}\, \psi(0)| 0 \rangle \,.
\label{A-loop}
\eea
We will denote for conciseness $H^{\Gamma^{\alpha}}(y,\ell_{\perp})\equiv \text{tr}[H(y,\ell_{\perp})\Gamma^{\alpha}]$ the hard part of the amplitude and we denote $\tilde{H}^{\Gamma^{\alpha}}(y,x_{\perp})$ its Fourier transform in the transverse $x_T$ plane. The amplitude rewritten in terms of $\tilde{H}^{\Gamma^{\alpha}}(y,x_{\perp})$ reads 
\bea
\Phi^{\gamma^* \to \rho}&\!\!\!\!=& \!\!\!\!-\frac{1}{4} \int d y \!\int\! \frac{d^2\ell_{\perp}}{(2\pi)^2} \int d^2 x_{\perp}\, \tilde{H}^{\Gamma^{\alpha}}(y,x_{\perp})\, e^{-i x_{\perp} \cdot \ell_{\perp}}\!\! \int \frac{d\lambda}{2\pi} e^{-i\lambda y} \nonumber \\
&\times& \int d^2 z_{\perp} e^{-i \ell_{\perp} \cdot z_{\perp}} \, \langle \rho(p)|
\bar\psi(\lambda n +z_{\perp}) \,\Gamma_{\alpha}\,\psi(0)| 0 \rangle\,.
\label{A-Fourier}
\eea
The collinear approximation corresponds to a Taylor expansion of the term $e^{-i \ell_{\perp} \cdot x_{\perp}}$ around the collinear direction, i.e around $\ell_{\perp}=0$. Up to twist 3, we need to expand up to the first order:
$e^{-i \ell_{\perp} \cdot x_{\perp}}=1-i \ell_{\perp} \cdot x_{\perp}+ o(\text{twist 3})\,.$
We denote by $
\Phi^{\gamma^* \to \rho(k)}_{{\rm 2-parton}}$  the contribution of the $k^{\rm th}$ term in the Taylor series.

The first term of the Taylor expansion gives terms of twist 2 and 3, as
\bea
\Phi^{\gamma^* \to \rho(0)}_{\rm 2-parton}=-\frac{1}{4} \!\int\! \!d y \! \!\int d^2 x_{\perp}\, \tilde{H}^{\Gamma^\alpha}(y,x_{\perp}) \!\!\int\! \frac{d\lambda}{2\pi} e^{-i\lambda y} \langle \rho(p)|
\bar\psi(\lambda n) \,\Gamma_{\alpha}\, \psi(0)| 0 \rangle\,.\,
\label{A-Lowest-Taylor}
\eea
The corresponding matrix elements of the relevant non-local correlators with a light-like separation between quark fields are parametrised by a set of distribution amplitudes (DAs) \cite{Anikin:2009hk, Anikin:2009bf}
\bea
\hspace{-.15cm}\int \!\!\frac{d\lambda}{2 \pi} e^{-i \lambda y} \langle \rho(p)|\bar\psi(\lambda n) \,\gamma_{\mu}\, \psi(0)| 0 \rangle \!\!\!\!&=& \!\!\!\!m_{\rho} f_{\rho} [\varphi_1(y)\,(e_{\rho}^* \cdot n) \,p_{\mu}+\varphi_3(y) \,e^*_{\rho T \mu}] \,,\,\,
\label{gamma_mu}\\
\hspace{-2cm}\int \!\! \frac{d\lambda}{2 \pi} e^{-i \lambda y} \langle \rho(p)|\bar\psi(\lambda n) \,\gamma_5 \gamma_{\mu}\, \psi(0)| 0 \rangle\!\!\!\! &=& \!\!\!\!m_{\rho} f_{\rho} \, i \, \varphi_A(y)\, \varepsilon_{\mu \alpha \beta \gamma} \,e_{\rho T}^{*\alpha} \,p^{\beta} \,n^{\gamma}\,.
\label{gamma5-gamma_mu}
\eea
The second term of the Taylor expansion gives only terms appearing at twist~3 and reads
\bea
\Phi^{\gamma^* \to \rho(1)}_{{\rm 2-parton}}&=&-\frac{1}{4} \int d y \int \frac{d^2\ell_{\perp}}{(2\pi)^2} \int d^2 x_{\perp}\, (-i\ell_{\perp} \cdot x_{\perp}) \tilde{H}^{\Gamma^\alpha}(y,x_{\perp}) \int \frac{d\lambda}{2\pi} e^{-i\lambda y} \nonumber \\
&\times& \int d^2z_{\perp}\, e^{-i\ell_{\perp} \cdot z_{\perp}} \langle \rho(p)|
\bar\psi(\lambda n+z_{\perp}) \,\Gamma_\alpha\, \psi(0)| 0 \rangle \nonumber \\
&=& -\frac{1}{4} \int d y \int \frac{d^2\ell_{\perp}}{(2\pi)^2} \int d^2 x_{\perp}\, x_{\perp}^{\alpha} \tilde{H}^{\Gamma^\beta}(y,x_{\perp}) \int \frac{d\lambda}{2\pi} e^{-i\lambda y} \nonumber \\
&\times&\int d^2z_{\perp}\, \frac{\partial}{\partial z_{\perp}^{\alpha}}(e^{-i\ell_{\perp} \cdot z_{\perp}}) \langle \rho(p)|
\bar\psi(\lambda n+z_{\perp}) \,\Gamma_\beta\, \psi(0)| 0 \rangle \,.
\label{Phi-2-body}
\eea
An integration by parts, denoting $\stackrel{\longleftrightarrow}{\partial^{\perp}_{\alpha}}=\frac{1}{2}( \stackrel{\longrightarrow}{\partial^{\perp}_{\alpha}}-\stackrel{\longleftarrow}{\partial^{\perp}_{\alpha}})$, and using the identity
\beq
 - \frac{\partial}{\partial z_{\perp}^{\alpha}} \langle \rho(p)|
\bar\psi(\lambda n+z_{\perp}) \,\Gamma_\beta\, \psi(0)| 0 \rangle
= \langle \rho(p_{\rho}) |\bar{\psi}(z)\Gamma_\beta \stackrel{\longleftrightarrow}{\partial^T_{\alpha}}\psi(0)|0\rangle 
\,,
\label{prep-der}
\eq
finally leads to
\beqa
\Phi^{\gamma^* \to \rho(1)}_{{\rm 2-parton}}&=& -\frac{1}{4} \int d y \int d^2 x_{\perp} x_{\perp}^{\alpha} \tilde{H}^{\Gamma^\beta}(y,x_{\perp})\nonumber\\
&\times& \int \frac{d\lambda}{2\pi} e^{-i\lambda y} \langle \rho(p_{\rho}) |\bar{\psi}(\lambda n)\Gamma_\beta \stackrel{\longleftrightarrow}{\partial^{\perp}_{\alpha}}\psi(0)|0\rangle\,.
\label{A-Taylor-1}
\eqa
The corresponding matrix elements of the relevant non-local correlators involved in Eq.~(\ref{A-Taylor-1}) are parametrised by \cite{Anikin:2009hk, Anikin:2009bf}
\bea
\hspace{-1.5cm}\int \frac{d\lambda}{2 \pi} e^{-i \lambda y} \langle \rho(p)|\bar\psi(\lambda n) \,\gamma_{\mu}\stackrel{\longleftrightarrow}{\partial^{\perp}_{\alpha}}\, \psi(0)| 0 \rangle &&\hspace{-.6cm}= -i m_{\rho} f_{\rho} \, \varphi_1^T(y) \, p_{\mu}\, e^*_{\rho T \alpha}\,,
\label{def-phi1T}
\\
\hspace{-.8cm}\int \frac{d\lambda}{2 \pi} e^{-i \lambda y} \langle \rho(p)|\bar\psi(\lambda n) \,\gamma_5 \gamma_{\mu}\stackrel{\longleftrightarrow}{\partial^{\perp}_{\alpha}}\, \psi(0)| 0 \rangle &&\hspace{-.6cm}= m_{\rho} f_{\rho}  \,\varphi_A^T(y) \,p_{\mu}\, \varepsilon_{\alpha \beta \delta \gamma} \, e_{\rho T}^{*\beta}\, p^{\delta} \,n^{\gamma}\,.
\label{def-phiAT}
\eea
To summarize our approach,  the impact factor up to twist 3, after performing the collinear approximation, has the form
\bea
\sum^{1}_{k=0}\Phi^{\gamma^* \to \rho(k)}_{{\rm 2-parton}}
&=&-\frac{1}{4}\int dy \int d^2 x_{\perp}\, \tilde{H}^{\Gamma^\beta}(y,\xb)\nonumber\\
&&\hspace{-3.2cm}\times\sum^{1}_{k=0} \, \frac{(-x_{\perp}\cdot \partial_{\perp})^k}{k!}\left(\int \frac{d\lambda}{2 \pi} e^{-i \lambda y}\langle \rho(p)|
\bar\psi(\lambda n+z_{\perp}) \,\Gamma_\beta\, \psi(0)| 0 \rangle\right)_{z_{\perp}=0_{\perp}} \,.
\label{systematic-twist}
\eea
Using the  hard and soft parts together, we thus obtain, for the vector part of the 2-parton impact factor $\Phi^{\gamma^* \to \rho}_V$ and axial-vector part $\Phi^{\gamma^* \to \rho}_A$
\bea
&&\hspace{-.7cm}\Phi^{\gamma^* \to \rho}_V\!\!=\hspace{-.07cm}-\frac{1}{4}m_{\rho} f_{\rho}\! \!\int \!\!\!dy \!\!\!\int\!\! d^2 x_{\perp}\,[\varphi_3(y)e_{\rho \perp \mu}^*-i\varphi_1^T(y)p_{1\mu}(e_{\rho \perp}^* \cdot x_{\perp})] \tilde{H}^{\gamma^\mu}(y,x_{\perp}),
\label{A-2body-V}
\\
&&\hspace{-.7cm}\Phi^{\gamma^* \to \rho}_A\!\!=\hspace{-.07cm}-\frac{1}{4}m_{\rho} f_{\rho}\!\! \int\! \!\!dy \!\!\!\int\!\! d^2 x_{\perp}\,[i\varphi_A(y)\varepsilon_{\mu e^*_{\rho \perp} p_1 n}\!+\!\varphi_A^T(y)p_{1\mu}\varepsilon_{x_{\perp} e^*_{\rho \perp}p_1 n}]\tilde{H}^{\gamma_5\gamma^\mu}(y,x_{\perp})\,.\nonumber \\
\label{A-2body-A}
\eea
In the following part we will compute the 2-parton hard parts $H^{\gamma^\mu}(y,\ell_{\perp})$ and $H^{\gamma_5\gamma^\mu}(y,\ell_{\perp})$, and then we will derive the expressions of their Fourier transforms in the transverse coordinate space $\tilde{H}^{\gamma^\mu}(y,x_{\perp})$ and $\tilde{H}^{\gamma_5\gamma^\mu}(y,x_{\perp})$.
\subsection{Calculations of the hard parts in transverse coordinate space}
\label{SubSec_hard-part-2body}

We now compute the hard parts in momentum space $H^{\gamma^\mu}(y,\ell_{\perp})$ and $H^{\gamma_5\gamma^\mu}(y,\ell_{\perp})$ in a kinematics extended to the transverse degrees of freedom $\ell_{\perp}$ of the partons when compared to the kinematics of Ref.~\cite{Anikin:2009hk,Anikin:2009bf,Anikin:2011sa}. 
The partons are kept on the mass-shell and in the collinear limit ($\ell_{\perp}\to0$), this kinematics is the same than in the previous references, 
 which allows to compare the final results after integration over $\xb$ of the Fourier transforms.
 
The Sudakov decomposition of the momenta are\footnote{In this paper, overlined letters like $\bar{y}$ denote $1-y.$}
\bea
\ell_q &=& y p_1+ \ell_{q\perp}+ \frac{\lb^2}{y s} p_2 \nonumber \\
\ell_{\bar q}&=& \yb p_1 - \ell_{q\perp}+ \frac{\lb^2}{\yb s} p_2 \nonumber\\
p_{\rho}&=& p_1 + \frac{\lb^2}{s}\frac{1}{y \yb} p_2=p_1 + \frac{p_{\rho}^2}{s} p_2\,.
\label{Kinematics-2body}
\eea
The momentum of the incoming photon is
\beq
q=p_1-\frac{Q^2}{s}p_2\,,
\label{def-q}
\eq
while the momenta of the gluons in $t$-channel are 
\bea
k_1&=&\frac{\kappa + \kb^2+ Q^2}{s}p_2+k_{\perp} \, ,\nonumber
\\
k_2&=& \frac{\kappa+\kb^2-p_{\rho}^2}{s} p_2+k_{\perp}\,.
\label{def-k1k2}
\eea
%
\begin{figure}[h]
 \scalebox{1}{\begin{tabular}{cccc}
\psfrag{u}{$\phantom{-}\lb$}
\psfrag{d}{$-\lqqb$}
 \hspace{0.cm}\raisebox{0cm}{\epsfig{file=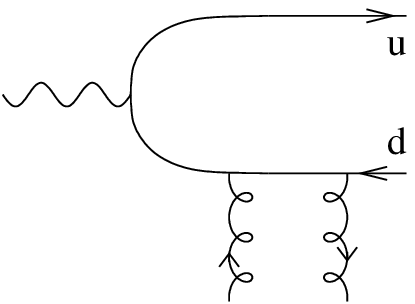,width=\li}} & \si
\psfrag{u}{}
\psfrag{d}{}
 \raisebox{0cm}{\epsfig{file=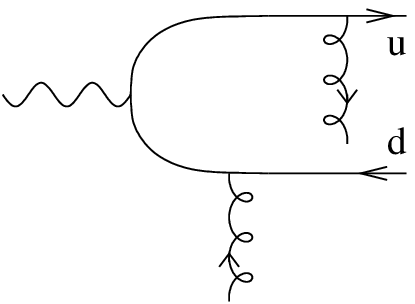,width=\li}} & \si
\psfrag{u}{}
\psfrag{d}{}
\raisebox{\ra}{\epsfig{file=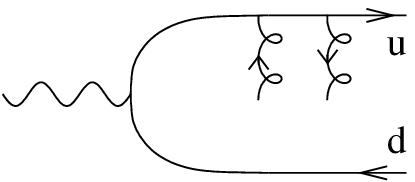,width=\li}}  & \si
\psfrag{u}{}
\psfrag{d}{}
 \raisebox{0cm}{\epsfig{file=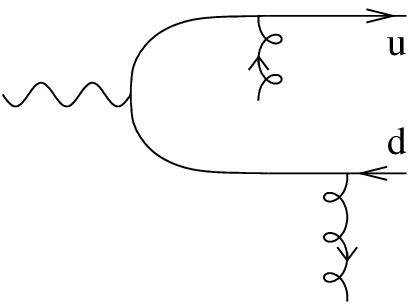,width=\li}}
\\
\\
\hspace{1.2cm} (a) & \sci (b) & \sci (c) & \sci (d)
\\
\\
\\
\psfrag{u}{}
\psfrag{d}{}
 \raisebox{0cm}{\epsfig{file=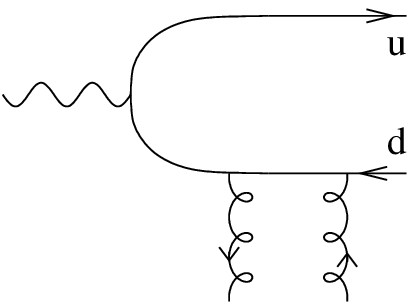,width=\li}}  & \si
\psfrag{u}{}
\psfrag{d}{}
 \raisebox{\ra}{\epsfig{file=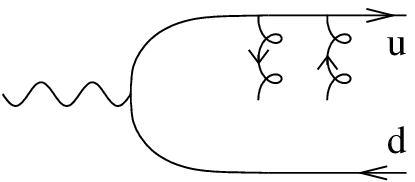,width=\li}} \\
\\
 \hspace{1.2cm}  (e) & \sci (f)
 \end{tabular}}
\caption{The 6 hard diagrams attached to the 2-parton correlators, which contribute to the   $\gamma^* \to \rho$ impact factor, with momentum flux of external line, along $p_1$ direction.
These drawing implicitly 
assume that the two right-hand side spinor lines are closed on  
 the various possible Fierz structures $\Gamma$, here 
$\slashchar{p}$
or $\slashchar{p} \, \gamma^5$.
}
\label{Fig:NoDer2}
\end{figure}
We use the same labelling of diagrams as in Ref.~\cite{Anikin:2009bf}. 
After computing all the 6 diagrams (a), (b), (c), (d), (e), (f), shown in Fig.~\ref{Fig:NoDer2}, we perform the integral over $\kappa$ by the method of residues to get the contribution to the impact factor, according to Eq.~(\ref{imfac}). Four poles in $\kappa$ appear, which are
%
\bei
\item Diagram (a) and (e) : $\kappa_1= \frac{(\lb-y\kb)^2}{y\yb}-i\eta$
\item Diagram (b) and (c) : $\kappa_2= \frac{(\lb+\yb\kb)^2}{y\yb}-i\eta$
\item Diagram (b) and (e) : $\kappa_3= \frac{-1}{\yb}((\kb+\lb)^2-\lb^2+\yb (\kb^2+Q^2)) +i\eta$
\item Diagram (d) and (f) : $\kappa_4= \frac{-1}{y}((\kb-\lb)^2-\lb^2+ y (\kb^2+Q^2)) +i\eta$\,.
\ei
The amplitudes associated to each diagrams (a), (b), (c), (e), that will give a non-zero contribution when we integrate over $\kappa$, closing the contour in the lower $\kappa$ plane, denoted by $\mathcal{C}_{-}$, reads for the different structures $ \Gamma^\alpha~\equiv~\{\gamma^\mu, \gamma^\mu\gamma^5\}$ 
\bei
\item diagram (a):
\bea
i\mathcal{M}^{\Gamma^\alpha}_a &=& \frac{i e }{\sqrt{2}} \frac{g^2 \delta^{ab}}{2 N_c} \frac{2}{s} \frac{\text{tr}[\sla e_{\gamma} (-\yb \sla p_1 + \frac{\lb^2+yQ^2}{y s}\sla p_2+\sla \ell_{\perp})\sla p_2(-\yb)\sla p_1 \sla p_2 \Gamma^\alpha]}{-\frac{\yb}{y} (\lb^2+\mu^2)(\kappa-\kappa_1)} \nonumber\\
&=& \frac{2 i e }{\sqrt{2}} \frac{g^2 \delta^{ab}}{2 N_c} y \frac{\text{tr}[\sla e_{\gamma} (-\yb \sla p_1 +\sla \ell_{\perp}) \sla p_2 \Gamma^{\mu}]}{(\lb^2+\mu^2)(\kappa-\kappa_1)} \,.
\label{diag-a}
\eea
\item diagram (b)
\bea
i\mathcal{M}^{\Gamma^\alpha}_b &=& -\frac{i e }{\sqrt{2}} \frac{g^2 \delta^{ab}}{2 N_c} \frac{2}{s} \frac{\text{tr}[\sla p_2 (y \sla p_1 +\sla k_{\perp}+\sla \ell_{\perp})\sla e_{\gamma} (-\yb \sla p_1 +\sla k_{\perp}+\sla \ell_{\perp}) \sla p_2 \Gamma^\alpha]}{y \yb (\kappa-\kappa_2)(\kappa-\kappa_3)} \nonumber\\
&=& \frac{2 i e }{\sqrt{2}} \frac{g^2 \delta^{ab}}{2 N_c}\frac{1}{y \yb (\kappa-\kappa_2)(\kappa-\kappa_3)} \nonumber\\
&&\times \left\{\yb \text{tr}(\sla p_2 (\sla k_{\perp}+\sla \ell_{\perp}) \sla e_{\gamma} \Gamma^\alpha) -y \text{tr}(\sla e_{\gamma} (\sla k_{\perp}+ \sla \ell_{\perp}) \sla p_2 \Gamma^\alpha)\right\}\,.
\label{diag-b}
\eea
\item diagram (c)
\bea
i\mathcal{M}^{\Gamma^\alpha}_c &=& -\frac{2 i e }{\sqrt{2}} \frac{g^2 \delta^{ab}}{2 N_c} \yb \frac{\text{tr}[\sla p_2 (y \sla p_1 +\sla \ell_{\perp}) \sla e_{\gamma} \Gamma^\alpha]}{(\lb^2+\mu^2)(\kappa-\kappa_2)} \,.
\label{diag-c}
\eea
\item diagram (e)
\bea
i\mathcal{M}^{\Gamma^\alpha}_{e}&=& - \frac{2 i e }{\sqrt{2}} \frac{g^2 \delta^{ab}}{2 N_c}\frac{1}{y \yb (\kappa-\kappa_1)(\kappa-\kappa_4)}\nonumber\\
&& \!\!\!\!\times \left\{\yb \, \text{tr}(\sla p_2 (\sla k_{\perp}-\sla \ell_{\perp}) \sla e_{\gamma} \Gamma^\alpha) -y \, \text{tr}(\sla e_{\gamma} (\sla k_{\perp}- \sla \ell_{\perp}) \sla p_2 \Gamma^\alpha)\right\}\,.
\label{diag-e}
\eea
\ei
In the above expressions, the scale $\mu^2$ is defined by 
\beq
\mu^2 = y \, \yb \, Q^2\,.
\label{def-mu2}
\eq
The impact factor is proportional to the integral  over $\kappa$ of the hard part, according to the definition (\ref{imfac}), which we symbolically write
\beq
\frac{1}{2s}\int_{\mathcal{C}_{-}} \frac{d\kappa}{2 \pi} i \mathcal{M}^{\Gamma^\alpha}=-\frac{i}{2 s} \mathcal{R}es_{\kappa}(i\mathcal{M}^{\Gamma^\alpha})
\label{impact-M}
\eq
with $i\mathcal{M}^{\Gamma^\alpha}=\sum^{4}_{k=1}i\mathcal{M}^{\Gamma^\alpha}_i$ is the sum over the 4 contributing diagrams, and since the integration is performed along the contour $\mathcal{C}_{-}$, the r.h.s of (\ref{impact-M}) means the sum of the residue at $\kappa_1$ and $\kappa_2$. We 
note that
\beqa
\kappa_1-\kappa_4&=&\frac{1}{y\yb} ((\lb-\kb)^2+\mu^2)\,,
\label{relations-kappa-1-4}
\\
\kappa_2-\kappa_3&=&\frac{1}{y\yb} ((\lb+\kb)^2+\mu^2)\,.
\label{relations-kappa-2-3}
\eqa
Thus,
\beas
H^{\Gamma^\alpha}_{\gamma^*_T}(y,\lb)&=&\frac{e g^2\delta^{ab}}{\sqrt{2}2 N_c s} \left\{ \frac{y\,\text{tr}[\sla e_{\gamma} (\sla \ell_{\perp}-\yb \sla p_1)\sla p_2 \Gamma^\alpha] - \yb \, \text{tr} [\sla p_2 (y\sla p_1 + \sla \ell_{\perp}) \sla e_{\gamma} \Gamma^\alpha]}{\lb^2+\mu^2}\right.\\
&& \left. - \frac{y\,\text{tr}[\sla e_{\gamma} (\sla \ell_{\perp}-\sla k_{\perp})\sla p_2 \Gamma^\alpha] - \yb \,\text{tr} [\sla p_2 (\sla \ell_{\perp}-\sla k_{\perp}) \sla e_{\gamma} \Gamma^\alpha]}{(\lb-\kb)^2+\mu^2} \right.\\
&&\left. -\frac{y\,\text{tr}[\sla e_{\gamma} (\sla \ell_{\perp}+\sla k_{\perp})\sla p_2 \Gamma^\alpha] - \yb \,\text{tr} [\sla p_2 (\sla \ell_{\perp}+\sla k_{\perp}) \sla e_{\gamma} \Gamma^\alpha]}{(\lb+\kb)^2+\mu^2} \right\}\,.
\eeas
The vector and the axial-vector hard part read respectively
\bea
H^{\gamma^\mu}_{\gamma^*_T}(y,\lb)
&=& -4\frac{e g^2}{\sqrt{2}} \frac{\delta^{ab}}{2 N_c} e_{\gamma}^{\mu}\frac{y \yb}{\lb^2+\mu^2}-4\frac{e g^2}{s\sqrt{2}} \frac{\delta^{ab}}{2 N_c}(y-\yb)\,p_2^{\mu}\nonumber\\
&\times& \left\{ \frac{\eb_{\gamma} \cdot \lb}{\lb^2+\mu^2}-\frac{\eb_{\gamma} \cdot (\lb+\kb)}{(\lb+\kb)^2+\mu^2}-\frac{\eb_{\gamma} \cdot (\lb-\kb)}{(\lb-\kb)^2+\mu^2} \right\} \label{H-gamma_mu}
\eea
and
\bea
&&\hspace{-.7cm}H^{\gamma^\mu\gamma_5}_{\gamma^*_T}(y,\lb)
=4 i \frac{e g^2}{s\sqrt{2}}\frac{\delta^{ab}}{2 N_c} \nonumber\\
&\times & \varepsilon^{\mu \nu \rho \sigma}\left(
 \frac{e_{\gamma \nu}\ell_{\perp \rho} p_{2 \sigma}}{\lb^2+\mu^2}
 -\frac{e_{\gamma \nu}(\ell_{\perp \rho}+k_{\perp \rho}) p_{2 \sigma}}{(\lb+\kb)^2+\mu^2}-\frac{e_{\gamma \nu}(\ell_{\perp \rho}-k_{\perp \rho}) p_{2 \sigma}}{(\lb-\kb)^2+\mu^2} \right).
\label{H-gamma_mu-gamma5}
\eea
The Fourier transforms of propagators in (\ref{H-gamma_mu}, \ref{H-gamma_mu-gamma5}) are related to the modified Bessel functions $K_{\nu}(x)$
\bea
\frac{1}{\lb^2+\mu^2}&=&\int \frac{d^2 \xb}{2 \pi} K_0(\mu | \xb |)e^{i \lb \cdot \xb}\,,\label{TF-1}\\  
\frac{\lb}{\lb^2+\mu^2}&=&-i \int \frac{d^2 \xb}{2 \pi} \mu \frac{\xb}{| \xb |} K_1(\mu | \xb |)e^{i \lb \cdot \xb}\label{TF-2} \,.
\eea
The Fourier transforms of the vector and axial-vector hard part read thus
\bea
\hspace{-1cm}\tilde{H}^{\gamma^\mu}(y,\xb)&\!\!\!\!=&\!\!\!\! 4\frac{e g^2}{(2\pi)\sqrt{2}} \frac{\delta^{ab}}{2 N_c} \left(- y \yb K_0(\mu | \xb |) e_{\gamma}^{\mu}\right. \nonumber\\
&&\hspace{-2.6cm}\left.+\,p_{2}^{\mu}(y-\yb) i \mu \frac{\eb_{\gamma} \cdot \xb}{| \xb |} K_1(\mu | \xb |)\left[ (1-e^{i \kb \cdot \xb})(1-e^{-i \kb \cdot \xb}) - 1 \right]\right) \,,\,
\label{TF-H-gamma_mu}\\
\hspace{-.5cm}\tilde{H}^{\gamma^\mu \gamma_5}(y,\xb)&\!\!\!\!=& \nonumber\\
&&\hspace{-2.9cm} 4 \frac{e g^2}{s (2\pi)\sqrt{2}}\frac{\delta^{ab}}{2 N_c} \mu K_1(\mu | \xb |)\!\left[\varepsilon^{\mu \nu \rho \sigma}\,e_{\gamma \nu}\frac{x_{\perp \rho}}{| \xb |} p_{2 \sigma}\right]\![(1-e^{i \kb \cdot \xb})(1-e^{-i \kb \cdot \xb})-1]\,.\,
\label{TF-H-gamma_mu-gamma5}
\eea
\subsection{The 2-parton $\gamma^*_T\to\rho_T$ impact factor}
\label{SubSec_2-parton}

Substituting in the expressions of the impact factors, Eqs.~(\ref{A-2body-V}, \ref{A-2body-A}), respectively the vector (\ref{TF-H-gamma_mu}) and axial-vector (\ref{TF-H-gamma_mu-gamma5}) hard parts, 
 we get 
\bea
\Phi^{\gamma^* \to \rho}_V&=&-\frac{C^{ab} Q^2}{2}  \int dy \int \frac{d^2 \xb}{2 \pi}\left\{-2y\yb \varphi_3(y) K_0(\mu | \xb |)\, \Pu\cdot\eu\right.\nonumber \\
&&\hspace{-2.2cm}\left.+\,(y-\yb)\varphi_1^T(y)(\eu \cdot \xb)\,\frac{\xb\cdot \Pu}{| \xb |}  \mu K_1(\mu | \xb |) 
\!\left[(1-e^{i \kb \cdot \xb})(1-e^{-i \kb \cdot \xb})-1\right]\!\!\right\}
\label{A-2body-V-final}
\eea
and
%
\bea
\Phi^{\gamma^* \to \rho}_A&\hspace{-.3cm}=&\hspace{-.3cm}\frac{-C^{ab}Q^2}{s} \int dy \int \frac{d^2 x_{\perp}}{2 \pi}\varphi_A^T(y) \frac{2}{s} \varepsilon_{x_{\perp} e^*_{\rho \perp}p_1 p_2}\varepsilon_{x_{\perp} e_{\gamma \perp}p_1 p_2} \nonumber \\
&\times& \!\! \mu K_1(\mu | \xb |)\,((1-e^{i \kb \cdot \xb})(1-e^{-i \kb \cdot \xb})-1) \,,
\label{A-2body-A-1}
\eea
where we define 
\beq
C^{ab}=-\frac{e g^2}{\sqrt{2}}m_{\rho} f_{\rho}\frac{\delta^{ab}}{2 N_c}\frac{1}{Q^2}\,.
\eq
Note that the term with $\varphi_A$ in Eq.~(\ref{A-2body-A}) vanishes due to the structure of the expression (\ref{TF-H-gamma_mu-gamma5}).
Using $\varepsilon_{x_{\perp} e^*_{\rho \perp}p_1 p_2}\varepsilon_{x_{\perp} e_{\gamma \perp}p_1 p_2}=\frac{s^2}{4} (x_{\perp}^2 e_{\gamma \perp} \cdot e^*_{\rho \perp}-(e_{\gamma \perp} \cdot x_{\perp})(x_{\perp} \cdot e^*_{\rho \perp}))$, the axial-vector contribution takes the form
\bea
&&\hspace{-.6cm}\Phi^{\gamma^* \to \rho}_A = -\frac{C^{ab} Q^2}{2} \int \! dy \! \int \frac{d^2 \xb}{2 \pi}\varphi_A^T(y) \left[ \eu \cdot \Pu-\frac{(\eu\cdot \xb)\,(\xb \cdot \Pu)}{| \xb |^2}\right] \nonumber \\
&&\times \,\mu | \xb | K_1(\mu | \xb |) ((1-e^{i \kb \cdot \xb})(1-e^{-i \kb \cdot \xb})-1) \,.
\label{A-2body-A-final}
\eea
The whole 2-parton contribution thus reads
\bea
\Phi^{\gamma^* \to \rho}_{{\rm 2-parton}}&=& \Phi^{\gamma^* \to \rho}_V+\Phi^{\gamma^* \to \rho}_A\\
&=& -\frac{C^{ab} Q^2}{2}  \int dy \int \frac{d^2 \xb}{2 \pi}\left\{ -2y\yb \varphi_3(y) K_0(\mu | \xb |) \,\Pu\cdot \eu \right.\nonumber \\
&&\hspace{-2cm}\left.+\left[\left((y-\yb)\varphi_1^T(y)-\varphi_A^T(y)\right) \frac{(\eu \cdot \xb)\,(\xb\cdot \Pu)}{| \xb |^2}+\varphi_A^T(y) \, \eu \cdot \Pu\right]\right.\nonumber \\
&\times&\left.\mu | \xb | K_1(\mu | \xb |) ((1-e^{i \kb \cdot \xb})(1-e^{-i \kb \cdot \xb})-1) \right\}\,.
\label{A-2body}
\eea
This result does not seem to be proportional to the familiar dipole factor 
\beq
\mathcal{N}(\xb,\kb)=(1-e^{i\kb\cdot\xb})(1-e^{-i\kb\cdot\xb})
\label{DipoleFactor}
\eq
describing the 
coupling to the two $t-$channel gluons. The dipole structure (\ref{DipoleFactor}) is recovered by the use of the QCD EOM, which relate the various DAs appearing in Eq.~(\ref{A-2body}). The solution for a DA $\varphi(y)$ of these equations can be split in $\varphi(y)=\varphi^{WW}(y)+\varphi^{gen}(y)$, where $\varphi^{WW}(y)$ is solution of EOM in the Wandzura-Wilczek approximation, i.e. in the limit of vanishing 3-parton DAs. These equations \cite{Anikin:2009hk,Anikin:2009bf} lead in the WW approximation to the relation
\beq
2 y \, \yb \, \varphi_3^{WW}(y)+(y-\yb)\varphi_1^{T\, WW}+\varphi_A^{T\, WW}(y)=0\,,
\label{QCD-EOM-2-body}
\eq
which appears in Eq.~(\ref{A-2body}) rewritten in the form
\bea
\label{A-2body-1}
\Phi^{\gamma^* \to \rho}_{{\rm 2-parton}}&\! \!=&\! \! -\frac{C^{ab} Q^2}{2} \int dy \int \frac{d^2 \xb}{2 \pi}  \\
&&\hspace{-2cm}
 \times \left\{\left[-(2 y \,\yb \, \varphi_3(y)+(y-\yb)\varphi_1^{T}+\varphi_A^{T}(y))\right] K_0(\mu | \xb |)\,\Pu\cdot \eu\!
\right.\! \!\!\!\nonumber \\
&&
\hspace{-2cm}+\!\!\left.\left[\left[(y-\yb)\varphi_1^{T}(y)-\varphi_A^{T}(y)\right]\frac{\eu\cdot\xb\,\xb\cdot \Pu}{| \xb |^2}+\varphi_A^{T}(y)  \eu\cdot \Pu \!\right] \right.\nonumber \\
&&\hspace{-2cm}\times \left.\mu | \xb | K_1(\mu | \xb |)\mathcal{N}(\xb,\kb) \right\} \!,\nonumber
\eea
in which we used the following integral relations
\bea
\label{relation-Bessel-K0-K1a}
\int \frac{d^2 \xb}{2 \pi} \mu| \xb |K_1(\mu | \xb |)&=& 2 \int \frac{d^2 \xb}{2 \pi} K_0(\mu | \xb |)=\frac{2}{\mu^2}
\\
\int \frac{d^2 \xb}{2 \pi} \eu \cdot \xb \, \frac{\xb\cdot \Pu}{| \xb |}\mu K_1(\mu | \xb |)&=&\Pu \cdot \eu \int \frac{d^2 \xb}{2 \pi} K_0(\mu | \xb |)\,.\,
\label{relation-Bessel-K0-K1b}
\eea
\subsection{The Wandzura-Wilczek approximated impact factor in the dipole picture}
\label{SubSec_WW-dipole}

Using Eq.~(\ref{QCD-EOM-2-body}), the 2-parton WW contribution (\ref{A-2body-1})
exhibits
the interaction term $(1-e^{i \kb \cdot \xb})(1-e^{-i \kb \cdot \xb})$ between the dipole of transverse size $\xb$ and the gluons in $t$-channel of transverse momenta $\kb$ as 
\bea
\label{A-2body-final}
&&\hspace{-1.2cm}\Phi^{\gamma^* \to \rho \,WW}_{{\rm 2-parton}}= -\frac{C^{ab} Q^2}{2} \int dy \int \frac{d^2 \xb}{2 \pi}\left\{\left[(y-\yb)\varphi_1^{T\, WW}(y)-\varphi_A^{T\, WW}(y)\right] \right.\nonumber \\
 &&\hspace{-1.4cm}\left.\times \, \frac{(\eu\cdot\xb)\,(\xb \cdot\Pu)}{| \xb |^2}+ \varphi_A^{T\, WW}(y) \eu \cdot \Pu \!\right\}
 \mu | \xb | K_1(\mu | \xb |) \, \mathcal{N}(\xb,\kb)\,.\, 
\eea
This result will be extended in part \ref{SubSec_dipole-3-body-EOM} to the genuine DAs $\varphi^{gen}$ solutions of EOM in the presence of the 3-parton DAs.
Similarly to the momentum space analysis, one can split the result (\ref{A-2body-final}) into spin non-flip $\Phi^{\gamma^* \to \rho}_{\rm 2-parton,\, n.f.}$ and  spin flip $\Phi^{\gamma^* \to \rho}_{\rm 2-parton, \, f.}$ contributions
\bea
\Phi^{\gamma^* \to \rho\,WW}_{\rm 2-parton, \,n.f.}&=&-\frac{C^{ab} Q^2}{2} \int dy \left(\varphi_A^{T\, WW}+(y-\yb)\varphi_1^{T\, WW} \right) \nonumber\\
 &&\hspace{-2cm}\times\int \frac{d^2\xb}{2\pi}\,\frac{1}{2} \,\eu\cdot \Pu  \, \mu | \xb |K_1(\mu | \xb |) \,\mathcal{N}(\xb,\kb) \label{2body-non-flip}
\eea
and
\bea
\Phi^{\gamma^* \to \rho\,WW}_{\rm 2-parton, \,f.}&=&-\frac{C^{ab} Q^2}{2}  \int dy \left(\varphi_A^{T\, WW}-(y-\yb)\varphi_1^{T\, WW} \right)\nonumber\\
 &&\hspace{-2cm}\times \int \frac{d^2\xb}{2\pi}\left(\frac{1}{2}\eu\cdot\Pu -\frac{(\eu\cdot \xb) (\Pu\cdot \xb)}{| \xb |^2} \right)  \mu | \xb | K_1(\mu | \xb |) \, \mathcal{N}(\xb,\kb)\,.
\label{2body-flip}
\eea

\begin{figure}
\psfrag{gam}[cc][cc]{$\Pu$, $\lambda_{\gamma}$}
\psfrag{fl}[cc][cc]{$\xb$}
\psfrag{hq}[cc][cc]{$\lambda_q$}
\psfrag{X}[cc][cc]{$\!\times \quad \xb \ $}
\psfrag{hqb}[cc][cc]{} 
\psfrag{Soft}[cc][cc]{$\quad \phi^{T}_{\rho \,\lambda_{\rho} \lambda_{q}}(y)$}
\psfrag{hqbs}[cc][cc]{}
\psfrag{hqs}[cc][cc]{$\lambda_q$}
\psfrag{gami}[cc][cc]{{\footnotesize $\Gamma$}}
\centerline{\epsfig{file=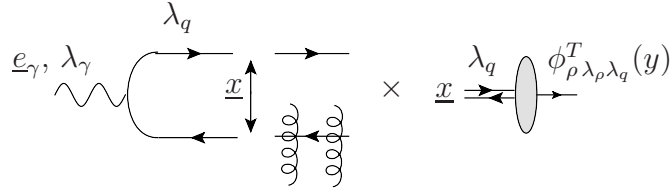,width=8cm}}
\caption{The 2-parton contribution to the $\gamma^* \to \rho_T$ impact factor in the dipole factorised form.}
\label{Fig:2-parton-dipole}
\end{figure}

Let us note that after the introduction of quark helicites as characteristics of the states (here quark and antiquark are massless, therefore the quark and antiquark helicities are opposite), the final formula has a form (see e.g. Ref.~\cite{Bartels:2003yj}) of a sum over intermediate quark helicity ($\lambda$) states of the products of photon wave functions, dipole factor and the specific combinations of DAs for each $\rho$ meson helicity state, according to
\bea
\label{A-2body-final-non-flip}
\Phi^{\gamma^* \to \rho\, WW}_{\rm 2-parton, \, n.f.}&=& \frac{m_{\rho}\, f_{\rho}}{\sqrt{2}}\int dy\ \int d^2\xb\,g^2 \,\delta^{ab}\, \mathcal{N}(\xb,\kb)\,\sum_{\lambda} \phi_{i \lambda}^{WW}
\, \Psi_{i \lambda}^{\gamma^*_{T}} \,, \\
\label{A-2body-final-flip}
\Phi^{\gamma^* \to \rho\,WW}_{\rm 2-parton, \, f.}&=& \frac{m_{\rho}\, f_{\rho}}{\sqrt{2}} \int dy\ \int d^2\xb \,g^2 \, \delta^{ab} \,\mathcal{N}(\xb,\kb)\,\sum_{\lambda, i\neq j} \phi^{WW}_{i \lambda} 
\, \Psi_{j \lambda}^{\gamma^*_T} \,.
\eea
The coefficient $\frac{1}{\sqrt{2}}$, explicitly factorised out, is related to the partonic content of the $\rho^0-$meson as a $\frac{u\bar{u}-d\bar{d}}{\sqrt{2}}$ state such that the
meson involved below is understood as a one flavour quark--antiquark state.
We now define the following combinations 
\bea
\phi^{WW}_{+ +}&=& -i(\eb_{\rho}^{+ *}\cdot \xb) 
\frac{1}{8 N_c} (\varphi^{T\,WW}_A+\varphi^{T\,WW}_{1})\nonumber\\
&=& \frac{ x^*}{\sqrt{2}} 
\frac{1}{8 N_c} (\varphi^{T\,WW}_{A}+\varphi^{T\,WW}_{1})\,,\\ 
\phi^{WW}_{+ -}&=&-i (\eb_{\rho}^{+ *}\cdot \xb) 
 \frac{1}{8 N_c} (\varphi^{T\,WW}_{A}-\varphi^{T\,WW}_{1})\nonumber\\
 &=& \frac{ x^*}{\sqrt{2}}  
\frac{1}{8 N_c}(\varphi^{T\,WW}_{A}-\varphi^{T\,WW}_{1})\,,\\
\phi^{WW}_{- +}&=&-i(\eb_{\rho}^{- *}\cdot \xb) 
\frac{1}{8 N_c} (\varphi^{T\,WW}_{A}-\varphi^{T\,WW}_{1})\nonumber\\
&=&  - \frac{ x}{\sqrt{2}}   
\frac{1}{8 N_c}(\varphi^{T\,WW}_{A}-\varphi^{T\,WW}_{1})\,,\\
\phi^{WW}_{ - -}&=& -i(\eb_{\rho}^{- *}\cdot \xb) 
\frac{1}{8 N_c} (\varphi^{T\,WW}_{A}+\varphi^{T\,WW}_{1})\nonumber\\
 &=& - \frac{ x}{\sqrt{2}}  
 \frac{1}{8 N_c}(\varphi^{T\,WW}_{A}+\varphi^{T\,WW}_{1})\,,
\eea
and for the $\gamma^*$ \cite{Ivanov:1998jw}
\bea
\Psi_{++}(y,\xb)&=&  i\frac{e}{\pi} y \frac{ \eb^{+}_{\gamma}\cdot \xb}{| \xb |} \mu K_1(\mu | \xb |)= \frac{e}{\pi \sqrt{2}} y \frac{ x}{| \xb |} \mu K_1(\mu | \xb |)\,,
\label{Psi++}
\\
\Psi_{+-}(y,\xb)&=&  i\frac{e}{\pi} \yb \frac{ \eb^{+}_{\gamma}\cdot \xb}{| \xb |} \mu K_1(\mu | \xb |)=\frac{e}{\pi \sqrt{2}} \yb \frac{ x}{| \xb |} \mu K_1(\mu | \xb |) \,,
\label{Psi+-}
\\
\Psi_{-+}(y,\xb)&=&i \frac{e}{\pi} \yb \frac{ \eb^{-}_{\gamma}\cdot \xb}{| \xb |} \mu K_1(\mu | \xb |)= - \frac{e}{\pi \sqrt{2}} \yb \frac{ x^*}{| \xb |} \mu K_1(\mu | \xb |)\,,
\label{Psi-+}
\\
\Psi_{--}(y,\xb)&=& i\frac{e}{\pi} y \frac{ \eb^{-}_{\gamma}\cdot \xb}{| \xb |} \mu K_1(\mu | \xb |)= - \frac{e}{\pi \sqrt{2}} y \frac{ x^*}{| \xb |} \mu K_1(\mu | \xb |)\,,
\label{Psi--}
\eea
with, for the polarisations of the $\gamma^*$ and the $\rho$ meson, 
\beq
 \epsilon^{\pm}=\mp\frac{i}{\sqrt{2}}\left( \begin{array}{c}
0  \\
1  \\
\pm i\\
0  \end{array} \right)\,,
\label{polarisation}
\eq
and 
\beq
x = x_1 + i x_2\,.
\label{def-z}
\eq
The obtained factorised structures (\ref{A-2body-final-non-flip}) and (\ref{A-2body-final-flip}) are illustrated in Fig.~\ref{Fig:2-parton-dipole}.
We believe that the combinations of DAs appearing in these formulae are universal and related to the first derivative of the unknown $\rho$ meson wave function. It would be interesting to have a deeper understanding of the appearance of those combinations of DAs.

Performing the integral over  $\xb$ of the results (\ref{2body-non-flip}) and (\ref{2body-flip}) using
\bea
\int \frac{d^2\xb}{2\pi} \mu \frac{ \xb_i\,\xb_j}{| \xb |}K_1(\mu | \xb |)e^{i\lb\cdot\xb}&=& \frac{1}{\lb^2+\mu^2}\left(\delta_{ij}-2 \frac{\lb_i\,\lb_j}{\lb^2+\mu^2}\right)\,,
\eea
we recover the same results in momentum space of Ref.~\cite{Anikin:2009bf} for the impact factors $\Phi^{\gamma^* \to \rho}_{\rm 2-parton, \,n.f.}$ and $\Phi^{\gamma^* \to \rho}_{\rm 2-parton, \,f.}$
\bea
\Phi^{\gamma^* \to \rho\,WW}_{\rm 2-parton, \,n.f.}&=&-\frac{C^{ab}Q^2}{2} \int dy \left(\varphi^{T\,WW}_{A}+(y-\yb)\varphi^{T\,WW}_{1} \right) \nonumber\\
&&\hspace{1.5cm}\times \,T_{\rm n.f.}\,\frac{2}{\mu^2} \frac{\kb^2(\kb^2+2\mu^2)}{(\kb^2+\mu^2)^2}\,,\\
\Phi^{\gamma^* \to \rho\,WW}_{\rm 2-parton, \,f.}&=&-\frac{C^{ab}Q^2}{2}  \int dy \left(\varphi^{T\,WW}_{A}-(y-\yb)\varphi^{T\,WW}_{1} \right)\nonumber\\
&&\hspace{1.5cm}\times \,T_{\rm f.} \,\frac{-4\kb^2}{(\kb^2+\mu^2)^2}\,,
\eea
with for the spin non-flip $T_{\rm n.f.}$ and spin flip $T_{\rm f.}$ tensors
\bea
T_{\rm n.f.}&=&\eu\cdot\Pu\,,\\
T_{\rm f.}&=&\frac{(\eu\cdot\kb)\,(\kb\cdot\Pu)}{\kb^2}-\frac{\eu\cdot\Pu}{2}\,.
\eea
This fact can be seen as a self-consistency check of our calculation.
Note that in the treatment in momentum space performed in Refs.~\cite{Anikin:2009hk,Anikin:2009bf}, the contribution up to twist 3 of the 2-body part involves 6  diagrams for $H^{\Gamma^{\mu}}(y)$, (a), .., (f) and 12 additional diagrams (a1), (a2), ..,(f2) corresponding to the first order term of the Taylor expansion around the longitudinal direction $p_1$ of the hard part 
$\frac{\partial H^{\Gamma^{\mu}}}{\partial \ell_{\alpha}}(\ell=y p_1)$. 

%
%
%

\subsection{$\gamma^*_L \to \rho_L$ impact factor in dipole picture}
\label{SubSec_L-to-T}

We end up this section by recalling for completeness the result for the $\gamma^*_L \to \rho_L$ impact factor. The $\gamma^*_L\to\rho_L$ impact factor starts at twist 2, and the following result holds up to twist 3 as all contributions of twist 3 are zero. The only contribution involves the twist 2 correlator,
\beq
\int\! \frac{d\lambda}{2\pi} e^{-i\lambda y} \langle \rho(p)|\bar\psi(\lambda n) \,\gamma_{\alpha}\, \psi(0)| 0 \rangle=m_{\rho} f_{\rho}\varphi_1(y)(e^*\cdot n) p_{1 \alpha}\,.
\eq
According to Eq.~(\ref{A-Lowest-Taylor}), the amplitude reads
\bea
\Phi^{\gamma^* \to \rho(0)}_{\rm 2-parton}&=&-\frac{1}{4} \!\int\! \!d y \! \int d^2 x_{\perp}\, \tilde{H}^{\gamma^\alpha}(y,x_{\perp}) \!\!\int\! \frac{d\lambda}{2\pi} e^{-i\lambda y} \langle \rho(p)|
\bar\psi(\lambda n) \,\gamma_{\alpha}\, \psi(0)| 0 \rangle \nonumber\\
&=&-\frac{m_{\rho} f_{\rho}}{4} \!\int\! \!d y \varphi_1(y)(e^*\cdot n) \!\int d^2 x_{\perp}\, \tilde{H}^{\sla p_1}(y,x_{\perp})\,.
\label{A-Lowest-Taylor-L}
\eea
 With the same kinematics, the hard part reads
\bea
p_{1\mu} H^{\gamma^\mu}_{\gamma^*_L}(y,\lb) \!=\! -\frac{4 Q e g^2 \delta^{ab}}{\sqrt{2} 2 N_c} y\,\yb \! \left(\! \frac{2}{\lb^2+\mu^2} -\frac{1}{(\lb+\kb)^2+\mu^2}-\frac{1}{(\lb-\kb)^2+\mu^2} \!\right)\!.
\label{hard-gammaL-rhoL}
\eea
The Fourier transform has thus the form
\bea
\tilde{H}^{\sla p_1}_{\gamma^*_L}(y,\xb) &=& -\frac{4 \,Q\, e}{(2\pi)\sqrt{2} \,2 N_c} \,y\,\yb\, K_0(\mu | \xb |)  g^2\, \delta^{ab}\mathcal{N}(\xb,\kb)\nonumber\\
&=& -\frac{g^2 \delta^{ab}\mathcal{N}(\xb,\kb)}{N_C \,2 \,\sqrt{2}} 
 \sum_{\lambda}\Psi_{0\,\lambda}^{\gamma^*}(\xb,y)\,,
\label{hard-Long}
\eea
where we identify the wave function of $\gamma^*_L$ 
\beq
\Psi_{0\,\lambda}^{\gamma^*}(\xb,y)=\frac{e}\pi \frac{\mu^2}Q K_{0}(\mu | \xb |)
\eq
 in accordance with Ref.~\cite{Ivanov:1998gk} (up to a normalization). The impact factor has thus the form
\bea
\Phi^{\gamma^*_L \to \rho_L}_{\rm 2-parton}&=&\!\!\frac{m_{\rho}f_{\rho}}{\sqrt{2}}\int\! \!d y \!\int d^2 \xb\, g^2 \delta^{ab}\mathcal{N}(\xb,\kb)\sum_{\lambda} \frac{(e^*_L\cdot n)\varphi_1(y) }{8 N_C} 
 \Psi_{0\,\lambda}^{\gamma^*}(\xb,y)\nonumber \\
&=& \!\frac{m_{\rho}f_{\rho}}{\sqrt{2}}\int\! \!d y \!\int d^2 \xb\, g^2 \delta^{ab}\mathcal{N}(\xb,\kb)\sum_{\lambda} \phi_{0 \lambda} 
 \Psi_{0\,\lambda}^{\gamma^*}(\xb,y)
\label{sum-long}
\eea
where
 the
meson involved is again considered as a one flavour quark--antiquark state, with a wave function defined as
\beq
\phi_{0 \lambda}= \frac{e^*_L \cdot n}{8 N_c} \varphi_1(y)\,.
\label{def-phi-rhoL}
\eq

%
Note that performing the integral over $\xb$ we recover, as expected, the same result as in momentum space
\beas
\Phi^{\gamma^*_L\to\rho_L}&=& \frac{e g^2 \delta^{ab}f_{\rho} Q}{\sqrt{2} 2 N_c}\int dy \,y\yb \varphi_1(y)\int \frac{d^2 \xb}{2\pi} K_0(\mu | \xb |) (1-e^{i \kb \cdot \xb})(1-e^{-i \kb \cdot \xb})\\
&=&  \frac{2 e g^2 \delta^{ab}f_{\rho}}{\sqrt{2} 2 N_c Q}\int dy \,\varphi_1(y) \frac{\kb^2}{\kb^2+\mu^2}\,.
\eeas

\section{The 3-parton impact factor}
\label{Sec_3-parton}
\subsection{Collinear factorisation beyond leading twist, in the transverse coordinate space}
\label{SubSec_Collinear-3-body}

To perform a full twist 3 computation of the $\Phi^{\gamma^*_T\to\rho_T}$ impact factor, we have to go beyond the WW aproximation and take into account the $\rho-$meson 3-parton Fock state contribution.
\psfrag{Ga}[cc][cc]{$\gamma^*$}
\psfrag{Hqqg}[cc][cc]{$H^{\alpha}$}
\psfrag{Sqqg}[cc][cc]{$S_{\alpha}$}
\psfrag{G}[cc][cc]{}
\psfrag{Gm}[cc][cc]{}
\psfrag{HqqgG}[cc][cc]{$H^{\alpha\Gamma^{\beta}}$}
\psfrag{SqqgG}[cc][cc]{$S^{\Gamma_{\beta}}_{\alpha}$}
\psfrag{rho}[cc][cc]{$\rho$}
\begin{figure}[htb]
\begin{tabular}{ccc} \hspace{1cm}\epsfig{file=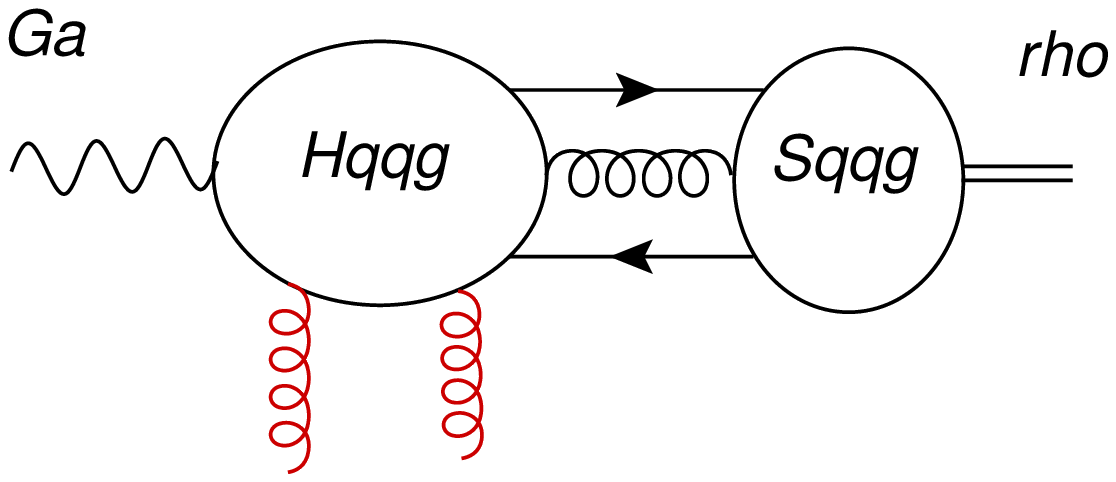,width=\widss} &\raisebox{1cm}{$\longmapsto$}& \epsfig{file=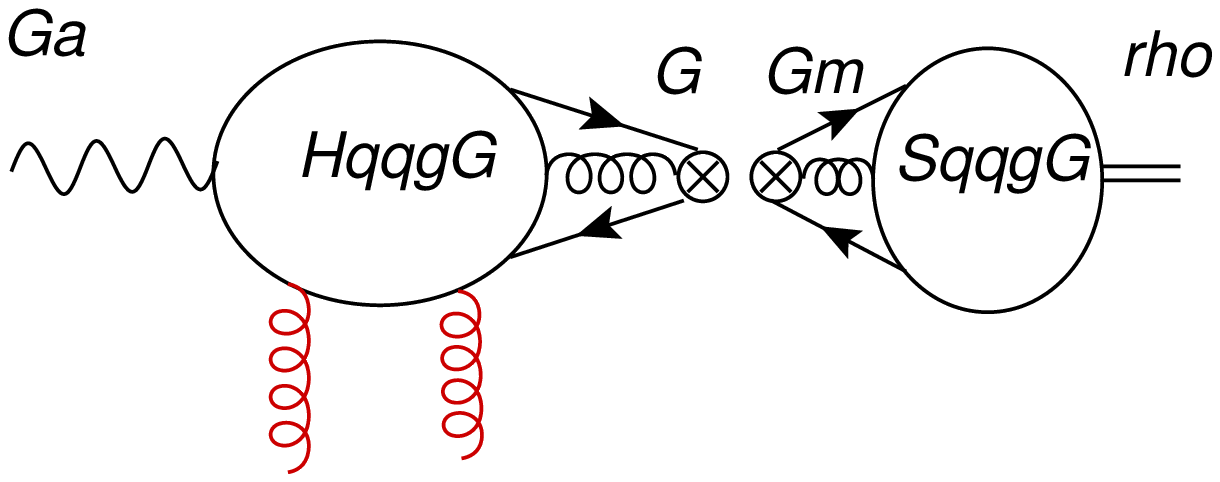,width=\widss}
\end{tabular}
\caption{Factorization with Fierz identity of the 3-parton $\gamma^*\to\rho$ impact factor.}
\label{Fig:Fierz3parton}
\end{figure}
The treatment that we will follow below for the 3-parton amplitude, goes globally in the same way as in the 2-parton case.

Using the Fierz identity we first factorise spinor and colour indices linking the hard part to the fields of the non-local correlator, see Fig.~\ref{Fig:Fierz3parton}, as
\bea
\label{phi-3-body}
\Phi^{\gamma^* \to \rho}_{3-parton}&=&\int \frac{d^4\ell_1}{(2 \pi)^4}  \frac{d^4\ell_g}{(2 \pi)^4} \, \text{tr}[H^\alpha(\ell_1,\ell_g) S_\alpha(\ell_1,\ell_g)] \nonumber\\
&=& 
-\frac{1}{4} \int \frac{d^4\ell_1}{(2 \pi)^4} \frac{d^4\ell_g}{(2 \pi)^4} \, \text{tr}[H^\alpha(\ell_1,\ell_g) \Gamma^{\beta}] \,S_\alpha^{\Gamma_{\beta}}(\ell_1,\ell_g)\,.
\eea
%
We use the Sudakov decomposition of the momenta of partons $i$, $\ell_i = \alpha_i p +\beta_i n +\ell_{i\perp}$ and for the Fourier conjugate coordinates $z_i=\alpha_{z_i}p+\beta_{z_i}n+z_{i \perp}$ in the argument of the non-local correlator defining the soft part $S.$
We factorise the amplitude in the momentum space, and reduce it to a convolution on the longitudinal fractions $y_i$ of the $\rho$ meson momentum $p$   carried by the partons. It reads 
\bea
\Phi^{\gamma^* \to \rho}_{3-parton}&=&
-\frac{1}{4} \int d y_1 d y_g \int \frac{d^4\ell_1}{(2 \pi)^4}
\frac{d^4\ell_g}{(2 \pi)^4}
 \,\text{tr}[H^\alpha(\ell_{1},\ell_{g}) \Gamma^{\beta}]\nonumber\\
&&\hspace{-2cm} \times   \int \frac{d\lambda_1}{2  \pi} e^{i\lambda_1(\ell_1\cdot n-y_1)}\int \frac{d\lambda_g}{2\pi}e^{i\lambda_g (\ell_g\cdot n-y_g)} \,\int d^4 z_1 e^{-i\ell_1\cdot z_1} \int d^4 z_g e^{-i\ell_g\cdot z_g}\nonumber\\
&&\hspace{-2cm}\times  
\langle \rho(p)|
\bar\psi(z_1) \,i \Gamma_{\beta}\,g A^{\perp}_{\alpha}(z_g) \psi(0)| 0 \rangle\,.
\eea
Using the Sudakov decompositions of the momenta $\ell_i$ and of the coordinates $z_i$
we obtain the expression
\bea
\Phi^{\gamma^* \to \rho}_{3-parton}&=&
 -\frac{1}{4} \int d y_1 d y_g \int \frac{d^2\ell_{1\perp}}{(2 \pi)^2} \frac{d^2\ell_{g\perp}}{(2 \pi)^2} \,\text{tr}[H^\alpha(y_1,y_g,\ell_{1\perp},\ell_{g\perp}) \Gamma^{\beta}]\nonumber\\
&&\hspace{-2cm}\times \int \frac{d\alpha_1 \,d\beta_1}{(2 \pi)^2} \int \frac{d\alpha_g \,d\beta_g}{(2 \pi)^2}
 \int \frac{d\lambda_1}{2  \pi} \frac{d\lambda_g}{2\pi} e^{-i\lambda_1 y_1-i\lambda_g y_g} \nonumber\\
&&\hspace{-2cm}\times \int d \alpha_{z_1} d \beta_{z_1} \int d \alpha_{z_g} d\beta_{z_g}\,e^{-i \alpha_1 (\beta_{z_1}-\lambda_1)-i\alpha_g(\beta_{z_g}-\lambda_g)} e^{-i\beta_1 \alpha_{z_1}-i\beta_{g} \alpha_{z_g}}\nonumber\\
&&\hspace{-2cm}\times \int d^2 z_{1 \perp} d^2 z_{g \perp} e^{-i\ell_{1\perp}\cdot z_{1\perp} -i\ell_{g\perp}\cdot z_{g\perp}} \langle \rho(p)|
\bar\psi(z_1) \,i \Gamma_{\beta}\,g A^{\perp}_{\alpha}(z_g) \psi(0)| 0 \rangle \,,\,
 \label{phi3zl-0}
\eea
in which after simple integrations over $\alpha_i$ and $\beta_i$ we obtain the following formula in which the hard
part has still a dependence on the transverse momenta of partons
%
%
\bea
\Phi^{\gamma^* \to \rho}_{3-parton}
&=&
 -\frac{1}{4} \int d y_1 d y_g \int \frac{d^2\ell_{1\perp}}{(2 \pi)^2} \frac{d^2\ell_{g\perp}}{(2 \pi)^2} \,\text{tr}[H^\alpha(y_1,y_g,\ell_{1\perp},\ell_{g\perp}) \Gamma^{\beta}]\nonumber\\
 &&\hspace{-1cm}\times  \int \frac{d\lambda_1}{2  \pi}  e^{-i\lambda_1 y_1} \,\int \frac{d\lambda_g}{2\pi} e^{-i\lambda_g y_g}\,\int d^2 z_{1 \perp} e^{-i\ell_{1\perp}\cdot z_{1\perp}}\,\int d^2 z_{g \perp} e^{-i\ell_{g\perp}\cdot z_{g\perp}}\nonumber\\
&&\hspace{-1cm} \times 
 \langle \rho(p)|\bar\psi(\lambda_1 n + z_{1\perp}) \,i \Gamma_{\beta}\,g A^{\perp}_{\alpha}(\lambda_g n +z_{g\perp}) \psi(0)| 0 \rangle \,.
 \label{phi3zl}
\eea
Let us denote by $H^{\alpha, \Gamma^{\mu}}(y_i,y_j,\ell_{i\perp},\ell_{j\perp})$ the trace which involves the hard scattering amplitude
\beq
H^{\alpha, \Gamma^{\beta}}(y_i,y_j,\ell_{i\perp},\ell_{j\perp})=\text{tr}[H^\alpha(y_i,y_j,\ell_{i\perp},\ell_{j\perp}) \Gamma^{\beta}]\,,
\eq
and let us introduce its Fourier transform $\tilde{H}^{\Gamma^{\beta}}(y_i,y_j,x_{i\perp},x_{j\perp})$  defined as
\bea
&&H^{\alpha, \Gamma^{\beta}}(y_i,y_j,\ell_{i\perp},\ell_{j\perp})\nonumber\\
&&=\int d^2 x_{i \perp} d^2 x_{j \perp} \tilde{H}^{\alpha,\Gamma^{\beta}}(y_i,y_j,x_{i\perp},x_{j\perp})\, e^{-i(\ell_{i \perp} \cdot x_{i\perp}+\ell_{j\perp}\cdot x_{j\perp})}\,.
\eea
%
 At twist 3, the 3-parton hard part contribution is given by the first term of the Taylor expansion around the collinear direction. Hence the 3-parton contribution to twist 3 is given by
\beq
H^{\alpha, \Gamma^{\beta}}(y_i,y_j,0_{\perp},0_{\perp})
 = \int d^2 x_{1 \perp} d^2 x_{g \perp} \tilde{H}^{\alpha,\Gamma^{\beta}}(y_1,y_g,x_{1\perp},x_{g\perp})\,.
\label{def-H-tilde}
\eq
%
Substituting the representation (\ref{def-H-tilde}) of the hard part in Eq.~(\ref{phi3zl}) and performing the integral over $\ell_{i\perp}$, we thus obtain
\bea
&&\Phi^{\gamma^* \to \rho}_{3-parton}=
 -\frac{1}{4} \int d y_1 d y_g \,\int d^2 x_{1 \perp} d^2 x_{g \perp} \tilde{H}^{\alpha,\Gamma^{\beta}}(y_1,y_g,x_{1\perp},x_{g\perp})\nonumber\\
 &\times & \int \frac{d\lambda_1}{2  \pi}  e^{-i\lambda_1 y_1} \,\int \frac{d\lambda_g}{2\pi} e^{-i\lambda_g y_g}
 \langle \rho(p)|\bar\psi(\lambda_1 n ) \,i \Gamma_{\beta}\,g A^{\perp}_{\alpha}(\lambda_g n ) \psi(0)| 0 \rangle \,.
 \label{phi3b}
\eea
%
The soft correlators appearing in Eq.~(\ref{phi3b}) are parametrised by
\bea
\label{def-B}
&&\int \frac{d\lambda_1}{2  \pi} \frac{d\lambda_g}{2\pi} e^{-i\lambda_1 y_1-i\lambda_g y_g} \langle \rho(p)|\bar\psi(\lambda_1 n) \,i \gamma_{\mu}\,g A^{\perp}_{\alpha}(\lambda_g n) \psi(0)| 0 \rangle \nonumber \\
&& \qquad = -i m_{\rho}  \, f_{\rho} \, \zeta^{V}_{3\rho} \, B(y_1,y_2)  \, p_{\mu} \,  e_{\rho \perp \alpha}\,,\\
\label{def-D}
&& \int \frac{d\lambda_1}{2  \pi} \frac{d\lambda_g}{2\pi} e^{-i\lambda_1 y_1-i\lambda_g y_g} \langle \rho(p)|\bar\psi(\lambda_1 n) \,i \gamma_5 \gamma_{\mu}\,g A^{\perp}_{\alpha}(\lambda_g n) \psi(0)| 0 \rangle \nonumber \\
&& \qquad = -i m_{\rho}  \, f_{\rho} \,\zeta^{A}_{3\rho}  \, i  \, D(y_1,y_2)  \, p_{\mu} \,  \varepsilon_{\alpha e_{\rho \perp} p n}\,.
\eea
Using these parametrisation in (\ref{phi3b}), we  obtain
\bea
\Phi^{\gamma^* \to \rho}_{3-parton}&=&
 \frac{i m_{\rho} f_{\rho}}{4} \int d y_1 d y_g \,\int d^2 x_{1 \perp} d^2 x_{g \perp} \nonumber\\
&\times &\left[\tilde{H}^{e_{\rho \perp},\sla p}(y_1,y_g,x_{1\perp},x_{g\perp})\, \zeta^{V}_{3\rho} B(y_1,y_1+y_g)\right.\nonumber\\
 &+&\left.\tilde{H}^{R_{\perp},\sla p \gamma_5}(y_1,y_g,x_{1\perp},x_{g\perp})\,i  \zeta^{A}_{3\rho} D(y_1,y_1+y_g)\right]\,,
\label{phi3c}
\eea
where we use the notations
\bea
\tilde{H}^{a,\Gamma^{\mu} b_{\mu}} &\equiv & \tilde{H}^{\alpha,\Gamma^{\mu}} \,a_{\alpha}\,b_{\mu}\,,\\
R_{\perp\alpha}&\equiv & \varepsilon_{\alpha \mu \nu \sigma}\, e_{\rho \perp}^{\mu}\, p^{\nu}\, n^{\sigma}\,.
\eea
It is more convenient to use the combinations $S(y_1,y_2)$ and $M(y_1,y_2)$ of 3-parton DAs defined in Ref.~\cite{Anikin:2011sa} as 
\bea
S(y_1,y_2)&=& \zeta^{V}_{3\rho} \, B(y_1,y_2)+\zeta^{A}_{3\rho} \,  D(y_1,y_2) 
\label{defS}\,,\\
M(y_1,y_2)&=& \zeta^{V}_{3\rho}  \, B(y_1,y_2)-\zeta^{A}_{3\rho} \,  D(y_1,y_2)\,.
\label{defM}
\eea
$S(y_1,y_2)$ and $M(y_1,y_2)$ satisfy the following relations
\bea
\label{symSM}
S(\yb_2,\yb_1)&=&-M(y_1,y_2)\,,\\
M(\yb_2,\yb_1)&=&-S(y_1,y_2)\,,\label{symMS}
\eea
due to the symmetry properties of the DAs $B(y_1,y_2)$ and $D(y_1,y_2)$ under $C-$parity transformations.
Using the combinations (\ref{defS}, \ref{defM}) of DAs, the 3-parton contribution (\ref{phi3c}) reads 
\bea
&&\Phi^{\gamma^* \to \rho}_{3-parton}=
 \frac{i m_{\rho} f_{\rho}}{4} \int d y_1 d y_g \,\int d^2 x_{1 \perp} d^2 x_{g \perp} \nonumber\\
&& \hspace{-1.3cm}\times\left[\frac{S(y_1,y_1+y_g)}{2}(\tilde{H}^{e_{\rho \perp},\sla p}(y_1,y_g,x_{1\perp},x_{g\perp})+i\,\tilde{H}^{R_{\perp},\sla p \gamma_5}(y_1,y_g,x_{1\perp},x_{g\perp}))\right.\nonumber\\
 &&\hspace{-1.3cm}+\left.\frac{M(y_1,y_1+y_g)}{2}(\tilde{H}^{e_{\rho \perp},\sla p}(y_1,y_g,x_{1\perp},x_{g\perp})-i\,\tilde{H}^{R_{\perp},\sla p \gamma_5}(y_1,y_g,x_{1\perp},x_{g\perp}))\right].\,
\label{phi3d}
\eea
The next section is mainly devoted to the computations of the Fourier transforms  $\tilde{H}^{e_{\rho \perp},\sla p}(y_1,y_g,x_{1\perp},x_{g\perp})\pm i\,\tilde{H}^{R_{\perp},\sla p \gamma_5}(y_1,y_g,x_{1\perp},x_{g\perp})$. 

\subsection{Diagrams and colour structure of the 3-parton hard part}
\label{SubSec_colour-3-body}

\subsubsection{
Kinematics}
\label{SubSubSec_kinematics}

To get information on the dependence of the Fourier transforms $\tilde H$'s on transverse coordinates, we need to extend the kinematics in order to include the transverse degrees of freedom of the momenta of the partons. The kinematics we use here, as in the 2-partons kinematics Eq.~(\ref{Kinematics-2body}), satisfies the  on-shell condition for each parton, and it reduces to the  kinematics of Ref.~\cite{Anikin:2009bf}  in the collinear limit.

\begin{figure}
\psfrag{q}[cc][cc]{$q$}
\psfrag{lq}[cc][cc]{ $\ell_1$}
\psfrag{lqb}[cc][cc]{$\ell_2$}
\psfrag{lg}[cc][cc]{ $\ell_g$}
\psfrag{k1}[cc][cc]{$k_1$}
\psfrag{k2}[cc][cc]{$k_2$}
\psfrag{HGa}[cc][cc]{$H^{a_{\perp},\sla b}$}
\centerline{\raisebox{-0.1cm}{\vspace{.2cm}
  \epsfig{file=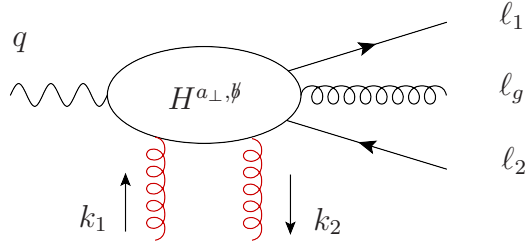,width=7cm}}}
 \caption{The kinematics for 3-parton contributions.}
\label{fig:kinematics-3-body}
\end{figure}
The Sudakov decomposition of the momenta of the partons are thus defined in analogy with Eq.~(\ref{Kinematics-2body}),
\bea
\ell_1 &=& y_1 p_1+ \ell_{1 \perp}+ \frac{\lb_1^2}{y_1 s} \, p_2 \,,
\label{kin-l1}\\
\ell_{2}&=& \yb_2 p_1 + \ell_{2 \perp}+ \frac{\lb_2^2}{\yb_2 s}\, p_2\,,
\label{kin-l2}\\
\ell_g &=& y_g p_1+ \ell_{g \perp}+ \frac{\lb_g^2}{y_g s} \,p_2 \,,
\label{kin-lg}
\eea
with $y_1+\yb_2+y_g=1$ and $\ell_{1\perp}+\ell_{2\perp}+\ell_{g\perp}= 0$. The resulting momentum of the $\rho$ meson is:
\bea
\label{p_rho-3-body}
p_{\rho}&=& p_1 + \frac{1}{s}\left(\frac{\lb_1^2}{y_1}+\frac{\lb_2^2}{\yb_2}+\frac{\lb_g^2}{y_g}\right) p_2 = p_1 + \frac{p_{\rho}^2}{s} p_2
\eea
and thus
\bea
p_{\rho}^2&=&\frac{\lb_1^2}{y_1}+\frac{\lb_2^2}{\yb_2}+\frac{\lb_g^2}{y_g}\,.
\eea
The momenta of the incoming photon and of the $t$-channel gluons are fixed according to Eqs.~(\ref{def-q}) and (\ref{def-k1k2}).

\subsubsection{Classification of the diagrams in colour dipole configurations} 
\label{SubSubSec_classification}

 Below we consider QCD with $SU(N_c)$ group, with $N_c$ {\em finite}.
The  factorised impact factor in momentum space  reads
\bea
&&\Phi^{\gamma^* \to \rho}_{3-parton,\, ext.}=
 \frac{i m_{\rho} f_{\rho}}{4} \int d y_1 d y_g \,  \nonumber\\
&& \hspace{-1.5cm}\times\left[\frac{S(y_1,y_1+y_g)}{2}(H^{e_{\rho \perp},\sla p}(y_1,y_g, \ell_{1\perp}, \ell_{g\perp})+i\, H^{R_{\perp},\sla p \gamma_5}(y_1,y_g,\ell_{1\perp}, \ell_{g\perp}))\right.\nonumber\\
 &&\hspace{-1.52cm}+\left.\frac{M(y_1,y_1+y_g)}{2}(H^{e_{\rho \perp},\sla p}(y_1,y_g, \ell_{1\perp},\ell_{g\perp})-i\, H^{R_{\perp},\sla p \gamma_5}(y_1,y_g, \ell_{1\perp}, \ell_{g\perp}))\right].\,
\label{phi3d-momentum}
\eea
Note that the above expression $\Phi^{\gamma^* \to \rho}_{3-parton, \, ext.}$ is an extension of the impact factor, in such a way that the hard part is treated in the  kinematics where the partons carry non zero transverse momenta. 
This $\Phi^{\gamma^* \to \rho}_{3-parton, \,ext.}$  mixes twist 3 terms (which are the only one remaining in the collinear limit $\ell_{1 \perp}=\ell_{g \perp}=0$), with higher twist terms introduced by the nonvanishing transverse momenta  $\ell_{1 \perp}$ and $\ell_{g \perp}$. 
%
%
\begin{figure}[h!]
\psfrag{u}{\hspace{-.1cm}$ y_1$}
\psfrag{d}{\raisebox{.05cm}{$\hspace{-.35cm} -\bar{y}_2$}}
\psfrag{m}{$y_g$}
\psfrag{i}{}
\def\widtlH{0.245\columnwidth}
\def\widtlS{0.24\columnwidth}
\scalebox{1}{\begin{tabular}{ccc}
\hspace{-0.5cm}\epsfig{file=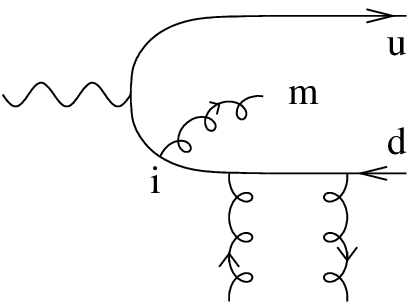,width=\widtlH}&\hspace{1cm}
\epsfig{file=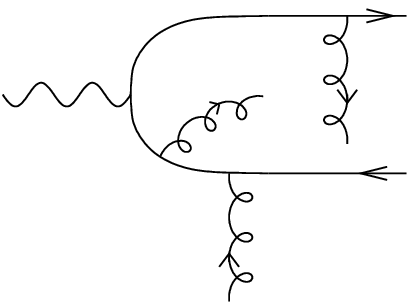,width=\widtlH}
& \hspace{1cm}\raisebox{.04cm}{\epsfig{file=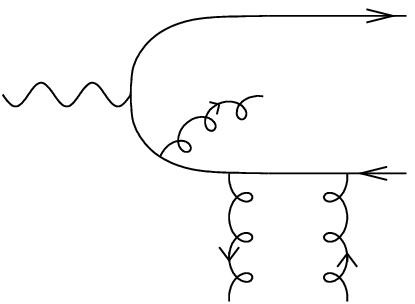,width=\widtlS}}\\
\\
\hspace{1.5cm} (aG1) &   \hspace{1.5cm} (bG1) &   \hspace{1.5cm} (eG1) \\
\\
\\
\psfrag{u}{}
\psfrag{d}{}
\psfrag{m}{}
\psfrag{i}{}
\hspace{-0.5cm}\raisebox{-.03cm}{\epsfig{file=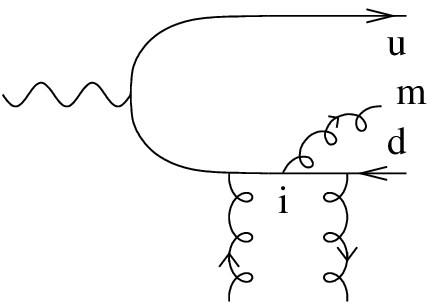,width=\widtlH}} &\hspace{1cm}
\raisebox{-.12cm}{\epsfig{file=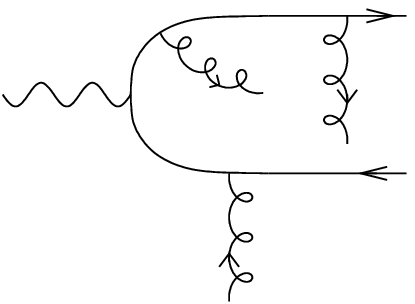,width=\widtlH}} & \hspace{1cm}\raisebox{-.07cm}{\epsfig{file=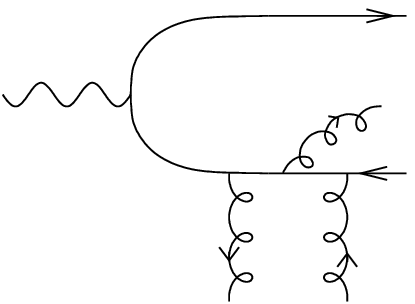,width=\widtlS}}
\\
\\
\hspace{1.5cm} (aG2) & \hspace{1.5cm}  (bG2)  &  \hspace{3cm} (eG2) \\
\\
\\
\hspace{-0.5cm}\epsfig{file=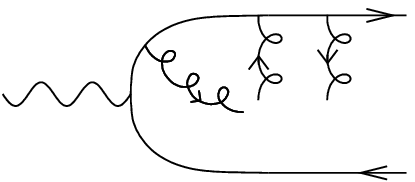,width=\widtlH} &\hspace{1cm}
\raisebox{-1.cm}{\epsfig{file=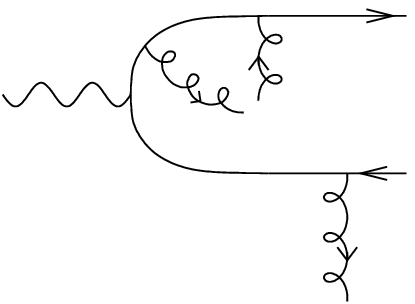,width=\widtlH}}
& \hspace{1cm}\raisebox{.02cm}{\epsfig{file=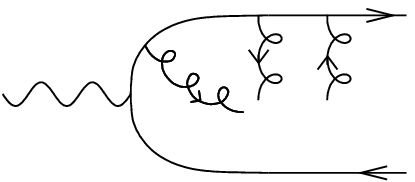,width=\widtlS}}\\
\\
\hspace{1.5cm} (cG1)  & \hspace{1.5cm} (dG1)  & \hspace{3cm} (fG1) \\
\\
\\
\hspace{-0.5cm}\epsfig{file=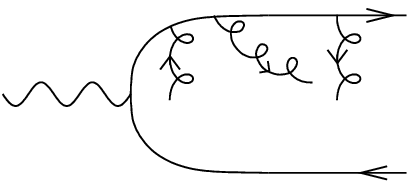,width=\widtlH} &\hspace{1cm}
\raisebox{-.98cm}{\epsfig{file=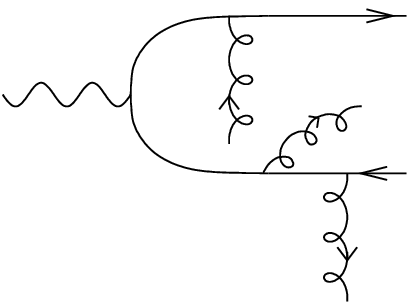,width=\widtlH}}  &\hspace{1cm}\raisebox{.03cm}{\epsfig{file=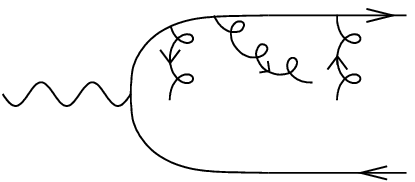,width=\widtlS}}\\
\hspace{1.5cm} (cG2)  & \hspace{1.5cm} (dG2)  &  \hspace{3cm} (fG2) \\
\end{tabular}}
\caption{The 12 ''Abelian`` (i.e. without triple gluon vertex) type contributions from  the hard scattering amplitude attached to the 3-parton correlators for the   $\gamma^* \to \rho$ impact factor, with momentum flux of external line, along $p_1$ direction.}
\label{Fig:3Abelian}
\end{figure}
%
%
%
%
%
%
%
\def\li{.17\columnwidth}
\def\si{\hspace{.8cm}}
\def\sci{\hspace{1.7cm}}
\psfrag{i}{}
\psfrag{u}{\raisebox{-.05cm}{\footnotesize\hspace{-.1cm}$ y_1$}}
\psfrag{d}{\raisebox{.05cm}{\footnotesize$\hspace{-.35cm} -\bar{y}_2$}}
\psfrag{m}{\footnotesize$\hspace{-.1cm}y_g$}
\begin{figure}[h!]
 \scalebox{1}{\begin{tabular}{cccc}
 \hspace{0.cm}\raisebox{0cm}{\epsfig{file=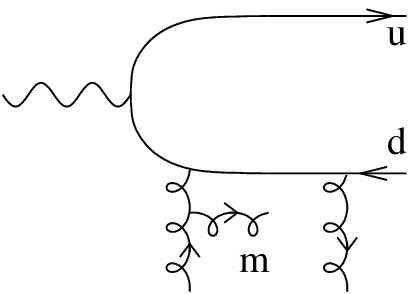,width=\li}} & \si
\psfrag{u}{}
\psfrag{d}{}
\psfrag{m}{}
 \raisebox{0cm}{\epsfig{file=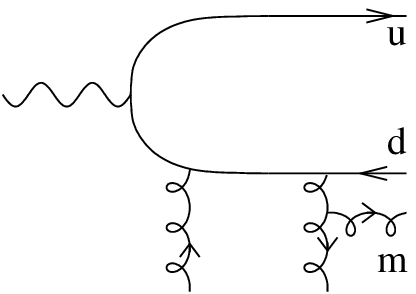,width=\li}} & \si
\psfrag{u}{}
\psfrag{d}{}
\psfrag{m}{}
\raisebox{0cm}{\epsfig{file=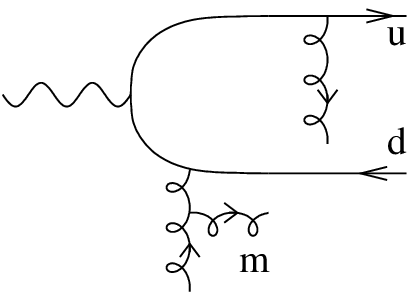,width=\li}}  & \si
\psfrag{u}{}
\psfrag{d}{}
 \psfrag{m}{}
 \raisebox{0.25cm}{\epsfig{file=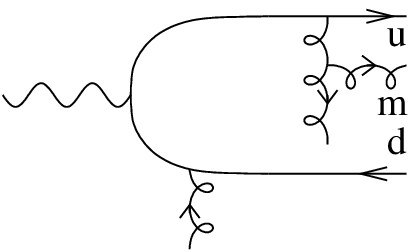,width=\li}}
\\
\\
\hspace{1.2cm} (atG1) & \sci (atG2) & \sci (btG1) & \sci (btG2)
\\
\\
\psfrag{u}{}
\psfrag{d}{}
\psfrag{m}{}
\hspace{0.cm}\raisebox{1cm}{\epsfig{file=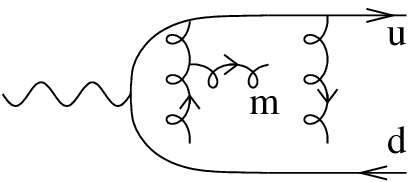,width=\li}} & \si
\psfrag{u}{}
\psfrag{d}{}
\psfrag{m}{}
 \raisebox{1cm}{\epsfig{file=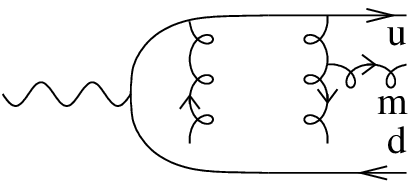,width=\li}} & \si
\psfrag{u}{}
\psfrag{d}{}
\psfrag{m}{}
\raisebox{0.61cm}{\epsfig{file=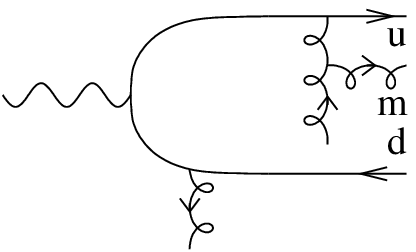,width=\li}}  & \si
\psfrag{u}{}
\psfrag{d}{}
\psfrag{m}{}
 \raisebox{0.37cm}{\epsfig{file=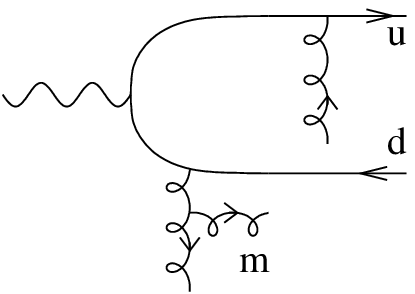,width=\li}}
\\
\hspace{1.2cm} (ctG1) & \sci (ctG2) & \sci (dtG1) & \sci (dtG2)
\\
\\
\psfrag{u}{}
\psfrag{d}{}
\psfrag{m}{}
\hspace{0.cm}\raisebox{0.26cm}{\epsfig{file=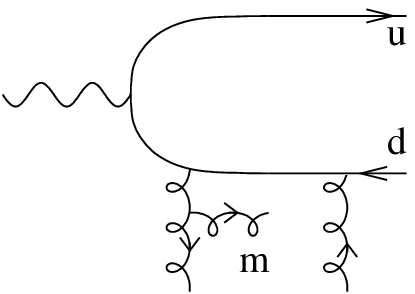,width=\li}} & \si
\psfrag{u}{}
\psfrag{d}{}
\psfrag{m}{}
 \raisebox{0.26cm}{\epsfig{file=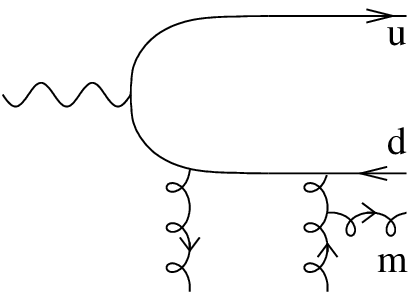,width=\li}} & \si
\psfrag{u}{}
\psfrag{d}{}
\psfrag{m}{}
\raisebox{.9cm}{\epsfig{file=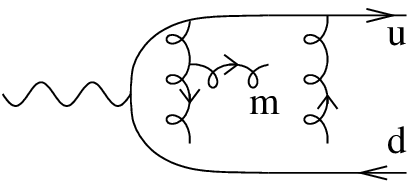,width=\li}}  & \si
\psfrag{u}{}
\psfrag{d}{}
\psfrag{m}{}
 \raisebox{.9cm}{\epsfig{file=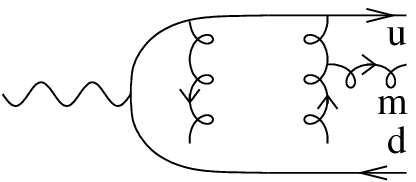,width=\li}}
\\
\hspace{1.2cm} (etG1) & \sci (etG2) & \sci (ftG1) & \sci (ftG2)
 \end{tabular}}
\caption{The 12 ''non-Abelian`` (with one triple gluon vertex) contributions from the hard scattering amplitude attached to the 3-parton correlators, for the   $\gamma^* \to \rho$ impact factor, with momentum flux of external line, along $p_1$ direction.}
\label{Fig:3NonAbelian}
\end{figure}

\def\li{.17\columnwidth}
\def\si{\hspace{.8cm}}
\def\sci{\hspace{1.7cm}}
\psfrag{i}{}
\psfrag{u}{\footnotesize$\hspace{-.1cm}y_1$}
\psfrag{d}{\footnotesize$\hspace{-.3cm}-\bar{y}_2$}
\psfrag{m}{\footnotesize$\hspace{0cm}y_g$}
\begin{figure}[htb]
 \scalebox{1}{\begin{tabular}{cccc}
 \hspace{0.cm}\raisebox{0cm}{\epsfig{file=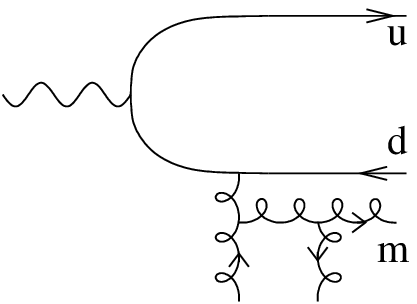,width=\li}} & \si
\psfrag{u}{}
\psfrag{d}{}
\psfrag{m}{}
 \raisebox{0cm}{\epsfig{file=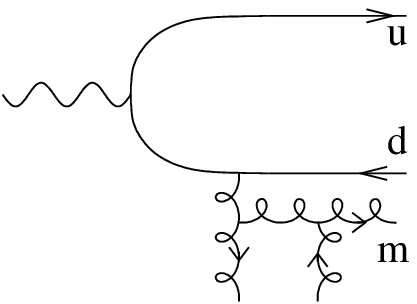,width=\li}} & \si
\psfrag{u}{}
\psfrag{d}{}
\psfrag{m}{}
\raisebox{.67cm}{\epsfig{file=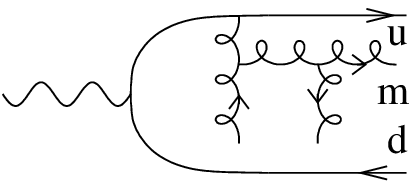,width=\li}}  & \si
\psfrag{u}{}
\psfrag{d}{}
\psfrag{m}{}
 \raisebox{.67cm}{\epsfig{file=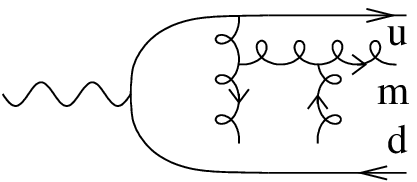,width=\li}}
\\
\\
\hspace{1.2cm} (gttG1) & \sci (gttG2) & \sci (httG1) & \sci (httG2)
 \end{tabular}}
\caption{The 4 ''non-Abelian`` (with two triple gluon vertices) contributions from the hard scattering amplitude attached to the 3-parton correlators, for the   $\gamma^* \to \rho$ impact factor, with momentum flux of external line, along $p_1$ direction.}
\label{Fig:3NonAbelianTwo}
\end{figure}



%
The panel of diagrams contributing to Eq.~(\ref{phi3d-momentum})
 is composed of 12 'Abelian' diagrams, shown in Fig.~\ref{Fig:3Abelian} and 12 'Non-abelian' diagrams with one gluon vertex, see Fig.~\ref{Fig:3NonAbelian} and 4 'Non-Abelian' diagrams with 2 triple-gluon vertices, see Fig.~\ref{Fig:3NonAbelianTwo}. Similarly to the 2-parton hard part computation, we perform the integral over $\kappa$, according to Eq.~(\ref{imfac}),
 using the residues method applied to the contour $\mathcal{C}_-$. 
 
 These calculations proceed in full analogy with the ones performed in \cite{Anikin:2009bf}. For completeness of the present paper, we present in the Appendix A in more details the calculation of the diagrams aG1 (of the Abelian types), atG1 (the non-Abelian type with one triple gluon vertex) and gttG1 (the non-Abelian type with 2 triple gluon vertex).
\begin{figure}[htb]
\psfrag{g}[cc][cc]{$g$}
\psfrag{qb}[cc][cc]{$\qb$}
\psfrag{kb}[cc][cc]{$\kb$}
\psfrag{q}[cc][cc]{$q$}
\psfrag{Ga}[cc][cc]{ $\gamma^*$}
\begin{tabular}{ccccc}
\hspace{1cm}\raisebox{-1cm}{\epsfig{file=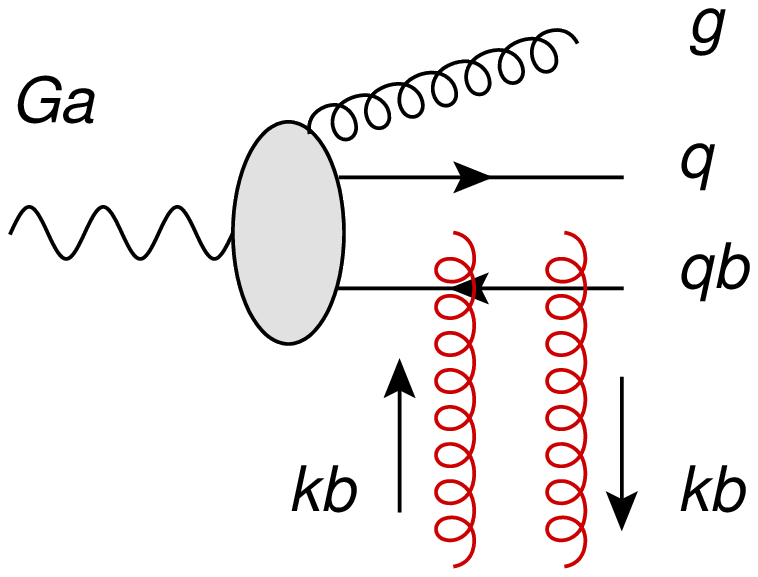,width=2.9cm}}& \raisebox{0cm}{+} &\raisebox{-1cm}{\epsfig{file=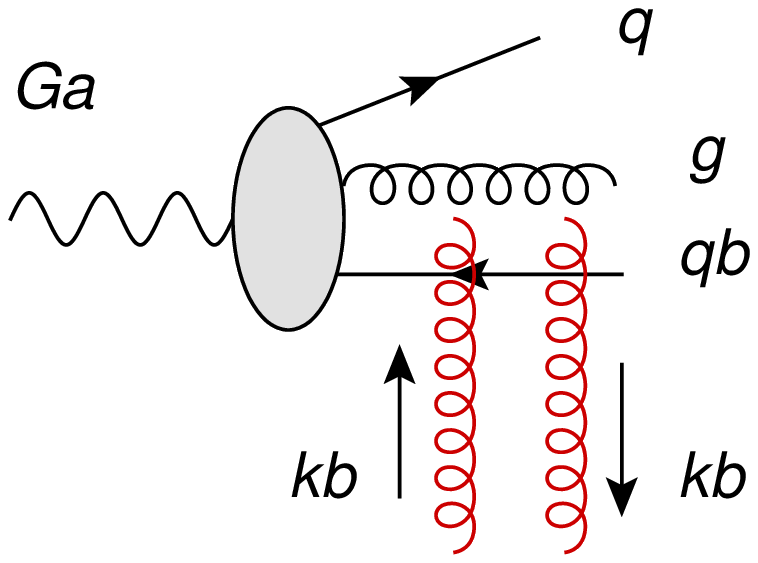,width=2.9cm}}& \raisebox{0cm}{+} &\raisebox{-1cm}{\epsfig{file=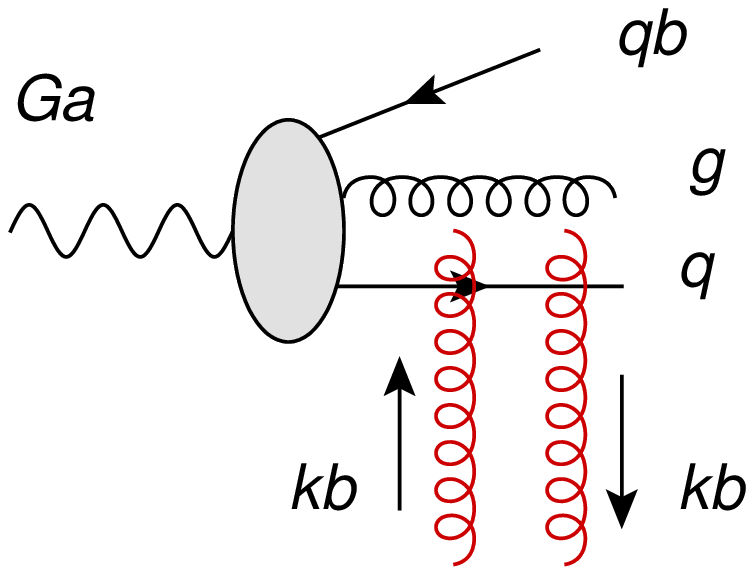,width=2.9cm}}
\end{tabular}
\caption{The $\{q \qb\}$, $\{q g\}$  and $\{\qb g\}$  dipoles interacting with the $t-$channel gluons.}
\label{fig:dipole-int}
\end{figure}
The degrees of freedom which we want to consider now are colour dipoles, made of a colour-anticolour singlet pair in the fundamental representation, constructed either from a $q \bar{q}$ pair, denoted by$\{q \qb\}$,  or from a quark and 
the ''antiquark part" of the gluon (denoted by $\{q g\}$), or from an antiquark and 
the ''quark part" of the gluon (denoted by $\{\qb g\}$).
We now group, see Fig.~\ref{fig:dipole-int}, the diagrams involving independently the 3 colour dipoles constituted by the pairs $\{q \qb\}$, $\{q g\}$  and $\{\qb g\}$. Due to the topology of the  associated diagrams, the dipole $\{q \qb\}$ is suppressed by $1/N_c^2$, the corresponding diagram being non-planar. We will justify in Section \ref{SubSubSec_DipoleColour} that these diagrams contain only colour dipole interactions with the $t-$channel gluons in colour space.

\begin{figure}[h!]
\psfrag{g}[cc][cc]{$g$}
\psfrag{qb}[cc][cc]{$\qb$}
\psfrag{q}[cc][cc]{$q$}
\psfrag{kb}[cc][cc]{$\kb$}
\psfrag{pkb}[cc][cc]{$\!\!\!\!\pm \kb$}
\psfrag{mkb}[cc][cc]{$\pm \kb$}
\psfrag{Ga}[cc][cc]{ $\gamma^*$}
\begin{tabular}{ccccc}
\hspace{1cm}\raisebox{-1cm}{\epsfig{file=QbGsystem.eps,width=2.9cm}}& \raisebox{0.2cm}{=} &\raisebox{-1cm}{\epsfig{file=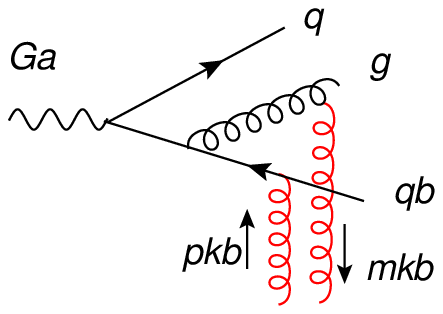,width=2.9cm}}& \raisebox{0.2cm}{+} &\raisebox{-1cm}{\epsfig{file=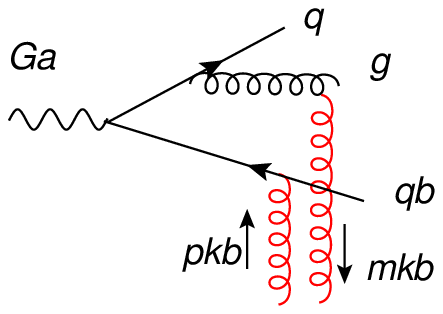,width=2.9cm}}\\
\\
& \raisebox{-.3cm}{+} &\raisebox{-1.5cm}{\epsfig{file=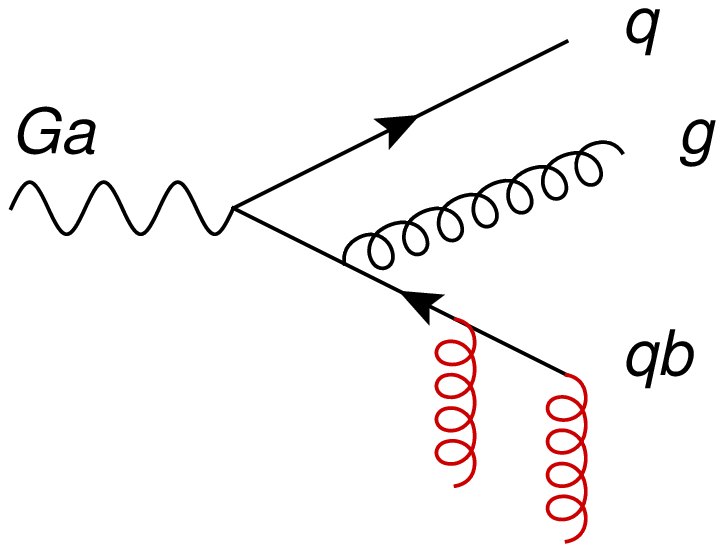,width=2.9cm}}& \raisebox{-.3cm}{+} &\raisebox{-1.5cm}{\epsfig{file=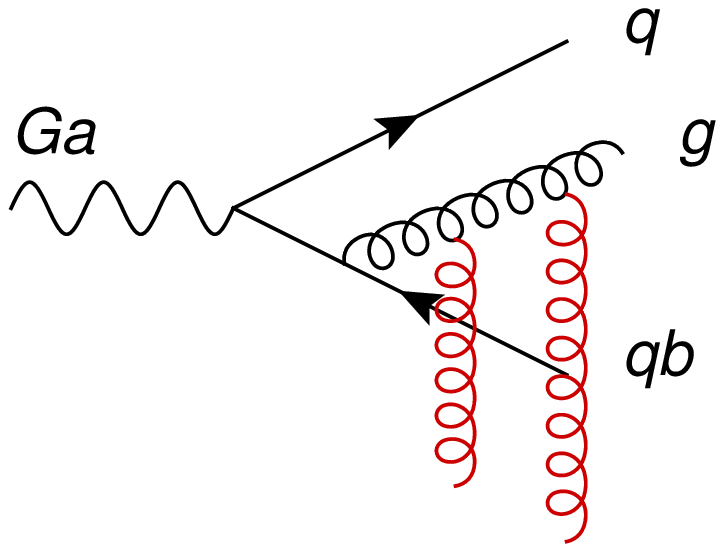,width=2.9cm}}
\end{tabular}
\caption{The  $\{\qb g\}$  dipole content.}
\label{fig:dipole-qb_g}
\end{figure}
\begin{figure}[h!]
\psfrag{g}[cc][cc]{$g$}
\psfrag{qb}[cc][cc]{$\qb$}
\psfrag{q}[cc][cc]{$q$}
\psfrag{kb}[cc][cc]{$\kb$}
\psfrag{Ga}[cc][cc]{ $\gamma^*$}
\begin{tabular}{ccccc}
\hspace{1cm}\raisebox{-1.2cm}{\epsfig{file=QQbsystem.eps,width=2.9cm}}& \raisebox{-.1cm}{=} &\raisebox{-1cm}{\epsfig{file=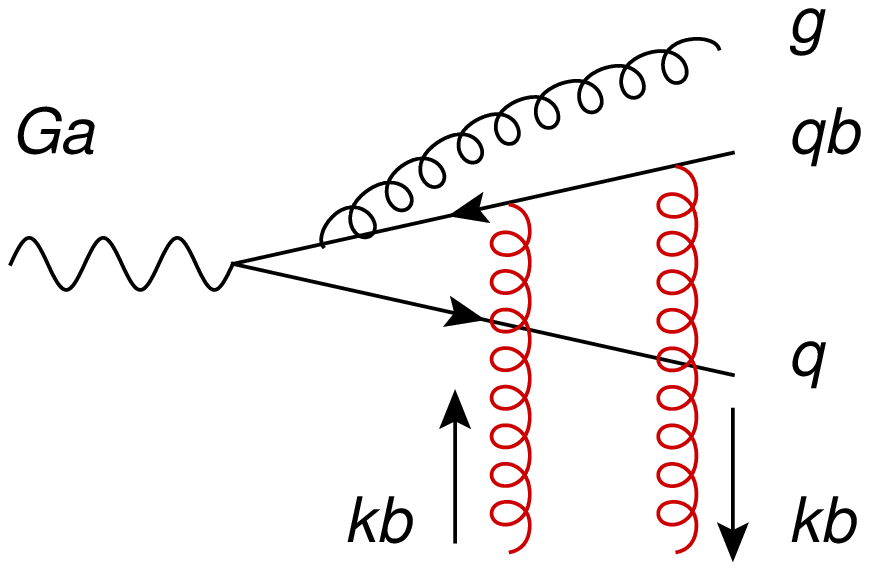,width=2.9cm}}& \raisebox{-.1cm}{+} &\raisebox{-1cm}{\epsfig{file=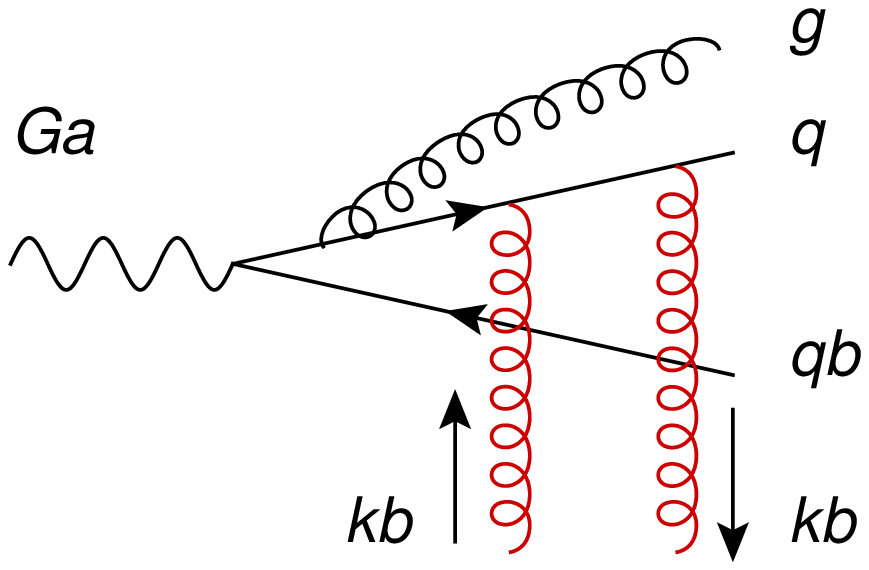,width=2.9cm}}\\
\\
& \raisebox{-.6cm}{+} &\psfrag{kb}[cc][cc]{$\pm \kb$}\raisebox{-1.5cm}{\epsfig{file=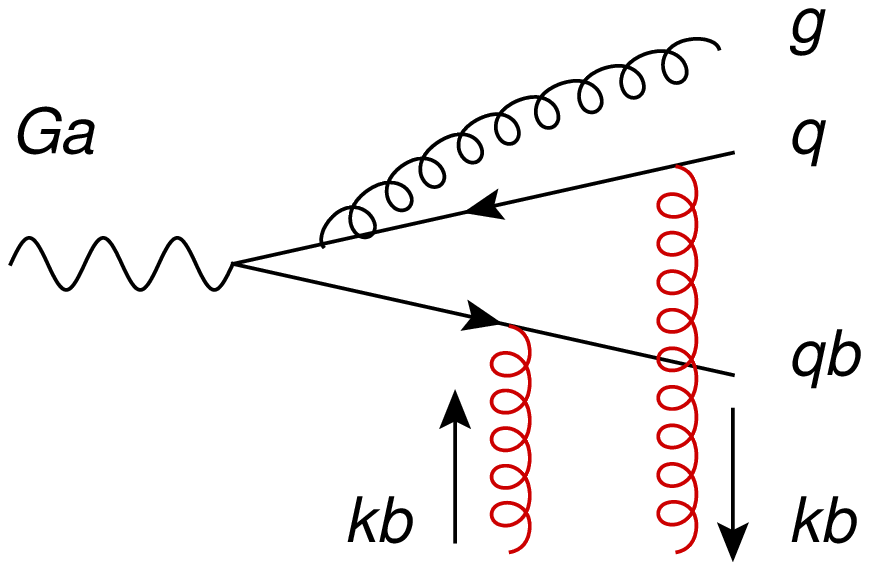,width=2.9cm}}& \raisebox{-.6cm}{+} &\raisebox{-1.5cm}{\epsfig{file=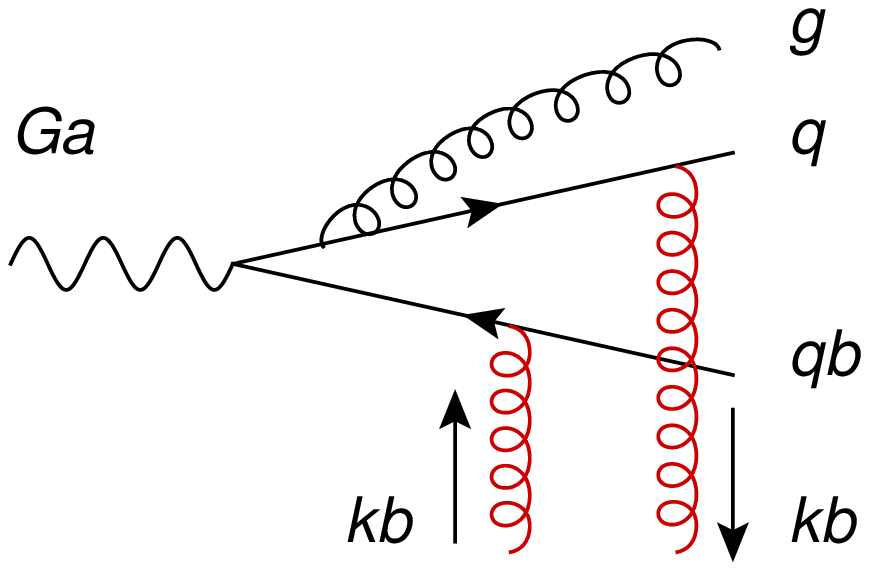,width=2.9cm}}
\end{tabular}
\caption{The  $\{q \qb \}$  dipole content.}
\label{fig:dipole-q_qb}
\end{figure}

The 6 diagrams corresponding to the interaction of the $\{\qb g\}$ system with the $t-$channel gluons are the contribution in $\frac{\delta^{ab}}{2 N_C}\frac{N_C}{C_F}$ of aG1, httG1, atG1, etG1, dtG1, btG2, shown in Fig.~\ref{fig:dipole-qb_g}. The results of the diagrams associated to the $\{qg\}$ system, cG1, gttG1, ctG1, ftG1, btG1, dtG2 are obtained from the diagrams of the $\{\qb g\}$ system by exchanging the role of the quark and the anti-quark.
The diagrams associated to the $\{q \qb\}$ system are the contribution in $\frac{\delta^{ab}}{2N_C} \left(\frac{N_C}{C_F}-2\right)$ of aG1, bG2, dG1 and the symmetric diagrams under exchange of the quark and the anti-quark, cG1, dG2, bG1, shown in Fig.~\ref{fig:dipole-q_qb}.

\subsubsection{Colour structure of the hard part}
\label{SubSubSec_DipoleColour}

In colour space, after factorisation of the hard part and of the soft part, two types of diagrams appear, the abelian diagrams corresponding to the $\{q\qb\}$ system and the non-abelian diagrams corresponding to the $\{qg\}$ and $\{\qb g\}$ systems. Using Fierz identity we can show that in colour space both type of diagrams are equivalent to a colour dipole interacting with the $t-$channel gluons.
For the abelian one, one gets
\beq
\psfrag{Ga}[cc][cc]{ $\gamma^*$}
\centerline{\begin{tabular}{ccc}
\raisebox{-1cm}{\epsfig{file=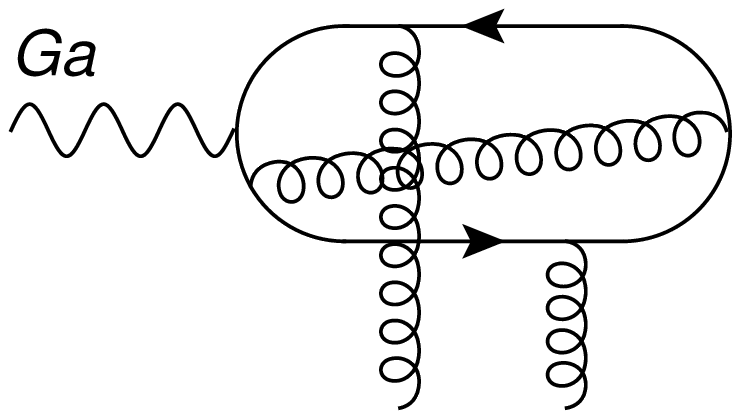,width=3cm}}& \raisebox{0cm}{= \hspace{0cm}{ $-\frac{1}{2}\frac{1}{N_C}$}} & \raisebox{-1cm}{\epsfig{file=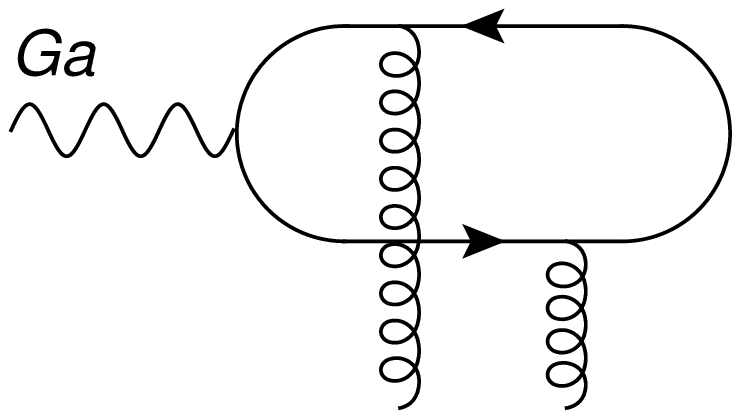,width=3cm}}
\end{tabular}}
\label{FactoFierzColourQQb}
\eq
while the non-abelian structure reduces to
\beq
\psfrag{Ga}[cc][cc]{ $\gamma^*$}
\centerline{\begin{tabular}{ccc}
\raisebox{-1cm}{\epsfig{file=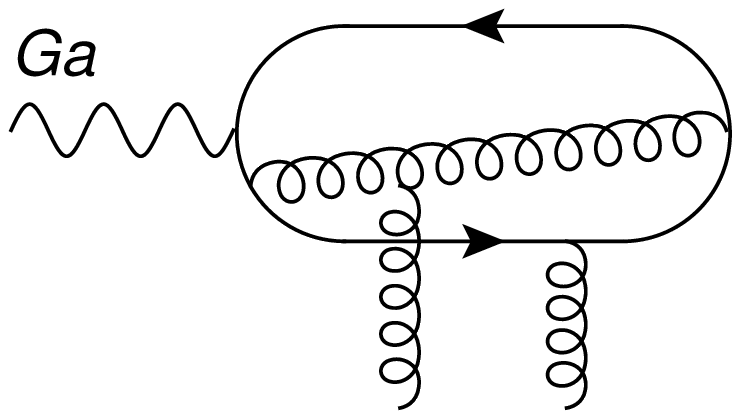,width=3cm}}& \raisebox{0cm}{= \hspace{0cm}$N_C$} & \raisebox{-1cm}{\epsfig{file=FactoFierzCouleurQG2.eps,width=3cm}}\,.
\end{tabular}}
\vspace{0cm}
\label{FactoFierzColourQG}
\vspace{0cm}
\eq
This second identity can be easily derived based on the relation
\beq
\label{fabctr}
 Tr ([t^a \,, t^ b] \,t^c) = \frac{i}{2} f_{abc}\,,
\eq
which can be represented graphically as
\beq
\label{fabctrG}
 \frac{i}{2} f_{abc}=
\raisebox{-0.3 \totalheight}{
\epsfysize=1.3cm{
\psfrag{a}[cc][cc]{$\,a$}
\psfrag{b}[cc][cc]{$b$}
\psfrag{c}[cc][cc]{$c$}
\epsfbox{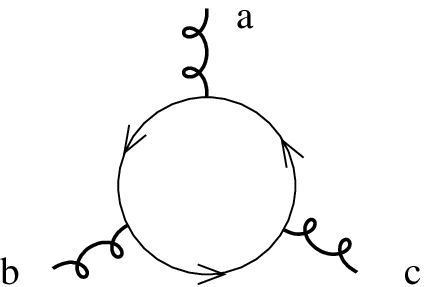}}}
-
\raisebox{-0.3 \totalheight}{
\epsfysize=1.3cm{
\psfrag{a}[cc][cc]{$\,a$}
\psfrag{b}[cc][cc]{$b$}
\psfrag{c}[cc][cc]{$c$}
\epsfbox{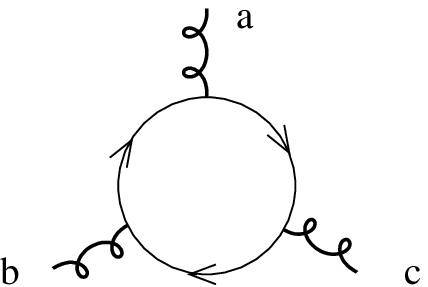}}}\,,
\eq
thus allowing to pass from the adjoint representation to the fundamental one.
We thus conclude from Eqs.~(\ref{FactoFierzColourQQb}, \ref{FactoFierzColourQG}) that in colour space we expect only dipole colour contributions even at finite $N_C$.
\subsection{Fourier transforms of the 3-parton diagrams in the collinear limit}
\label{SubSec_Fourier3partons}

In what follows we denote $\xb_i$ and $\lb_i$ respectively the transverse position and momentum of the parton $i$. In order to get connection with the dipole picture involving three partonic states and interacting with two $t-$channel gluons, it is important to identify the relative momentum $\lb_{ij}$ of the partons forming the dipole $\{ij\}$ interacting with these gluons. This will result for example in the appearance in the formulae (\ref{atG1V}, \ref{atG1A})   of the dependence on the center of mass momentum of the partons $\Lb_{ij}$ forming the interacting dipole. As we show later, this dependence in $\Lb_{ij}$ will affect the final form of formulae (\ref{atG1V}, \ref{atG1A})  after Fourier transform with respect to $\lb_{ij}$ and then taking the collinear limit. 
The 3-parton system carries two colour dipoles of momenta $\Lb_{ij}$ and $\lb_{ij}$. The interaction of the $t-$channel gluons having transverse momentum $\kb$ with the dipole ${ij}$ is encoded in the dependence on gluonic momenta having the form of the shift $\lb_{ij}+\kb$. 
Thus in our approach only information about a dipole of momentum $\lb_{ij}$ interacting with the $t-$channel
gluons survives the collinear limit. 

Following Ref.~\cite{Susskind:1967rg},
we identify the proper definitions of $\lb_{ij}$ and $\Lb_{ij}$ in a way inspired by the analogy that exists in the infinite momentum frame between the transformations of the sub-group F of Lorentz group leaving invariant the hypersurface orthogonal to the direction $p_1$ and the Galilei tranformations in nonrelativistic 2-dimensional mechanics. In Ref.~\cite{Susskind:1967rg}, a dictionary is established between the generators of the Galilean transformations of a 2-dimensional system of nonrelativistic particles of masses $m_i$, positions $x_i$ and momenta $\ell_i$, and the generators of the sub-group F of the system of particles of longitudinal fractions of momentum $y_i$, transverse positions $\xb_i$ and transverse momenta $\lb_i$. The analog of the mass $m_i$ of the parton $i$ is proportional to $y_i Q$ in our computation, where $y_i$ is respectively $y_1$, $\yb_2$ and $y_g$ for the quark, the antiquark and the gluon.



 
%

\begin{figure}[h!]
\centerline{\epsfig{file=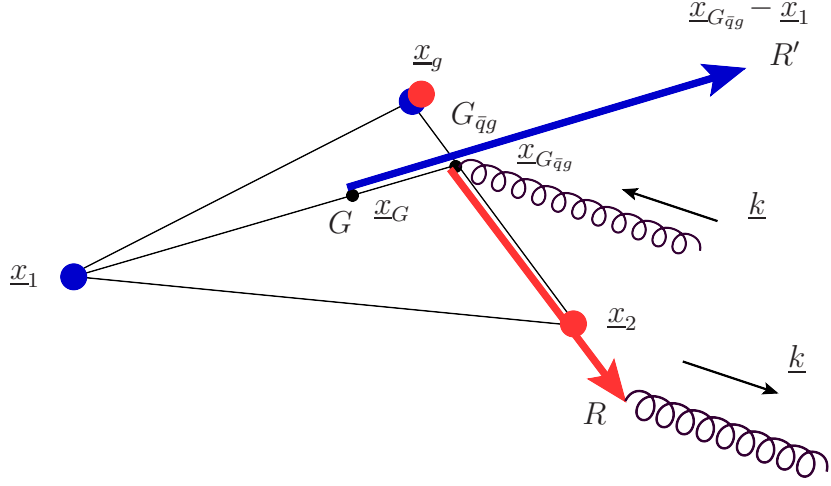,width=11cm}}
\caption{The dynamics of the colour dipoles, illustrated for the diagram atG1. Here $G$ is the center of mass of the 3-parton system, $G_{\qb g}$ is the center of mass of the system $\qb g$, $R'$ and $R$ are the respective reduced particles of the 2-body system $\{q \,G_{\qb g}\}$ and $\{\qb g\}$. The blue (black) dots symbolise the color $c_q$ and anticolor $\bar{c}_q$ charges, which make the spectator dipole. The red (grey) dots symbolise the color $c_\qb$ and anticolor $\bar{c}_\qb$ charges, which make the dipole interacting with the two $t-$channel gluons.}
\label{Fig:Dynamics-Dipoles}
\end{figure}

\begin{table}[htbp]
    \centering
 \begin{tabular}{|c|c|c|}
\hline
$\{\qb g\}$ & Center of mass $G_{\qb g}$ & Reduced particle $R_{\qb g}$\\ \hline
 & &\\
momenta & $\Lb_{\qb g}=\lb_2+\lb_g=-\lb_1$& $\lb_{\qb g}= \frac{y_g \, \lb_2-\yb_2 \, \lb_g}{\yb_2+y_g}=\frac{y_g \, \lb_2-\yb_2 \, \lb_g}{\yb_1}$\\ 
& & \\ \hline
& & \\
positions & $\xb_{G_{\qb g}} = \frac{\yb_2 \xb_2+y_g \xb_g}{\yb_1}$ & $\xb = \xb_2-\xb_g$ \\ 
& & \\ \hline
& &\\
masses & $ m_{G_{\qb g}} = (\yb_2+y_g) Q$ & $m_{R_{\qb g}}=\frac{\mu_{\qb g}^2}{Q}=\frac{\yb_2 y_g Q}{\yb_1}$ \\ 
& & \\ \hline
\end{tabular}
\caption{Kinematical variables of the center of mass $G_{\qb g}$ and of the reduced particle $R_{\qb g}$ for the system $\left\{\qb g\right\}$.}
\label{TableQbg}
\end{table}
Let us focus on the diagram atG1 as an example. The diagram atG1 corresponds to the interaction of the antiquark-gluon 2-parton system $\{\qb g\}$ with the $t$-channel gluons. 
 According to the above discussion, we treat the 2-parton system $\{\qb g\}$ as a 2-body non-relativistic system which is characterized by its center of mass $G_{\qb g}$ of position $\xb_{G_{\qb g}}$ and momentum $\Lb_{\qb g}$ and its reduced particle $R_{\qb g}$ of relative position $\xb$ from the center of mass. 
 We show in Table~\ref{TableQbg} the kinematical variables associated with $G_{\qb g}$ and $R_{\qb g}$ for the system $\left\{\qb g\right\}$. 
This is illustrated in Fig.~\ref{Fig:Dynamics-Dipoles}.
\psfrag{xg}[cc][cc]{$\xb_g$}
\psfrag{x2}[cc][cc]{$\xb_2$}
\psfrag{xG}[cc][cc]{$\xb_G$}
\psfrag{xGqg}[cc][cc]{$\xb_{G_{\qb g}}$}
\psfrag{x1xG}[cc][cc]{$\xb_{G_{\qb g}}\!-\xb_1$}
\psfrag{kb}[cc][cc]{$\kb$}
\psfrag{x1}[cc][cc]{$\xb_1$}
\psfrag{Gq}[cc][cc]{$G_{\qb g}$}
\psfrag{G}[cc][cc]{$G$}
\psfrag{R'}[cc][cc]{$R'$}
\psfrag{R}[cc][cc]{$R$}
%
%
The Fourier transform of the hard part associated to the 2-parton system ${\{\qb g\}}$ reads in this kinematics
\bea
\tilde{H}^{\{\qb g\}}(\xb_1,\xb_2,\xb_g)&=&\int \frac{d^2\Lb}{(2\pi)^2}\frac{d^2\lb_2}{(2\pi)^2}\frac{d^2\lb_g}{(2\pi)^2} H^{\{\qb g\}}(\Lb,\lb_2,\lb_g)\delta^2(\Lb)\nonumber\\
&&\times\exp(-i((\Lb-\lb_2-\lb_g)\cdot \xb_1 +\lb_2\cdot\xb_2+\lb_g\cdot\xb_g))\nonumber\\
&=&\int \frac{d^2\Lb}{(2\pi)^2}\frac{d^2\lb_2}{(2\pi)^2}\frac{d^2\lb_g}{(2\pi)^2} H^{\{\qb g\}}(\Lb,\lb_2,\lb_g)\delta^2(\Lb)\nonumber\\
&&\times\exp(-i(\Lb\cdot\xb_{G}+\Lb_{\qb g}\cdot(\xb_{G_{\qb g}}-\xb_1)+\lb_{\qb g}\cdot \xb))\nonumber\\
&=&\int \frac{d^2\lb_2}{(2\pi)^2}\frac{d^2\lb_g}{(2\pi)^2} H^{\{\qb g\}}(\lb_2,\lb_g)\nonumber\\
&&\times\exp(-i(\Lb_{\qb g}\cdot(\xb_{G_{\qb g}}-\xb_1)+\lb_{\qb g}\cdot \xb))
\eea
with $\Lb=\lb_1+\lb_2+\lb_g$ and $\xb_G=y_1\,\xb_1+\yb_2\,\xb_2+y_g\,\xb_g$ the momentum and the position of the center of mass $G$ of the 3-parton system. We now perform the change of variables ($\lb_2$, $\lb_g$) $\to$ ($\lb_{\qb g}$, $\Lb_{\qb g}$), which involves the Jacobian  $(\yb_2+y_g)/\yb_1=1$, leading to
\beq
\tilde{H}^{\{\qb g\}}(\xb_1,\xb_2,\xb_g)\!\!=\!\!\!\int \!\!\frac{d^2\lb_{\qb g}}{(2\pi)^2}\frac{d^2\Lb_{\qb g}}{(2\pi)^2} H^{\{\qb g\}}\!(\lb_{\qb g},\Lb_{\qb g})\!\exp(-i(\Lb_{\qb g}\cdot(\xb_{G_{\qb g}}-\xb_1)+\lb_{\qb g}\cdot \xb))\,.
\eq
%
%
Let us denote $F^{\{ij\}}_{\text{Diagr},V}(\lb_{ij},\Lb_{ij},y_i,y_j)$ and $F^{\{ij\}}_{\text{Diagr},A}(\lb_{ij},\Lb_{ij},y_i,y_j)$ respectively the vector and axial-vector contributions of the diagram 'Diagr' related to the 2-parton system ${\{ij\}}$. They read
\bea
\label{def-Fdiag-V}
F^{\{ij\}}_{\text{Diagr},V}(\lb_{ij},\Lb_{ij},y_i,y_j)&=&\frac{i m_{\rho} f_{\rho}}{4} \frac{1}{2}\left(S(y_1,y_2)+M(y_1,y_2)\right)\nonumber\\
&\times & H_{\text{Diagr}}^{{\{ij\}}\,e_{\rho \perp}, \,\sla p}(y_i,y_j,\lb_{ij},\Lb_{ij})\,,\\
F^{\{ij\}}_{\text{Diagr},A}(\lb_{ij},\Lb_{ij},y_i,y_j)&=&\frac{i m_{\rho} f_{\rho}}{4}\frac{1}{2}\left(S(y_1,y_2)-M(y_1,y_2)\right)\nonumber\\
&\times &  H_{\text{Diagr}}^{{\{ij\}}\,R_{\rho \perp},\, \sla p \gamma_5}(y_i,y_j,\lb_{ij},\Lb_{ij})\,.
\label{def-Fdiag-A}
\eea
The sum over all the diagrams associated to the 2-parton system $\{ij\}$ of all the 
$F^{\{ij\}}_{\text{Diagr},V}(\lb_{ij},\Lb_{ij},y_i,y_j)$ and $F^{\{ij\}}_{\text{Diagr},A}(\lb_{ij},\Lb_{ij},y_i,y_j)$ will be denoted by
 $F^{\{ij\}}(\lb_{ij},\Lb_{ij},y_i,y_j)$.
Note that after using the EOM, as shown in Section~\ref{SubSec_dipole-3-body-EOM} the ${\{ij\}}$ 2-parton system 
will be identified with the dipole ${\{ij\}}$.

For the diagram atG1 chosen as an example, we compute the  diagrams associated to $H_{\text{atG1}}^{\{\qb g\}\,e_{\rho \perp}, \,\sla p}(\yb_2,y_g,\lb_{\qb g},\Lb_{\qb g})$ and $H_{\text{atG1}}^{\{\qb g\}\,R_{\rho \perp}, \,\sla p \gamma_5}(\yb_2,y_g,\lb_{\qb g},\Lb_{\qb g})$ and we find respectively for the explicit expressions
\bea
F^{\{\qb g\}}_{atG1,V}(\lb_{\qb g},\Lb_{\qb g},\yb_2,y_g)&=&\frac{C^{ab} Q^2}{2} \frac{N_C}{C_F}\frac{1}{2}\left(S(y_1,y_2)+M(y_1,y_2)\right)\frac{1}{4\yb_1 Q}\nonumber\\
&&\hspace{-3cm}\times \left\{\eu\cdot \Pu \left( \frac{\Lb_{\qb g }^2}{2\mu_1^2/Q}+\frac{Q}2+\frac{\Lb_{\qb g}\cdot(\lb_{\qb g}+\kb)}{2\yb_1 \yb_2 Q}\right)\right.\nonumber\\
&&\hspace{-3cm}\left.-\frac{(\Lb_{\qb g}\cdot \Pu) (\eu\cdot (\lb_{\qb g}+\kb)))}{2\yb_2 y_g Q/(\yb_1+\yb_2)}-\frac{(\Lb_{\qb g}\cdot \eu) ( \Pu\cdot (\lb_{\qb g}+\kb))}{2\yb_2 Q}\right\}\nonumber\\
&&\hspace{-3cm}\times\frac{1}{\left(\frac{\Lb_{\qb g}^2}{2\mu_1^2/Q}+\frac{Q}2\right) \left( \frac{\Lb_{\qb g}^2}{2\mu_1^2/Q}+\frac{Q}2+\frac{(\lb_{\qb g}+\kb)^2}{2\mu_{\qb g}^2/Q} \right)}\,,\label{atG1V}
\eea
\bea
F^{\{\qb g\}}_{atG1,A}(\lb_{\qb g},\Lb_{\qb g},\yb_2,y_g)&=&\frac{C^{ab} Q^2}{2} \frac{N_C}{C_F}\frac{1}{2}\left(S(y_1,y_2)-M(y_1,y_2)\right)\frac{1}{4\yb_1 Q}\nonumber\\
&&\hspace{-3cm}\times \left\{\eu\cdot \Pu \left( \frac{\Lb_{\qb g }^2}{2\mu_1^2/Q}+\frac{Q}2-\frac{\Lb_{\qb g}\cdot(\lb_{\qb g}+\kb)}{2\yb_2 y_g Q/(\yb_1+\yb_2)}\right)\right.\nonumber\\
&&\hspace{-3cm}\left.+\frac{(\Lb_{\qb g}\cdot \eu) (\Pu\cdot (\lb_{\qb g}+\kb))}{2\yb_2 y_g Q/(\yb_1+\yb_2)}+\frac{(\Lb_{\qb g}\cdot \Pu) ( \eu\cdot (\lb_{\qb g}+\kb))}{2\yb_2 Q}\right\}\nonumber\\
&&\hspace{-3cm}\times\frac{1}{\left(\frac{\Lb_{\qb g}^2}{2\mu_1^2/Q}+\frac{Q}2\right) \left( \frac{\Lb_{\qb g}^2}{2\mu_1^2/Q}+\frac{Q}2+\frac{(\lb_{\qb g}+\kb)^2}{2\mu_{\qb g}^2/Q} \right)}\,,\label{atG1A}
\eea
with $\mu_1^2=y_1\yb_1 Q^2$.

It is interesting to note that  results (\ref{atG1V}, \ref{atG1A})
exhibit the appearance of two colour dipole dynamics, as shown in Fig.~\ref{Fig:Dynamics-Dipoles}.

The first dipole is made of the colour charge $c_\qb$ of the antiquark and the conjuguated charge present in the gluon $\bar{c}_\qb$. In momentum space, the colour $c_\qb$ is at rest and the conjuguate colour $\bar{c}_\qb$ is carried by the reduced particle $R_{\qb g}$ of the system $\{\qb g\}$ with momentum $\lb_{\qb g}$ and mass $m_{R_{\qb g}} =\mu_{\qb g}^2/Q$ giving a dipole of size $\xb=\xb_2-\xb_g$. This first dipole interacts with the $t-$channel gluon as the momentum $\lb_{\qb g}$ is shifted in $\lb_{\qb g}+\kb$.

 The second dipole is made of the colour charge $c_q$ of the quark and the anti-colour charge present in the gluon wave function $\bar{c}_q$. We can interpret that $c_q$ is carried by a particle at rest and $\bar{c}_q$ by the reduced particle $R'$ of the system constituted of the quark and the center of mass $G_{\qb g}$ of the $\{\qb g\}$ system. The particle $R'$ has an analoguous mass 
\beq
m_{R'}=\frac{\mu_1^2}Q= \left(\frac{1}{y_1}+\frac{1}{\yb_2+y_g}\right)^{-1} \frac{Q}{2}\,,
\label{def_mR'}
\eq
 a momentum 
\beq
\Lb_{R'}=\frac{(\yb_2+y_g)\Lb_{\qb g}-y_1\lb_1}{y_1+\yb_2+y_g}=\Lb_{\qb g}\,,
\label{def_LR'}
\eq
 and a transverse size $\xb_{R'}=\xb_G-\xb_1$. This dipole does not interact with the $t-$channel gluons in this diagram and thus its contribution will disappear in the collinear limit. 
 
\begin{figure}[h!]
\begin{tabular}{ccc}
\raisebox{-2cm}{\epsfig{file=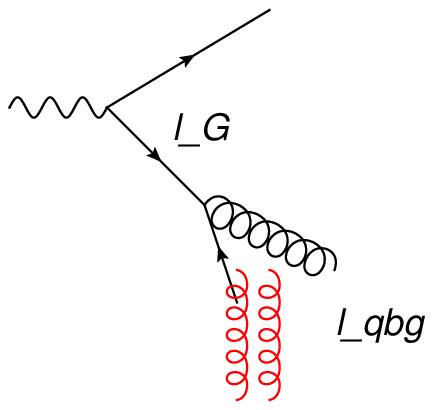,width=3.5cm}}& \raisebox{0cm}{$\longmapsto$} &\raisebox{-1cm}{\hspace{1cm}\epsfig{file=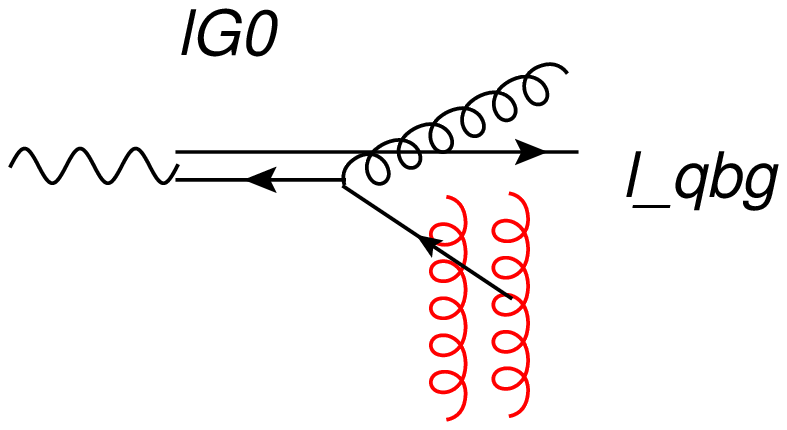,width=4cm}}
\end{tabular}
\caption{The relevant momentum $\lb_{\qb g}$ of the interation with $t-$channel gluons in the collinear approximation.}
\label{fig:CollinearLqbg}
\end{figure}

The kinetic energies of the two reduced particles entering the dynamics of the dipoles
\bea
E_{c,R'}&=&\frac{\Lb_{\qb g}^2}{2 \mu_1^2/Q}\,,\\
E_{c,R_{\qb g}}&=&\frac{(\lb_{\qb g}+\kb)^2}{2 \mu_{\qb g}^2/ Q}\,,
\eea
appear explicitely in the two last denominators of Eqs.~(\ref{atG1V}, \ref{atG1A}).
In the collinear limit up to twist 3 accuracy only information about a dipole formed by the reduced
particle interacting with $t-$channel gluons survives in the transverse collinear space.
%
We can then simplify the results of Eqs.~(\ref{atG1V}, \ref{atG1A}), by taking the limit $\Lb_{\qb g}=\underline{0}$, as illustrated Fig.~\ref{fig:CollinearLqbg}, which gives 
%
%
%
\bea
F^{\{\qb g\}}_{atG1,V}(\lb_{\qb g},\underline{0},\yb_2,y_g)&=&\frac{C^{ab}}{2} \frac{N_C}{C_F}\frac{1}{2}\left(S(y_1,y_2)+M(y_1,y_2)\right)\nonumber\\
&&\times \frac{\mu_{\qb g}^2}{\yb_1}\frac{\eu\cdot \Pu}{2  \left(\mu_{\qb g}^2+(\lb_{\qb g}+\kb)^2\right)}\,,\label{semiLimFV}\\
F^{\{\qb g\}}_{atG1,A}(\lb_{\qb g},\underline{0},\yb_2,y_g)&=&\frac{C^{ab}}{2} \frac{N_C}{C_F}\frac{1}{2}\left(S(y_1,y_2)-M(y_1,y_2)\right)\nonumber\\
&&\times \frac{\mu_{\qb g}^2}{\yb_1}\frac{\eu\cdot \Pu}{2  \left(\mu_{\qb g}^2+(\lb_{\qb g}+\kb)^2\right)}\,.\label{semiLimFA}
\eea
\psfrag{l_G}[cc][cc]{$\Lb_{\qb g}$}
\psfrag{l_qbg}[cc][cc]{$\lb_{\qb g}$}
\psfrag{lG0}[cc][cc]{$\Lb_{\qb g}=\underline{0}$}
We can now express Eqs.~(\ref{semiLimFV},~\ref{semiLimFA}) in terms of their Fourier transforms 
\bea
&&F^{\{\qb g\}}_{atG1,\,V}(\lb_{\qb g},\underline{0},\yb_2,y_g)=\frac{C^{ab}}{2} \frac{N_C}{C_F}\frac{1}{2}\left(S(y_1,y_2)+M(y_1,y_2)\right)\nonumber\\
&&\times\frac{\eu\cdot \Pu}{2 }\int \frac{d^2\xb}{(2 \pi)} \frac{\mu_{\qb g}^2}{\yb_1} K_0(\mu_{\qb g}\left|\xb \right|)\, e^{i (\lb_{\qb g}+\kb)\cdot \xb}\,,\\
&&F^{\{\qb g\}}_{atG1,\,A}(\lb_{\qb g},\underline{0},\yb_2,y_g)=\frac{C^{ab}}{2} \frac{N_C}{C_F}\frac{1}{2}\left(S(y_1,y_2)-M(y_1,y_2)\right)\nonumber\\
&&\times\frac{\eu\cdot \Pu}{2 }\int \frac{d^2\xb}{(2 \pi)} \frac{\mu_{\qb g}^2}{\yb_1} K_0(\mu_{\qb g}\left|\xb \right|)\,e^{i (\lb_{\qb g}+\kb)\cdot \xb}\,.
\eea
Finally, what remains in the collinear limit is 
\bea
&&F^{\{\qb g\}}_{atG1,V}(\underline{0},\underline{0},\yb_2,y_g)=\frac{C^{ab}}{2} \frac{N_C}{C_F}\frac{1}{2}\left(S(y_1,y_2)+M(y_1,y_2)\right)\nonumber\\
&&\times\frac{\eu\cdot \Pu}{2 }\int \frac{d^2\xb}{(2 \pi)} \frac{\mu_{\qb g}^2}{\yb_1} K_0(\mu_{\qb g}\left|\xb \right|)\,e^{i \kb\cdot \xb}\,,\\
&&F^{\{\qb g\}}_{atG1,A}(\underline{0},\underline{0},\yb_2,y_g)=\frac{C^{ab}}{2} \frac{N_C}{C_F}\frac{1}{2}\left(S(y_1,y_2)-M(y_1,y_2)\right)\nonumber\\
&&\times\frac{\eu\cdot \Pu}{2 }\int \frac{d^2\xb}{(2 \pi)} \frac{\mu_{\qb g}^2}{\yb_1} K_0(\mu_{\qb g}\left|\xb \right|)\,e^{i \kb\cdot \xb}\,.
\eea
The total contribution of the diagram atG1 is the sum of the above vector and  axial contributions
\bea
F^{\{\qb g\}}_{atG1}(\underline{0},\underline{0},\yb_2,y_g)=\frac{C^{ab}}{2} \frac{N_C}{C_F} S(y_1,y_2)\frac{\eu\cdot \Pu}{2 }\int \frac{d^2\xb}{(2 \pi)} \frac{\mu_{\qb g}^2}{\yb_1} K_0(\mu_{\qb g}\left|\xb \right|)\,e^{i \kb\cdot \xb}\,.\,
\eea

The computations of contributions of all the other
 3-parton diagrams go the same way. We first compute the diagrams associated to a dipole configuration, in terms of the center of mass and the reduced particle momenta and masses, to obtain $F^{\{ij\}}(\lb_{ij},\Lb_{ij},y_i,y_j)$. 
We compute then the Fourier transform $\tilde{f}^{\{ij\}}(\xb,y_i,y_j)$ of $F^{\{ij\}}(\lb_{ij},\Lb_{ij}=\underline{0},y_i,y_j)$ 
 as $\Lb_{ij}$ is never shifted by the $t-$channel gluon transverse momenta $\kb$. 
 Finally, the impact factor is expressed in terms of $\tilde{f}^{\{ij\}}(\xb,y_i,y_j)$, where $\xb=\xb_i-\xb_j$ is the size of the dipole $\{ij\}$, i.e. the variable conjugate to the momentum $\lb_{ij}$. 
The impact factor has thus the general form
\bea
\Phi^{\gamma^* \to \rho}_{3-parton}&=&\sum_{\{ij\}}\Phi^{\gamma^* \to \rho\,\{ij\}}_{3-parton}\\
&=&\sum_{\{ij\}}\int d y_i \,d y_j \int \frac{d^2 \xb}{(2\pi)^2}\tilde{f}^{\{ij\}}(\xb,y_i,y_j)\,.
\eea

The results of the diagrams in momentum space exhibit two kinds of structure denoted by $S_1(\lb,\mu)$ and $S_{2\,mn}(\lb,\mu_A,\mu_B)$
\bea
S_1(\lb,\mu)&=&\frac{\mu^2}{\lb^2+\mu^2}\,,\label{struct1}\\
S_{2\,mn}(\lb,\mu_A,\mu_B)&=&\frac{\lb_m\,\lb_n}{(\lb^2+\mu_A^2)(\lb^2+\mu_B^2)}\label{struct2}
\eea
where $m$ and $n$ are 2-dimensional euclidean indices and $\mu$, $\mu_A$, $\mu_B$ are the energies scales at stake. 
The Fourier transforms of formulae (\ref{struct1}) and (\ref{struct2}) are 
\beq
S_1(\lb,\mu)=\int \frac{d^2 \xb}{(2\pi)} \mu^2 K_0(\mu\left|\xb\right|)\,e^{i\lb\cdot\xb}\label{mu}
\eq
and
\bea
&&\hspace{-.8cm}S_{2\,mn}(\lb,\mu_A,\mu_B)\!\!=\!\!\int \frac{d^2 \xb}{(2\pi)} \frac{e^{-i\lb\cdot\xb}}{\mu_A^2-\mu_B^2} \frac{\partial}{\partial \xb_m}\frac{\partial}{\partial \xb_n} \left(K_0(\mu_A\left|\xb\right|)-K_0(\mu_B\left|\xb\right|)\right)\,e^{i\lb\cdot\xb}\nonumber\\
&&=\int \frac{d^2 \xb}{(2\pi)} \frac{1}{\mu_A^2-\mu_B^2}  \left\{\frac{\delta_{mn}}{2}\left[\mu_A^2 \left(\frac{K_0^{'}(\mu_A\left|\xb\right|)}{\mu_A\left|\xb\right|}+K_0^{''}(\mu_A\left|\xb\right|)\right)\right.\right.\nonumber\\
&&\left.\left.-\mu_B^2 \left(\frac{K_0^{'}(\mu_B\left|\xb\right|)}{\mu_B\left|\xb\right|}+K_0^{''}(\mu_B\left|\xb\right|)\right)\right]\right.\nonumber\\
&&+\left.\left(\frac{\delta_{mn}}{2}-\frac{\xb_m \xb_n}{\left|\xb\right|^2}\right) \left[\mu_A^2 \left(\frac{K_0^{'}(\mu_A\left|\xb\right|)}{\mu_A\left|\xb\right|}-K_0^{''}(\mu_A\left|\xb\right|)\right)\right.\right.\nonumber\\
&&\hspace{1.2cm}\left.\left.-\mu_B^2 \left(\frac{K_0^{'}(\mu_B\left|\xb\right|)}{\mu_B\left|\xb\right|}-K_0^{''}(\mu_B\left|\xb\right|)\right)\right]\right\}\,.
\label{int-l-l}
\eea
This expression can be simplified given that the modified Bessel function $K_{\nu}(\lambda)$ satisfies the equation
\beq
\lambda^2 K_{\nu}^{''}(\lambda)+\lambda K_{\nu}^{'}(\lambda)=(\lambda^2+\nu^2) K_{\nu}(\lambda)\,.
\label{Bessel-eq}
\eq
The expression (\ref{int-l-l}) thus reads
\bea
&&S_{2\,mn}(\lb,\mu_A,\mu_B)=\int \frac{d^2 \xb}{(2\pi)} \frac{e^{i\lb\cdot\xb}}{\mu_A^2-\mu_B^2} \nonumber\\
&&\left\{\frac{\delta_{mn}}{2}\left[\mu_A^2 K_0(\mu_A\left|\xb\right|)-\mu_B^2 K_0(\mu_B\left|\xb\right|)\right]\right.\nonumber\\
&&-\left.\left(\frac{\delta_{mn}}{2}-\frac{\xb_m \xb_n}{\left|\xb\right|^2}\right) \left[\mu_A^2 \, K_2(\mu_A\left|\xb\right|)-\mu_B^2\, K_2(\mu_B\left|\xb\right|)\right]\right\}\,,
\label{lilj}
\eea
where we have used the relation
\beq
-\frac{1}\lambda K_0'(\lambda) + K_0''(\lambda) = K_2(\lambda)
\label{cons-rec}
\eq
implied by the standard Bessel recursion formulas \cite{Gradshteyn}.
Let us note that the Fourier transforms in Eqs.~(\ref{mu}, \ref{lilj}) lead to the appearence of two functions in the
 3-parton impact factor, one associated with the spin non-flip transition $\Psi_{nf}$ and one associated to the spin flip transition $\Psi_{f}$:
\bea
\Psi_{\rm n.f.} &\propto&  \mu^2 K_0(\mu \left|\xb\right)\\
\Psi_{\rm f.} &\propto&  \mu^2 K_2(\mu \left|\xb\right)
\eea

\subsection{Spin non-flip and spin flip 3-parton impact factor}
\label{SubSec_SpinNonFlip}

In this section, we show the results we obtain for each colour dipole configuration interacting with the gluons in $t-$channel.

The sum of the contributions in $\frac{\delta^{ab}}{2N_C} \frac{N_C}{C_F}$ of the diagrams (cG1), (ctG1), (ftG1), (httG1), (btG1) and (dtG2), associated to the scattering amplitude of the $\{q g\}$ system on the $t-$channel gluons, 
leads to the impact factor
\bea
&&\hspace{-1cm}\Phi^{\gamma^* \to \rho}_{3-parton,\{qg\}}=\frac{C^{ab}}{2}\frac{N_C}{C_F}\int d y_1 d y_2 \,\int \frac{d^2 \xb}{(2\pi)} \mathcal{N}(\xb,\kb) \nonumber\\
&\times & \left\{\frac{\eu\cdot\Pu}{2}\left[\frac{M(y_1,y_2)}{y_2} \mu_{qg}^2 K_0(\mu_{qg}\left|\xb\right|)-\frac{S(y_1,y_2)}{\yb_1} \mu_1^2 K_0(\mu_1\left|\xb\right|) \right.\right.\nonumber\\
&+&\left.\left.\frac{\yb_1}{y_1\yb_2}M(y_1,y_2)\left[\mu_1^2K_0(\mu_1\left|\xb\right|)-\mu_{qg}^2K_0(\mu_{qg}\left|\xb\right|)\right]\right]\right.\nonumber\\
&+&\left.\left(\frac{\eu\cdot\Pu}{2}-\frac{\eu\cdot\xb\,\xb\cdot\Pu}{\left|\xb\right|^2}\right) \, \left( \frac{S(y_1,y_2)}{y_1}-\frac{M(y_1,y_2)}{\yb_2}\right)\right.\nonumber\\
&&\left.\hspace{1.5cm}\times\left[\mu_{qg}^2K_2(\mu_{qg}\left|\xb\right|)-\mu_1^2K_2(\mu_1\left|\xb\right|)\right]\right\}\nonumber\\
&+&\frac{C^{ab}}{2}\frac{N_C}{C_F}\int d y_1 d y_2 \,\int \frac{d^2 \xb}{(2\pi)} T_{\rm n.f.}\frac{S(y_1,y_2)}{\yb_1} \mu_1^2 K_0(\mu_1\left|\xb\right|)\,,
 \label{phi3h}
\eea
with $\mu_2^2=y_2\yb_2 Q^2$. Note that $\frac{\mu_2^2}{Q}=y_2\yb_2 Q$ is associated to the analoguous reduced mass of the 2-body system constituted by the antiquark and the center of mass of the quark and the gluon.
%
 We show in the Table~\ref{TableQg}, the kinematical variables associated to the center of mass $G$ and the reduced particle $R$ of the system $\{q g\}$ that we use to obtain, after simplifications described previously, the result (\ref{phi3h}). 
\begin{table}[htbp]
    \centering
 \begin{tabular}{|c|c|c|}
\hline
$\{q g\}$ & Center of mass $G$ & Reduced particle $R$\\ \hline
 & &\\
momenta & $\Lb_{qg}=\lb_1+\lb_g=-\lb_2$& $\lb_{q g}= \frac{y_1 \,\lb_g-y_g \,\lb_1}{y_g+y_1}=\frac{y_1 \,\lb_g-y_g \,\lb_1}{y_2}$\\ 
& & \\ \hline
& & \\
positions & $\xb_G = \frac{y_1 \xb_1+y_g \xb_g}{y_2}$ & $\xb = \xb_g-\xb_1$ \\ 
& & \\ \hline
& &\\
masses & $ m_G = y_2 Q$ & $m_R=\frac{\mu_{qg}^2}{Q}=\frac{y_1 y_g Q}{y_2}$ \\ 
& & \\ \hline
\end{tabular}
\caption{Kinematical variables of the center of mass $G$ and of the reduced particle $R$ of the system $\left\{q g\right\}$}
\label{TableQg}
\end{table}


The result for the $\{\qb g\}$ dipole is straightforward by exchanging the role of the quark and the antiquark i.e. exchanging $y_1$ and $\yb_2$ in (\ref{phi3h}). Adding the results for the $\{q g\}$ and for the $\{\qb g\}$ dipoles and using the symmetry properties of $S(y_1,y_2)$ and $M(y_1,y_2)$ given by Eqs.~(\ref{symSM}) and (\ref{symMS}), the spin non-flip part $\Phi^{\gamma^* \to \rho,\,\text{n.f.}}_{3-parton,\{qg\}+\{\qb g\}}$ and the spin flip part $\Phi^{\gamma^* \to \rho,\,\text{f.}}_{3-parton,\{qg\}+\{\qb g\}}$ read 
\bea
&&\hspace{-1cm}\Phi^{\gamma^* \to \rho,\,\text{n.f.}}_{3-parton,\{qg\}+\{\qb g\}}=\frac{-C^{ab}}{2}\frac{N_C}{C_F}\int d y_1 d y_2 \,\int \frac{d^2 \xb}{(2\pi)} \mathcal{N}(\xb,\kb) \frac{\eu\cdot\Pu}{2}\nonumber\\
&\times & \left\{ \frac{S(y_1,y_2)}{\yb_1}\left[\mu_1^2K_0(\mu_1\left|\xb\right|)+\mu_{\qb g}^2K_0(\mu_{\qb g}\left|\xb\right|)\right]\right.\nonumber\\
&-&\left. \frac{M(y_1,y_2)}{y_2}\left[\mu_2^2K_0(\mu_2\left|\xb\right|)+\mu_{qg}^2K_0(\mu_{qg}\left|\xb\right|)\right]\right.\nonumber\\
&+&\left.\left(\frac{y_2\yb_1}{y_1\yb_2}\right) \frac{S(y_1,y_2)}{\yb_1}\left[\mu_2^2K_0(\mu_2\left|\xb\right|)-\mu_{\qb g}^2K_0(\mu_{\qb g}\left|\xb\right|)
\right]\right.\nonumber\\
&-&\left.\left(\frac{y_2\yb_1}{y_1\yb_2}\right)\frac{M(y_1,y_2)}{y_2}\left[\mu_1^2K_0(\mu_1\left|\xb\right|)-\mu_{qg}^2K_0(\mu_{qg}\left|\xb\right|)\right]\right\}\nonumber\\
&+&\frac{C^{ab}}{2}\frac{N_C}{C_F}\int d y_1 d y_2 \,\int \frac{d^2 \xb}{(2\pi)} \frac{\eu\cdot\Pu}{2}\nonumber\\
&\times &\left[\frac{S(y_1,y_2)}{\yb_1}\mu_1^2K_0(\mu_1\left|\xb\right|)-\frac{M(y_1,y_2)}{y_2}\mu_2^2K_0(\mu_2\left|\xb\right|)\right]\,,
\label{phi3QQbarGNf}
\eea
and
\bea
&&\hspace{-1cm}\Phi^{\gamma^* \to \rho,\,\text{f.}}_{3-parton,\{qg\}+\{\qb g\}}=\frac{C^{ab}}{2}\frac{N_C}{C_F}\int d y_1 d y_2 \,\int \frac{d^2 \xb}{(2\pi)} \mathcal{N}(\xb,\kb) (\frac{\eu\cdot\Pu}{2}-\frac{\eu\cdot\xb\,\xb\cdot\Pu}{\left|\xb\right|^2})\nonumber\\
&\times & \left(\frac{S(y_1,y_2)}{y_1}-\frac{M(y_1,y_2)}{\yb_2}\right)\left[\mu_{qg}^2K_2(\mu_{qg}\left|\xb\right|)-\mu_1^2K_2(\mu_1\left|\xb\right|)
\right.\nonumber\\
&&\hspace{-.1cm}\left.+\mu_{\qb g}^2K_2(\mu_{\qb g}\left|\xb\right|)-\mu_2^2K_2(\mu_2\left|\xb\right|)
\right]\,.
\label{phi3QQbarGF}
\eea
The spin non-flip and spin flip impact factors associated to the scattering amplitude of the dipole $\{q\qb\}$ on the $t-$channel gluons are given by the contributions of type $\frac{\delta^{ab}}{2N_C} (\frac{N_C}{C_F}-2)$ from the diagrams (aG1), (cG1), (bG1), (dG2), (bG2), (dG1).
The results read
%
%
%
\bea
&&\Phi^{\gamma^* \to \rho,\, \text{n.f.}}_{3-parton,\{q\qb\}}=\frac{C^{ab}}{2}\left(\frac{N_C}{C_F}-2\right)
\int d y_1 d y_2 \,\int \frac{d^2 \xb}{(2\pi)} \mathcal{N}(\xb,\kb) \frac{\eu\cdot\Pu}{2}\nonumber\\
&\times & \left\{ \frac{S(y_1,y_2)}{\yb_1}\,\mu_1^2K_0(\mu_1\left|\xb\right|)-\frac{M(y_1,y_2)}{y_2}\,\mu_2^2K_0(\mu_2\left|\xb\right|)\right.\nonumber\\
&&-\left.\frac{S(y_1,y_2)}{y_g}\left[\left(\mu_1^2K_0(\mu_1\left|\xb\right|)-\mu_{q\qb}^2K_0(\mu_{q\qb}\left|\xb\right|)\right)\right.\right.\nonumber\\
&&\left.\left.\hspace{4cm}+\frac{y_2}{\yb_2}\left(\mu_2^2K_0(\mu_2\left|\xb\right|)-\mu_{q\qb}^2K_0(\mu_{q \qb}\left|\xb\right|)
\right)\right]\right.\nonumber\\
&&+\left.\frac{M(y_1,y_2)}{y_g}\left[\left(\mu_2^2K_0(\mu_2\left|\xb\right|)-\mu_{q\qb}^2K_0(\mu_{q\qb}\left|\xb\right|)
\right)\right.\right.\nonumber\\
&&\left.\left.\hspace{4cm}+\frac{\yb_1}{y_1}\left(\mu_1^2K_0(\mu_1\left|\xb\right|)-\mu_{q\qb}^2K_0(\mu_{q\qb}\left|\xb\right|)
\right)\right]\right\}\nonumber\\
&-&\frac{C^{ab}}{2}\left(\frac{N_C}{C_F}-2\right)\int d y_1 d y_2 \,\int \frac{d^2 \xb}{(2\pi)} \frac{\eu\cdot\Pu}{2}\nonumber\\
&\times & \left(\frac{S(y_1,y_2)}{\yb_1}\,\mu_1^2K_0(\mu_1\left|\xb\right|)-\frac{M(y_1,y_2)}{y_2}\,\mu_2^2K_0(\mu_2\left|\xb\right|)\right)
\label{phi3QQbarNf}
\eea
and
\bea
&&\hspace{-1cm}\Phi^{\gamma^* \to \rho,\,\text{f.}}_{3-parton,\{q\qb\}}\!=\!\frac{C^{ab}}{2}\!\left(\frac{N_C}{C_F}-2\right)\!\!\int\! d y_1 d y_2 \!\!\int\! \frac{d^2 \xb}{(2\pi)} \mathcal{N}(\xb,\kb) \!\left(\frac{\eu\cdot\Pu}{2}-\frac{\eu\cdot\xb\,\xb\cdot\Pu}{\left|\xb\right|^2}\right)\!\!\nonumber\\
&\times & 
\left(\frac{S(y_1,y_2)}{y_1}-\frac{M(y_1,y_2)}{\yb_2}\right)\left[\frac{\yb_2}{y_g}\left(\mu_{q\qb}^2K_2(\mu_{q\qb}\left|\xb\right|)-\mu_1^2K_2(\mu_1\left|\xb\right|)
\right)\right.\!\!\nonumber\\
&&\hspace{-.2cm}\left.+\frac{y_1}{y_g}\left(\mu_{q\qb}^2K_2(\mu_{q\qb}\left|\xb\right|)-\mu_2^2K_2(\mu_2\left|\xb\right|)
\right)\right]\,.\!\!
\label{phi3QQbarF}
\eea
We show in Table \ref{TableQqb} the kinematical variables associated to the system $\{q\qb\}$.
\begin{table}[htbp]
    \centering
 \begin{tabular}{|c|c|c|}
\hline
$\{q\qb\}$ & Center of mass $G$ & Reduced particle $R$\\ \hline
 & &\\
momenta & $\Lb_{q\qb}=\lb_1+\lb_2=-\lb_g$& $\lb_{q\qb}= \frac{\yb_2\, \lb_1-y_1 \,\lb_2}{y_1+\yb_2}=\frac{\yb_2 \,\lb_1-y_1  \,\lb_2}{\yb_g}$\\ 
& & \\ \hline
& & \\
positions & $\xb_G = \frac{y_1 \xb_1+\yb_2 \xb_2}{\yb_g}$ & $\xb = \xb_1-\xb_2$ \\ 
& & \\ \hline
& &\\
masses & $ m_G = \yb_g Q$ & $m_R=\frac{\mu_{q\qb}^2}{Q}=\frac{y_1 \yb_2 Q}{\yb_g}$ \\ 
& & \\ \hline
\end{tabular}
\caption{Kinematical variables of the center of mass $G$ and of the reduced particle $R$ of the system $\left\{q \qb\right\}$}
\label{TableQqb}
\end{table}
The total 3-parton results for the spin non-flip amplitude is thus
\bea
&&\hspace{-1cm}\Phi^{\gamma^* \to \rho,\,\text{n.f.}}_{3-parton}=-\frac{C^{ab}}{2}
\int d y_1 d y_2 \,\int \frac{d^2 \xb}{(2\pi)} \mathcal{N}(\xb,\kb) \frac{\eu\cdot\Pu}{2}\nonumber\\
&\times & \left\{2 
 \left[ \frac{S(y_1,y_2)}{\yb_1}\,\mu_1^2K_0(\mu_1\left|\xb\right|)-\frac{M(y_1,y_2)}{y_2}\,\mu_2^2K_0(\mu_2\left|\xb\right|)\right] \right.\nonumber\\
&+&\left.\frac{N_C}{C_F}
\left[\frac{S(y_1,y_2)}{\yb_1}\,\mu_{\qb g}^2K_0(\mu_{\qb g}\left|\xb\right|)-\frac{M(y_1,y_2)}{y_2}\,\mu_{qg}^2 K_0(\mu_{qg}\left|\xb\right|)\right.\right.\nonumber\\
&&+\left.\left.\left(\frac{y_2 \yb_1}{\yb_2 y_1}\right)\times\left( \frac{S(y_1,y_2)}{\yb_1}\left[\mu_2^2K_0(\mu_2\left|\xb\right|)-\mu_{\qb g}^2K_0(\mu_{\qb g}\left|\xb\right|)
\right]\right.\right.\right.\nonumber\\
&&\left.\left.\left.\hspace{1.5cm}-\frac{M(y_1,y_2)}{y_2}\left[\mu_1^2K_0(\mu_1\left|\xb\right|)-\mu_{qg}^2K_0(\mu_{qg}\left|\xb\right|)
\right]\right)\right]\right.\nonumber\\
&+&\left.\left(\frac{N_C}{C_F}-2\right)
\left[\frac{S(y_1,y_2)}{y_g} \left(\left[\mu_1^2K_0(\mu_1\left|\xb\right|)-\mu_{q\qb}^2K_0(\mu_{q\qb}\left|\xb\right|)
\right]\right.\right.\right.\nonumber\\
&&\left.\left.\left.\hspace{3cm}+\frac{y_2}{\yb_2}\left[\mu_2^2K_0(\mu_2\left|\xb\right|)-\mu_{q\qb}^2K_0(\mu_{q\qb}\left|\xb\right|)
\right]\right)\right.\right.\nonumber\\
&&-\left.\left.\frac{M(y_1,y_2)}{y_g} \left(\left[\mu_2^2K_0(\mu_2\left|\xb\right|)-\mu_{q\qb}^2K_0(\mu_{q\qb}\left|\xb\right|)
\right]\right.\right.\right.\nonumber\\
&&\left.\left.\left.\hspace{3cm}+\frac{\yb_1}{y_1}\left[\mu_1^2K_0(\mu_1\left|\xb\right|)-\mu_{q\qb}^2K_0(\mu_{q\qb}\left|\xb\right|)
\right]\right)\right]\right\}\nonumber\\
&+&\frac{C^{ab}}{2}\eu\cdot\Pu\int d y_1 d y_2 \left(\frac{S(y_1,y_2)}{\yb_1}-\frac{M(y_1,y_2)}{y_2}\right)\,,
\label{3bodyAvEom}
\eea
while the spin flip 3-parton impact factor is
\bea
&&\hspace{-1cm}\Phi^{\gamma^* \to \rho,\,\text{f.}}_{3-parton}=\frac{C^{ab}}{2}
\int d y_1 d y_2 \,\int \frac{d^2 \xb}{(2\pi)} \mathcal{N}(\xb,\kb) \left(\frac{\eu\cdot\Pu}{2}-\frac{(\eu\cdot\xb)\,(\xb\cdot\Pu)}{\left|\xb\right|^2}\right)\nonumber\\
&\times & \left(\frac{S(y_1,y_2)}{y_1}-\frac{M(y_1,y_2)}{\yb_2}\right)\left\{
\frac{N_C}{C_F}
\left[\mu_{qg}^2K_2(\mu_{qg}\left|\xb\right|)-\mu_1^2K_2(\mu_1\left|\xb\right|)\right.\right.\nonumber\\
&&\hspace{4cm}\left.\left.+\mu_{\qb g}^2K_2(\mu_{\qb g}\left|\xb\right|)-\mu_2^2K_2(\mu_2\left|\xb\right|)
\right]\right.\nonumber\\
&&+\left.\left(\frac{N_C}{C_F}-2\right)
\left[\frac{\yb_2}{y_g}\left(\mu_{q\qb}^2K_2(\mu_{q\qb}\left|\xb\right|)-\mu_1^2K_2(\mu_1\left|\xb\right|)
\right)\right.\right.\nonumber\\
&&\hspace{-.0cm}\left.\left.+\frac{y_1}{y_g}\left(\mu_{q\qb}^2K_2(\mu_{q\qb}\left|\xb\right|)-\mu_2^2K_2(\mu_2\left|\xb\right|)
\right)\right]\right\}\,.
\eea
In the formula (\ref{3bodyAvEom}), the last line is not proportional to the dipole factor $\mathcal{N}(\xb,\kb)$. In the following part, we will show that putting together the 2-parton result (beyond WW approximation) and the 3-parton result, all parts of the impact factor which do not have the dipole form cancel each other using the QCD EOM.
%
%
%
 This will extend to the full twist 3 result, the reasoning leading in the WW approximation from Eq.~(\ref{A-2body-1}) to Eq.~(\ref{A-2body-final}).
\section{The twist 3 $\gamma^*_T\to\rho_T$ impact factor in the dipole picture}
\label{Sec_dipole-3-body}

\subsection{The dipole picture arising from the equation of motion of QCD}
\label{SubSec_dipole-3-body-EOM}


The relation between DAs due to the QCD EOM, at full twist 3 (it extends the Eq.~(\ref{QCD-EOM-2-body}) beyond the WW approximation) reads
\bea
&&\int \frac{d y}{y \yb}\left(2 y \yb \varphi_3(y)+(y-\yb)\varphi_1^T(y)+\varphi_A^T(y)\right)\nonumber\\
&&\hspace{2.5cm}+\int d y_1 \int d y_2 \left(\frac{S(y_1,y_2)}{\yb_1}-\frac{M(y_1,y_2)}{y_2}\right)=0\,,
\label{EOM-full-3}
\eea
with $\varphi(y)=\varphi^{WW}(y)+\varphi^{gen}(y)$ being the complete DAs, i.e. which include both the WW and the genuine twist 3 contributions.
The 2-parton impact factor (\ref{A-2body-1}) before using the relations due to QCD EOM reads,  
\bea
&&\Phi^{\gamma^* \to \rho}_{2-parton}\!\!=\! \! \frac{-C^{ab}Q^2}{2}\eu\cdot\Pu \int dy \int \frac{d^2 \xb}{2 \pi}\mu K_1(\mu | \xb |)\mathcal{N}(\xb,\kb)\nonumber \\
&\times &\!\!\left(\left[(y-\yb)\varphi_1^T(y)-\varphi_A^T(y)\right]\eu \cdot \xb\frac{\xb\cdot \Pu}{| \xb |}+\varphi_A^T(y) \frac{\xb^2}{| \xb |} \eu\cdot \Pu \!\right)\nonumber\\
&+& \! \! \frac{C^{ab}}{2}\eu\cdot\Pu \int \frac{dy}{y\yb}\! \left[(2 y \yb \, \varphi_3(y)+(y-\yb)\varphi_1^T+\varphi_A^T(y))\right]
\label{A-2bodygen}
\eea
Collecting all terms arising from Eqs.~(\ref{3bodyAvEom}) and (\ref{A-2bodygen}) which are 
not proportional to the dipole factor, we observe that they cancel
\bea
&&\eu\cdot\Pu\frac{C^{ab}}{2}\left[\int \frac{d y}{y \yb}\left(2 y \yb \varphi_3(y)+(y-\yb)\varphi_1^T(y)+\varphi_A^T(y)\right)\right.\nonumber\\
&&\hspace{2.5cm}\left.+\int d y_1 \int d y_2 \left(\frac{S(y_1,y_2)}{\yb_1}-\frac{M(y_1,y_2)}{y_2}\right)\right]=0\,,
\eea
due to the relation (\ref{EOM-full-3}).
The final 2-parton impact factor is thus
\bea
\label{A-2body-final-tot}
&&\hspace{-1.2cm}\Phi^{\gamma^* \to \rho }_{{\rm 2-parton}}= -\frac{C^{ab} Q^2}{2} \int dy \int \frac{d^2 \xb}{2 \pi}\left\{\left[(y-\yb)\varphi_1^{T}(y)-\varphi_A^{T}(y)\right] \right.\nonumber \\
 &&\hspace{-1.4cm}\left.\times \, \frac{(\eu\cdot\xb)\,(\xb \cdot\Pu)}{| \xb |^2}+ \varphi_A^{T}(y) \eu \cdot \Pu \!\right\}
 \mu | \xb | K_1(\mu | \xb |) \, \mathcal{N}(\xb,\kb)\,,
\eea
and it can be decomposed into the spin non-flip and the spin-flip parts
\bea
\Phi^{\gamma^* \to \rho}_{\rm 2-parton, \,n.f.}&=&-\frac{C^{ab} Q^2}{2} \int dy \left(\varphi_A^{T}+(y-\yb)\varphi_1^{T} \right) \nonumber\\
 &&\hspace{-2cm}\times\int \frac{d^2\xb}{2\pi}\,\frac{1}{2} \,\eu\cdot \Pu  \, \mu | \xb |K_1(\mu | \xb |) \,\mathcal{N}(\xb,\kb)\,, 
\label{2body-non-flip-tot}
\eea
and
\bea
\Phi^{\gamma^* \to \rho}_{\rm 2-parton, \,f.}&=&-\frac{C^{ab} Q^2}{2}  \int dy \left(\varphi_A^{T}-(y-\yb)\varphi_1^{T} \right)\nonumber\\
 &&\hspace{-2cm}\times \int \frac{d^2\xb}{2\pi}\left(\frac{1}{2}\eu\cdot\Pu -\frac{(\eu\cdot \xb) (\Pu\cdot \xb)}{| \xb |^2} \right)  \mu | \xb | K_1(\mu | \xb |) \, \mathcal{N}(\xb,\kb)\,.
\label{2body-flip-tot}
\eea
The results (\ref{A-2body-final-tot}, \ref{2body-non-flip-tot}, \ref{2body-flip-tot}) are the extension of the formulae (\ref{A-2body-final}, \ref{2body-non-flip}, \ref{2body-flip}) to the full solution of the QCD EOM for the DAs, including the genuine twist 3 solutions.

The 3-parton spin non-flip result after using the relation (\ref{EOM-full-3}) is thus
\bea
&&\hspace{-1cm}\Phi^{\gamma^* \to \rho,\, \text{n.f.}}_{3-parton}=-\frac{C^{ab}}{2}
\int d y_1 d y_2 \,\int \frac{d^2 \xb}{(2\pi)} \mathcal{N}(\xb,\kb) \frac{\eu\cdot\Pu}{2}\nonumber\\
&\times & \left\{2 
 \left( \frac{S(y_1,y_2)}{\yb_1}\,\mu_1^2K_0(\mu_1\left|\xb\right|)-\frac{M(y_1,y_2)}{y_2}\,\mu_2^2K_0(\mu_2\left|\xb\right|)\right) \right.\nonumber\\
&+&\left.\frac{N_C}{C_F}
\left[\frac{S(y_1,y_2)}{\yb_1}\,\mu_{\qb g}^2K_0(\mu_{\qb g}\left|\xb\right|)-\frac{M(y_1,y_2)}{y_2}\,\mu_{qg}^2 K_0(\mu_{qg}\left|\xb\right|)\right.\right.\nonumber\\
&&+\left.\left.\left(\frac{y_2 \yb_1}{\yb_2 y_1}\right)\times\left( \frac{S(y_1,y_2)}{\yb_1}\left[\mu_2^2K_0(\mu_2\left|\xb\right|)-\mu_{\qb g}^2K_0(\mu_{\qb g}\left|\xb\right|)
\right]\right.\right.\right.\nonumber\\
&&\left.\left.\left.\hspace{2cm}-\frac{M(y_1,y_2)}{y_2}\left[\mu_1^2K_0(\mu_1\left|\xb\right|)-\mu_{qg}^2K_0(\mu_{qg}\left|\xb\right|)
\right]\right)\right]\right.\nonumber\\
&+&\left.\left(\frac{N_C}{C_F}-2\right)
\left[\frac{S(y_1,y_2)}{y_g} \left(\left[\mu_1^2K_0(\mu_1\left|\xb\right|)-\mu_{q\qb}^2K_0(\mu_{q\qb}\left|\xb\right|)
\right]\right.\right.\right.\nonumber\\
&&\left.\left.\left.\hspace{3cm}+\frac{y_2}{\yb_2}\left[\mu_2^2K_0(\mu_2\left|\xb\right|)-\mu_{q\qb}^2K_0(\mu_{q\qb}\left|\xb\right|)
\right]\right)\right.\right.\nonumber\\
&&\hspace{2cm}-\left.\left.\frac{M(y_1,y_2)}{y_g} \left(\left[\mu_2^2K_0(\mu_2\left|\xb\right|)-\mu_{q\qb}^2K_0(\mu_{q\qb}\left|\xb\right|)
\right]\right.\right.\right.\nonumber\\
&&\left.\left.\left.\hspace{3cm}+\frac{\yb_1}{y_1}\left[\mu_1^2K_0(\mu_1\left|\xb\right|)-\mu_{q\qb}^2K_0(\mu_{q\qb}\left|\xb\right|)
\right]\right)\right]\right\}\,,
\label{3bodyApEom}
\eea
and it can be rewritten in a more compact way, using the symmetry properties of the DAs under exchange of $y_1$ and $\yb_2$, as
\bea
&&\hspace{-1cm}\Phi^{\gamma^* \to \rho,\,\text{n.f.}}_{3-parton}=-C^{ab}
\int d y_1 d y_2 \,\int \frac{d^2 \xb}{(2\pi)} \mathcal{N}(\xb,\kb) \frac{\eu\cdot\Pu}{2}S(y_1,y_2)\nonumber\\
&\times & \left\{\frac{1}{\yb_1}\left(2\,\mu_1^2K_0(\mu_1\left|\xb\right|)+\frac{N_C}{C_F}\left[\,\mu_{\qb g}^2K_0(\mu_{\qb g}\left|\xb\right|)\right.\right.\right.\nonumber\\
&&\hspace{1.5cm}+\left.\left.\left.\left(\frac{y_2 \yb_1}{\yb_2 y_1}\right)\times \left[\mu_2^2K_0(\mu_2\left|\xb\right|)-\mu_{\qb g}^2K_0(\mu_{\qb g}\left|\xb\right|)
\right]\right]\right)\right.\nonumber\\
&+&\left.\frac{1}{y_g}\left(\frac{N_C}{C_F}-2\right)
\left[\left[\mu_1^2K_0(\mu_1\left|\xb\right|)-\mu_{q\qb}^2K_0(\mu_{q\qb}\left|\xb\right|)
\right]\right.\right.\nonumber\\
&&\left.\left.\hspace{2.5cm}+\frac{y_2}{\yb_2}\left[\mu_2^2K_0(\mu_2\left|\xb\right|)-\mu_{q\qb}^2K_0(\mu_{q\qb}\left|\xb\right|)
\right]\right]\right\}\,.
\label{3bodyApEom2}
\eea
\subsection{Equivalence with the results performed in momentum space in the light-cone collinear factorisation scheme}
\label{SubSec_3-body-eq-mom}

The integration of the spin non-flip result over $\xb$ is straightforward by using the relation
\beq
\int \frac{d^2 \xb}{(2\pi)} \mathcal{N}(\xb,\kb) \,\mu^2 K_0(\mu\left|\xb\right|)=\frac{2 \kb^2}{\kb^2+\mu^2}\,.
\eq
The integrated result thus gives
\bea
&&\Phi^{\gamma^* \to \rho,\, \text{n.f.}}_{3-parton}=-C^{ab}\int d y_1 d y_2 \,\eu\cdot\Pu \,S(y_1,y_2)\nonumber\\
&\times&\left\{\left(2-\frac{N_C}{C_F}\right)\frac{\alpha}{\yb_g\alpha+y_1\yb_2}\left(\frac{y_1^2}{\alpha+y_1\yb_1}+\frac{y_2\,\yb_2}{\alpha+y_2\yb_2}\right)\right.\nonumber\\
&&\left.+\frac{N_C}{C_F}\frac{\alpha}{\alpha\,\yb_1+\yb_2\, y_g}\frac{\alpha}{\alpha+y_2\,\yb_2}+\frac{2}{\yb_1} \frac{\alpha}{\alpha+y_1\yb_1}\right\}
\label{ApIntegration}
\eea
with $\alpha=\kb^2/Q^2$\,.
The result (\ref{ApIntegration}) is, as expected, identical to the one obtained in Ref.~\cite{Anikin:2009bf} using the light-cone collinear factorisation.

For the spin flip result, using the symmetry of the amplitude under the exchange of  $y_1$ and $\yb_2$, we get
\bea
\label{prep-integration-3body-sym_f}
&&\Phi^{\gamma^* \to \rho,\,\text{f.}}_{3-parton}=C^{ab}\int d y_1 d y_2 \,\int \frac{d^2 \xb}{(2\pi)} \mathcal{N}(\xb,\kb) \left(\frac{\eu\cdot\Pu}{2}-\frac{(\eu\cdot\xb)(\xb\cdot\Pu)}{\left|\xb\right|^2}\right)\nonumber\\
&\times & \left(\frac{S(y_1,y_2)}{y_1}\right)\left\{
\frac{N_C}{C_F}
\left[\mu_{qg}^2K_2(\mu_{qg}\left|\xb\right|)-\mu_1^2K_2(\mu_1\left|\xb\right|)\right.\right.\nonumber\\
&&\hspace{4cm}\left.\left.+\mu_{\qb g}^2K_2(\mu_{\qb g}\left|\xb\right|)-\mu_2^2K_2(\mu_2\left|\xb\right|)
\right]\right.\nonumber\\
&&+\left.\left(\frac{N_C}{C_F}-2\right)
\left[\frac{\yb_2}{y_g}\left(\mu_{q\qb}^2K_2(\mu_{q\qb}\left|\xb\right|)-\mu_1^2K_2(\mu_1\left|\xb\right|)
\right)\right.\right.\nonumber\\
&&\hspace{4cm}\left.\left.+\frac{y_1}{y_g}\left(\mu_{q\qb}^2K_2(\mu_{q\qb}\left|\xb\right|)-\mu_2^2K_2(\mu_2\left|\xb\right|)
\right)\right]\right\}\,.
\eea
We now integrate over $\xb=(\left|\xb\right| cos(\theta),\left|\xb\right| sin(\theta))$,
with $\kb=(\left|\kb\right| cos(\phi),\left|\kb\right| sin(\phi))\,.$
Using the fact that only the spin flip contributions are non zero, and based on the following identities
\bea
\frac{[(\eb_{\rho}^{-})^*\cdot\xb]\,[\xb\cdot\eb^{+}_{\gamma}]}{\left|\xb\right|^2}&=&\left(-i \frac{\left|\xb\right|e^{i\theta}}{\sqrt{2}}\right)\left(-i \frac{\left|\xb\right|e^{i\theta}}{\sqrt{2}}\right)\frac{1}{\left|\xb\right|^2} =-\frac{1}{2}e^{i 2 \theta}\,,\\
\label{Tfxpm}
\frac{[(\eb_{\rho}^{+})^*\cdot\xb]\,[\xb\cdot\eb^{-}_{\gamma}]}{\left|\xb\right|^2}&=& \left(i \frac{\left|\xb\right|e^{-i\theta}}{\sqrt{2}}\right)\left(i \frac{\left|\xb\right|e^{-i\theta}}{\sqrt{2}}\right)\frac{1}{\left|\xb\right|^2}=-\frac{1}{2}e^{-i 2 \theta}\,,
\label{Tfxmp}
\eea 
resulting from the definition of the polarisations in Eq.~(\ref{polarisation}), we obtain
\bea
&&\int \frac{d^2 \xb}{(2\pi)} \mathcal{N}(\xb,\kb) \, \mu^2 K_2(\mu\left|\xb\right|) \left(\frac{((\eb_{\rho}^{\mp})^*\cdot\xb)(\xb\cdot\eb^{\pm}_{\gamma})}{\left|\xb\right|^2}\right)\nonumber\\
&=& -\int d \lambda\, \lambda \, K_2(\lambda) \int \frac{d\theta}{2 \pi} 2\left(1-\cos\left[\frac{k \lambda}{\mu} \cos(\theta-\phi)\right]\right)\frac{1}{2}e^{\pm i 2 \theta}\nonumber\\
&=&-\frac{1}{2}e^{\pm i 2 \phi} \int d \lambda\, \lambda \, K_2(\lambda) \int \frac{d\theta}{2 \pi} \,2 \left(1-\cos\left[\frac{k \lambda}{\mu} \cos(\theta)\right]\right)e^{\pm i 2 \theta}\nonumber\\
&=&-\frac{1}{2}e^{\pm i 2 \phi}  2 \int d \lambda\, \lambda \,K_2(\lambda) J_2\left(\frac{k \lambda}{\mu} \right)\nonumber\\
&=& \frac{((\eb_{\rho}^{\mp})^*\cdot\kb)\,(\kb\cdot\eb^{\pm}_{\gamma})}{\left|\kb\right|^2} \,\frac{2 \kb^2}{\kb^2+\mu^2}\,,
\eea
with $\lambda=\mu\left|\xb\right|$. Hence the spin flip impact factor integrated over $\xb$ reads
\bea
&&\Phi^{\gamma^* \to \rho,\text{f}}_{3-parton}=-2 C^{ab} T_f \int d y_1 \, d y_2 \, S(y_1,y_2)\nonumber\\
&\times &\left\{\frac{N_C}{C_F}
\left[\alpha \yb_2 \left(\frac{y_1}{(\alpha+y_1\yb_1)(y_2\alpha+y_1y_g)}+\frac{\yb_2}{(\alpha+y_2\yb_2)(\yb_1\alpha+\yb_2y_g)}\right)\right]\right.\nonumber\\
&&+\left.\left(\frac{N_C}{C_F}-2\right)\left[\frac{\alpha \yb_2}{\yb_g\alpha+y_1\yb_2}\left(\frac{y_1}{\alpha+y_1\yb_1}+\frac{\yb_2}{\alpha+y_2\yb_2}\right)\right]\right\}\,.\label{phi3FapIntx}
\eea
which is the same result as the one obtained in Ref.~\cite{Anikin:2009bf}.

\subsection{Complete twist 3 result of the $\gamma^*_T\to\rho_T$ impact factor}
\label{SubSec_final-result}

Combining all the 2-parton and 3-parton results for the spin non-flip and spin flip impact factors $\Phi^{\gamma^* \to \rho}_{\rm n.f.}$, $\Phi^{\gamma^* \to \rho}_{\rm f.}$ of the $\gamma^*_T \to \rho_T$ transition, we finally obtain
\bea
\label{PhiNf}
&&\Phi^{\gamma^* \to \rho,\,\text{n.f.}}=-C^{ab} \,\int \frac{d^2 \xb}{(2\pi)}\, \mathcal{N}(\xb,\kb) \,\frac{\eu\cdot\Pu}{2}\nonumber\\
&\times & \left\{\frac{Q^2}{2}\int d y\,\left(\varphi_{AT}(y)+(y-\yb)\varphi_{1T}(y) \right)\mu | \xb | K_1(\mu | \xb |)\right.\nonumber\\
&&\left.+\int d y_1 \, d y_2\,\frac{S(y_1,y_2)}{\yb_1}\left(2\,\mu_1^2K_0(\mu_1\left|\xb\right|)+\frac{N_C}{C_F}\left[\,\mu_{\qb g}^2K_0(\mu_{\qb g}\left|\xb\right|)\right.\right.\right.\nonumber\\
&&\hspace{1.5cm}+\left.\left.\left.\left(\frac{y_2 \,\yb_1}{\yb_2 \,y_1}\right)\times \left[\mu_2^2K_0(\mu_2\left|\xb\right|)-\mu_{\qb g}^2K_0(\mu_{\qb g}\left|\xb\right|)
\right]\right]\right.\right.\nonumber\\
&&+\left.\left.\frac{1}{y_g}\left(\frac{N_C}{C_F}-2\right)\left[\left[\mu_1^2K_0(\mu_1\left|\xb\right|)-\mu_{q\qb}^2K_0(\mu_{q\qb}\left|\xb\right|)
\right]\right.\right.\right.\nonumber\\
&&\left.\left.\left.\hspace{2.5cm}+\frac{y_2}{\yb_2}\left[\mu_2^2K_0(\mu_2\left|\xb\right|)-\mu_{q\qb}^2K_0(\mu_{q\qb}\left|\xb\right|)\right]\right]\right)\right\}\,,
\eea
and
\bea
\label{Phif}
&&\Phi^{\gamma^* \to \rho,\,\text{f.}}=C^{ab} \,\int \frac{d^2 \xb}{(2\pi)} \mathcal{N}(\xb,\kb) \left(\frac{\eu\cdot\Pu}{2}-\frac{(\eu\cdot\xb)\,(\xb\cdot\Pu)}{\left|\xb\right|^2}\right)\nonumber\\
&\times & \left\{ -\frac{Q^2}{2}\int d y\,\left(\varphi_{AT}(y)-(y-\yb)\varphi_{1T}(y) \right)\mu | \xb | K_1(\mu | \xb |)  \right.\nonumber\\
&&\left.+\int d y_1 d y_2\,\left(\frac{S(y_1,y_2)}{y_1}\right)\left[
\frac{N_C}{C_F}
\left[\mu_{qg}^2K_2(\mu_{qg}\left|\xb\right|)-\mu_1^2K_2(\mu_1\left|\xb\right|)\right.\right.\right.\nonumber\\
&&\hspace{1.5cm}\left.\left.\left.+\mu_{\qb g}^2K_2(\mu_{\qb g}\left|\xb\right|)-\mu_2^2K_2(\mu_2\left|\xb\right|)
\right]\right.\right.\nonumber\\
&&+\left.\left.\left(\frac{N_C}{C_F}-2\right)
\left[\frac{\yb_2}{y_g}\left(\mu_{q\qb}^2K_2(\mu_{q\qb}\left|\xb\right|)-\mu_1^2K_2(\mu_1\left|\xb\right|)
\right)\right.\right.\right.\nonumber\\
&&\hspace{1.5cm}\left.\left.\left.+\frac{y_1}{y_g}\left(\mu_{q\qb}^2K_2(\mu_{q\qb}\left|\xb\right|)-\mu_2^2K_2(\mu_2\left|\xb\right|)
\right)\right]\right]\right\}\,.\,
\eea
The Eqs.~(\ref{PhiNf}) and (\ref{Phif}) are the main results of the present paper. We have achieved the goal of presenting the $\gamma^*_{L,T} \to \rho_{L,T}$ impact factor, in the forward limit, in a factorised form involving explicitly the coupling of two $t-$channel gluons with a dipole of transverse size $\xb$ produced during the transition of the $\gamma^*$ to the $\rho-$meson. 
The colour structure, after factorising a global normalization $1/(N_c^2-1)$, has the form
\beq
\Phi^{\gamma^* \to \rho} = \frac{\delta^{ab}}{(2 N_c) (N_c^2-1)} \left[ A_{\rm planar}\, N_c^2 + A_{\rm non-planar}\right]\,,
\label {planar}
\eq
where $A_{\rm planar}$ and $A_{\rm non-planar}$ are both $N_c$ independent.

This form of the impact factor will permit to include effects of saturation in the proton for exclusive $\rho$ meson electroproduction both for longitudinally and transversally polarized meson \cite{inPrep:pheno_saturation}. The $\phi^{WW}_{i\,\lambda}$'s defined by Eqs.~(\ref{A-2body-final-non-flip}, \ref{A-2body-final-flip}) could be extended beyond WW approximation, using Eqs.~(\ref{PhiNf}, \ref{Phif}). For this we would need to disentangle the 3-body wave-function of the transversely polarised photon $\gamma^*_T\,,$ which, to the best of our knowledge, is unfortunately unknown.


\section{Conclusion}
\label{Sec_conclusion}

In this paper we have shown that the $\gamma^*_{L,T} \to \rho_{L,T}$ transitions,
which involve contributions up to twist 3, preserve the usual dipole form of the scattering amplitude at finite $N_c\,.$
Such a dipole picture  of the scattering process has been used successful to describe phenomenologically the effect
of gluonic saturation inside the target at asymptotical energies for inclusive as well as for exclusive processes involving twist 2 contributions treated in the collinear 
approximation. 

At first sight, twist 3 contributions either of kinematical origin or due to genuine 3-parton correlators have a form which does not exhibit any dipole structure.
It is only after the use of QCD equations of motions that all terms can be reorganized
in a way consistent with the dipole picture. In particular, our result shows that
no quadrupole contribution  is involved in this process at the twist 3 order. We expect that this remains valid also beyond the twist 3 approximation.
The two contributions which we have obtained in Eqs.~(\ref{PhiNf}) and (\ref{Phif}) have the form (\ref{planar}) of a planar (scaling like $N_c^2$) and a non-planar topology (scaling like 1). 

The obtained formula are a starting point for further investigation of the effect of gluonic saturation for exclusive  processes beyond leading twist.

\section*{Acknowledgements}

 We thank B.~Pire, K.~Golec-Biernat, M.~Sadzikowski and L.~Motyka for stimulating discussions. This work is supported in part by the Polish NCN grant 
DEC-2011/01/B/ST2/03915,
by the French-Polish Collaboration Agreement Polonium and by the P2IO consortium.

\section*{Appendix}

The contributions of 3-parton correlators are of two types: the first one being of "Abelian" type (without triple gluon vertex, see Fig.~\ref{Fig:3Abelian}) and the second involving non-Abelian coupling with one triple gluon vertex (see Fig.~\ref{Fig:3NonAbelian}) or two (see Fig.~\ref{Fig:3NonAbelianTwo}).
Let us first consider the "Abelian" diagrams. They involve two kinds of Casimir invariants:
\beqa
\label{colourAbelian}
&&\hspace{-.4cm}\frac{1}{N_c}\,Tr(t^c \, t^a \, t^b \, t^c)\!=\!C_F \, \frac{\delta^{ab}}{2 \, N_c}\equiv C_a \, \frac{\delta^{ab}}{2\, N_c}\!:  \mbox{(aG1), (cG1), (eG1), (fG1)} \\
&&\hspace{-.4cm}\frac{1}{N_c} \,Tr(t^c \, t^a \, t^c \, t^b)\!=\!\left(\!\!C_F-\frac{N_c}2\!\!\right) \! \frac{\delta^{ab}}{2\, N_c}\! \equiv C_b \, \frac{\delta^{ab}}{2\, N_c}\!:  \mbox{(bG1), (dG1), (aG2), (cG2),} \nonumber\\
&&\mbox{ (bG2), (dG2), (eG2), (fG2)},
\eqa
where the $1/N_c$ comes from the Fierz coefficient when factorising the quark--antiquark state in colour space.
\begin{figure}
\psfrag{u}{$\hspace{-.3cm}\ell_1$}
\psfrag{d}{$\hspace{-.3cm}-\ell_2$}
\psfrag{m}{$\hspace{-.4cm}\ell_g$}
\psfrag{a}{$k_1$}
\psfrag{b}{$\hspace{-1cm}\ell_1+\ell_g-q$}
\psfrag{c}{$\hspace{-.8cm}\ell_1-q$}
\psfrag{e}{$\hspace{-.6cm}-k_2 -\ell_2$}
\psfrag{f}{$k_2 $}
\psfrag{q}{$q$}
\psfrag{i}{}
 \raisebox{0cm}{\centerline{\epsfig{file=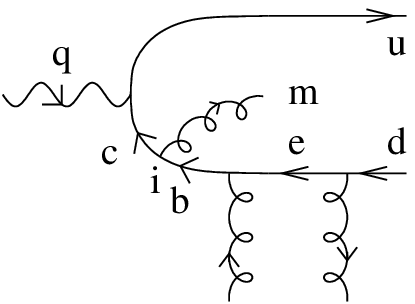,width=7cm}}}
\caption{The detailed structure of the diagram (aG1).}
\label{Fig:aG1}
\end{figure}
To illustrate the method, we consider the particuliar diagram (aG1) of Fig.~\ref{Fig:3Abelian}, illustrated in Fig.~\ref{Fig:aG1}. The vector contribution reads
\beqa
\label{SaG1V}
\Phi^V_{aG1}&=&-e_q \, \frac{1}4 \frac{2}s (i) \,  g^2  \,  f_\rho \, m_\rho \, \frac{\delta^{ab}}{2 \, N_c}\frac{1}{2s}\int\limits^1_0 \! dy_1 \, dy_2 \, B(y_1, \, y_2)\\
&\times &\int \frac{d \kappa}{2 \pi}\frac{Tr [\slashchar{e}_\gamma \, (\slashchar{\ell}_1-\slashchar{q}) \, \slashchar{e}^*_T \, ( \slashchar{\ell}_1+\slashchar{\ell}_g-\slashchar{q}) \,\slashchar{p}_2 \, (\slashchar{k}_2+\slashchar{\ell}_2) \, \slashchar{p}_2 \, \slashchar{p}_1]}{[ (\ell_1-q)^2+i \eta][(\ell_1+\ell_g-q)^2+i \eta] [ (k_2+\ell_2)^2+i \eta]}\,,\nonumber
\eqa
and it equals
\bea
\label{SaG1VRes}
\Phi^V_{aG1}&=&\frac{e_q\, g^2}2  \,  f_\rho \, m_\rho \, \frac{\delta^{ab}}{2 \, N_c} \int\limits^1_0 dy_1 \, dy_2\, B(y_1,\,y_2)\nonumber\\
&&\hspace{-1.9cm}\times\!\left\{ \eu\cdot\Pu \! \left(\yb_2\,\lb_1^2-y_1\,\lb_2\cdot\lb_1+y_1\,\yb_2Q^2\right)\right.\!+\!\left.y_1\!\left((\lb_1\cdot\Pu)(\lb_2\cdot\eu)\!+\!(\lb_1\cdot\eu)(\lb_2\cdot\Pu)\right)\right\}\!\!\nonumber\\
&& \hspace{-1cm}\times \frac{y_1 y_g}{(\lb_1^2+\mu_1^2)(y_g \yb_g \,\lb_1^2+2y_1y_g \,\lb_1\cdot\lb_g+y_1\yb_1\,\lb_g^2+y_1\yb_2y_g\, Q^2)}\,.
\eea
 The expression (\ref{SaG1VRes}) 
  is used in the description of two dipole configurations: $\{\qb g\}$ and $\{q\qb\}$ types. The changes of variables defined either in Table \ref{TableQbg} for the $\{\qb g\}$ system or in the Table \ref{TableQqb} for the $\{q\qb\}$ system leads respectively to explicit forms for $F^{\{\qb g\}}_{aG1V}(\lb_{\qb g}, \Lb_{\qb g}, y_g,\yb_2)$ 
  and $F^{\{q\qb\}}_{aG1V}(\lb_{q\qb}, \Lb_{q\qb}, y_1,\yb_2)$, defined by Eq.~(\ref{def-Fdiag-V}). 

The corresponding axial contribution reads
\beqa
\label{SaG1A}
\Phi^A_{aG1}&=&-e_q \, \frac{i}4
\frac{2}s (i) \,  g^2 \,  f_\rho \, m_\rho \, \frac{\delta^{ab}}{2 \, N_c}\frac{1}{2s}\int\limits^1_0 \! dy_1 \, dy_2\,\, \epsilon^\alpha_{\,\, e^*_T p n} \, D(y_1, \, y_2)\\
&\times&\int \frac{d \kappa}{2 \pi}\frac{Tr [\slashchar{e}_\gamma \, ( \slashchar{\ell}_1-\slashchar{q}) \,  \gamma_\alpha \, ( \slashchar{\ell}_1+\slashchar{\ell}_g-\slashchar{q}) \,\slashchar{p}_2 \, (\slashchar{k}_2+ \slashchar{\ell}_2) \, \slashchar{p}_2 \, \slashchar{p}_1]}
{[(\ell_1-q)^2+i \eta][(\ell_1+\ell_g-q)^2+i \eta] [ (k_2+\ell_2)^2+i \eta]}\,,\nonumber
\eqa
and equals
\bea
\label{SaG1ARes}
\Phi^A_{aG1}&=&\frac{e_q\, g^2}2  \,  f_\rho \, m_\rho \, \frac{\delta^{ab}}{2 \, N_c} \int\limits^1_0 dy_1 \, dy_2\, D(y_1,\,y_2)\nonumber\\
&&\hspace{-1.95cm}\times\!\left\{ \eu\cdot\Pu \! \left(\yb_2\,\lb_1^2+y_1\,\lb_2\cdot\lb_1+y_1\,\yb_2Q^2\right)\right.\!-\!\left.y_1\!\left[(\lb_1\cdot\Pu)\,(\lb_2\cdot\eu)\!+\!(\lb_1\cdot\eu)\,(\lb_2\cdot\Pu)\right]\!\right\}\!\!\nonumber\\
&&\hspace{-1.95cm}\times \frac{y_1 y_g}{(\lb_1^2+\mu_1^2)(y_g \yb_g \lb_1^2+2y_1y_g \lb_1\cdot\lb_g+y_1\yb_1\lb_g^2+y_1\yb_2y_g Q^2)}\,.\!\!\!
 \eea
 Similarly to the vector contribution (\ref{SaG1VRes}), 
  the expression (\ref{SaG1ARes}) is used again in the description 
  of two dipole configurations: $\{\qb g\}$ and $\{q\qb\}$ types. Again, the changes of variables defined either in Table \ref{TableQbg} for the $\{\qb g\}$ system or in the Table \ref{TableQqb} for the $\{q\qb\}$ system leads respectively to explicit forms for  $F^{\{\qb g\}}_{aG1A}(\lb_{\qb g}, \Lb_{\qb g}, y_g,\yb_2)$ and  $F^{\{q\qb\}}_{aG1A}(\lb_{q\qb}, \Lb_{q\qb}, y_1,\yb_2)$, defined by Eq.~(\ref{def-Fdiag-A}). %


Consider now the
 ''non-Abelian`` diagrams of Fig.~\ref{Fig:3NonAbelian}, involving a single triple gluon vertex.
They involve two kinds of colour structure:
\beqa
\label{colour3NonAbelian}
&&\hspace{-.4cm} \frac{2}{N_c^2-1} (-i)\,Tr(t^c \, t^b \, t^d) \, f^{cad}= \frac{N_c}2 \,\frac{1}{C_F} \frac{\delta^{ab}}{2\,N_c}\,:  \mbox{ (atG1), (dtG1), (etG1),}\nonumber\\
&& \mbox{(btG2), (ctG2), (ftG2)} \,,\\
&&\hspace{-.4cm}
\frac{2}{N_c^2-1} (-i)\,Tr(t^c \, t^d \, t^b) \, f^{cad}= -\frac{N_c}2 \,\frac{1}{C_F} \frac{\delta^{ab}}{2\,N_c}\,:  \mbox{ (ctG1), (btG1), (ftG1),}\nonumber\\
&&\mbox{ (atG2), (dtG2), (etG2)} \,,
\eqa
where the $2/(N_c^2-1)$ comes from the Fierz decomposition when factorising the quark--antiquark gluon state in colour space.
\begin{figure}
\psfrag{u}{$\hspace{-.3cm}\ell_1$}
\psfrag{d}{$\hspace{-.3cm}-\ell_2$}
\psfrag{m}{$\hspace{-.3cm}\ell_g$}
\psfrag{i}{}
\psfrag{a}{$k_1$}
\psfrag{b}{$\hspace{-1.cm}k_1-\ell_g$}
\psfrag{c}{$\hspace{-.8cm}\ell_1-q$}
\psfrag{e}{$\hspace{-.5cm}-k_2-\ell_2$}
\psfrag{f}{$k_2 $}
\psfrag{q}{$q$}
 \raisebox{0cm}{\centerline{\epsfig{file=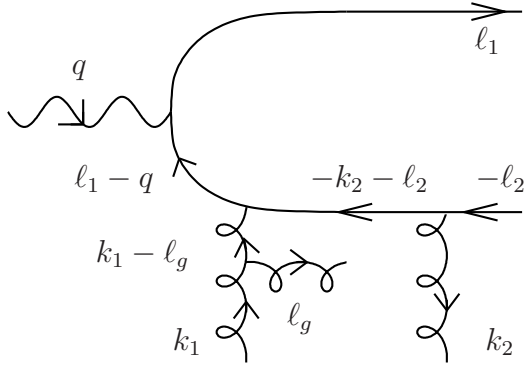,width=7cm}}}
\caption{The detailed structure of the ''non-Abelian`` (with one triple gluon vertex) diagram (atG1).}
\label{Fig:atG1}
\end{figure}
 Let us consider the diagram (atG1) of Fig.~\ref{Fig:3NonAbelian}, illustrated in Fig.~\ref{Fig:atG1}. We denote as
\beq
\label{defPropAxial}
d^{\nu \rho}(k)=g^{\nu \rho}-\frac{k^\nu n^\rho+k^\rho n^\nu}{k \cdot n}
\eq
the numerator of the gluon propagator in axial gauge, and
\beq
\label{defV}
V_{\mu_1 \, \mu_2\, \mu_3}(k_1,\, k_2,\, k_3)=(k_1-k_2)_{\mu_1} \, g_{\mu_1 \mu_2} + \cdots
\eq
the momentum part of the 3-gluon vertex, where $k_i$ are incoming, labeled in the counter-clockwise direction.
The contribution of the diagram (atG1) then reads, for the vector DA,
\beqa
\label{SatG1V}
\Phi^V_{atG1}&\!\!\!=&\!\!-e_q \, \frac{1}4 \frac{2}s \frac{(-i) N_c}{2 \, C_F} \,  g^2 \, m_\rho \, f_\rho \, \frac{\delta^{ab}}{2 \, N_c}\frac{1}{2s}\!\int\limits^1_0 \! dy_1 \, dy_2\, B(y_1, \, y_2)\nonumber\\
&\times& \!\!\int \!\frac{d \kappa}{2 \pi}Tr [\slashchar{e}_\gamma \, ( \slashchar{\ell}_1-\slashchar{q}) \, \gamma_\nu \, (\slashchar{k}_2+ \slashchar{\ell}_2) \,\slashchar{p}_2 \, \slashchar{p}_1] \nonumber \\
&\times& \frac{d^{\nu \rho}(k_1-\ell_g) \, V_{\rho \lambda \alpha}(-k_1+\ell_g,\,k_1,\,-\ell_g)}{[(\ell_1-q)^2+i \eta][(k_1-\ell_g)^2+i \eta] [ (k_2+\ell_2)^2+i \eta]}\,  p_2^\lambda \,e^{*\alpha}_T \,.
\eqa
Note that for this diagram, as well as for all ''non-Abelian`` diagrams, one can easily check that only the $g^{\nu \rho}$ part
 of  (\ref{defPropAxial}) contributes.

Closing the $\kappa$ contour above or below gives for the vector DA part of the diagram (atG1) the result
\bea
\label{SatG1VRes}
\Phi^V_{atG1}&=&\frac{e_q\, g^2}2\, m_\rho \, f_\rho \, \frac{\delta^{ab}}{2 \, N_c} \frac{N_c}{C_F} \int\limits^1_0 \! dy_1 \, dy_2\, B(y_1,\,y_2)\\
&&\hspace{-2cm}\times\left\{y_1\,\lb_1\cdot\Pu\left(y_g\,\lb_2\cdot\eu-2\,\yb_2\,\lb_g\cdot\eu+(\yb_1+\yb_2)\,\kb\cdot\eu\right)\right.\nonumber\\
&&\hspace{-2cm}+\left.y_g\left[\eu\cdot\Pu\left(\yb_2\,\lb_1^2-y_1\,\lb_1\cdot(\lb_2+\kb)+y_1\,\yb_2\,Q^2\right)+y_1\,\lb_1\cdot\eu\,\left(\lb_2+\kb\right)\cdot\Pu\right] \right\}\nonumber\\
&&\hspace{-2cm}\times\!\frac{y_1}{(\lb_1^2+\mu_1^2)(\yb_2y_g\lb_1^2\!+\!y_1y_g\lb_2^2\!+\!y_1\yb_2\lb_g^2\!+\!y_1 \yb_1\kb^2\!+\!2 y_1(y_g\lb_2-\yb_2\lb_g)\cdot\kb+y_1y_g\yb_2Q^2)} .\nonumber
 \eea
%
Similarly, the contribution of the diagram (atG1) reads, for the axial DA,
\beqa
\label{SatG1A}
&&\hspace{-2cm}\Phi^A_{atG1}=-e_q \, \frac{i}4 \frac{2}s \frac{(-i) N_c}{2 \, C_F} \,  g^2 \,  m_\rho \, f_\rho \, \frac{\delta^{ab}}{2 \, N_c}\frac{1}{2s}\!\int\limits^1_0 \! dy_1 \, dy_2\, D(y_1, \, y_2)\nonumber\\
 &&\hspace{0cm}\times \int \!\frac{d \kappa}{2 \pi}Tr [\slashchar{e}_\gamma \, ( \slashchar{\ell}_1-\slashchar{q}) \, \gamma_\nu \, (\slashchar{k}_2+ \, \slashchar{\ell}_2) \,\slashchar{p}_2 \, \slashchar{p}_1\, \gamma_5] \nonumber \\
&&\times \frac{d^{\nu \rho}(k_1-\ell_g) \, V_{\rho \lambda \alpha}(-k_1+\ell_g,\,k_1,\,-\ell_g)}{[(\ell_1-q)^2+i \eta][(k_1-\ell_g )^2+i \eta] [ (k_2+  \ell_2)^2+i \eta]} \,  p_2^\lambda \,\epsilon^\alpha_{\,\, e^*_T p n}  \,,
\eqa
and closing the $\kappa$ contour above or below gives
\bea
\label{SatG1ARes}
\Phi^A_{atG1}&=&\frac{e_q\, g^2}2 \, m_\rho \, f_\rho \, \frac{\delta^{ab}}{2 \, N_c} \frac{N_c}{C_F} \int\limits^1_0 \! dy_1 \, dy_2 \, D(y_1,\,y_2)\\
&&\hspace{-2cm}\times\left\{y_1\lb_1\cdot\eu\left(y_g\,\lb_2\cdot\Pu-2\yb_2\,\lb_g\cdot\Pu+(\yb_1+\yb_2)\,\kb\cdot\Pu\right)\right.\nonumber\\
&&\hspace{-2cm}-\left.(\eu\cdot\Pu)\left[y_1\lb_1\cdot(y_g\lb_2-2\yb_2\lb_g)+y_1(\yb_1+\yb_2)\,\kb\cdot\lb_1+y_g\yb_2\,\lb_1^2+y_1\yb_2y_gQ^2\right]\right.\nonumber\\
&&\hspace{-2cm}\left.+y_1\left[(\lb_1\cdot\eu)\left[\Pu\cdot(y_g\lb_2-2\yb_2\lb_g+(\yb_1+\yb_2)\kb)\right]+y_g(\lb_1\cdot\Pu) [\eu\cdot(\lb_2+\kb)]\right]\right\}\nonumber\\
&&\hspace{-2cm}\times\!\frac{y_1}{(\lb_1^2+\mu_1^2)(\yb_2y_g\lb_1^2\!+\!y_1y_g\lb_2^2\!\!+y_1\yb_2\lb_g^2\!+\!y_1\yb_1\kb^2\!+\!2y_1(y_g\lb_2-\yb_2\lb_g)\cdot\kb\!+\!y_1y_g\yb_2Q^2)} .\!\!\nonumber
\eea
 The expressions (\ref{SatG1VRes}), (\ref{SatG1ARes}), supplemented by the changes of the variables defined in Table \ref{TableQbg}, lead to the explicit forms for $F^{\{\qb g\}}_{atG1,V}(\lb_{\qb g}, \Lb_{\qb g}, \yb_2,y_g)$ and $F^{\{\qb g\}}_{atG1,A}(\lb_{\qb g}, \Lb_{\qb g}, \yb_2,y_g)\,,$ given 
 respectively by Eqs.~(\ref{atG1V}) and (\ref{atG1A}).

%
\begin{figure}
\psfrag{u}{$\hspace{-.3cm}\ell_1$}
\psfrag{d}{$\hspace{-.3cm}-\ell_2$}
\psfrag{m}{$\hspace{.2cm}\ell_g$}
\psfrag{a}{$k_1$}
\psfrag{b}{$\hspace{-1.7cm}\ell_1+\ell_2-q$}
\psfrag{c}{$\hspace{-.8cm}\ell_1-q$}
\psfrag{e}{$\hspace{-.1cm}k_2 +\ell_g$}
\psfrag{f}{$k_2 $}
\psfrag{q}{$q$}
\psfrag{i}{}
 \raisebox{0cm}{\centerline{\epsfig{file=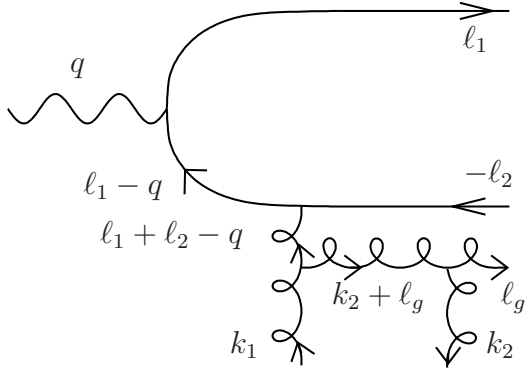,width=7cm}}}
\caption{The detailed structure of the diagram (gttG1).}
\label{Fig:gttG1}
\end{figure}
We
consider now the
 ''non-Abelian`` diagrams of Fig.~\ref{Fig:3NonAbelianTwo}, involving two triple gluon vertices.
They all involve the colour structure
\beq
\label{colour3NonAbelianTwo}
-\frac{2}{N_c^2-1}Tr[t^c \, t^d] f^{cea} \, f^{edb}= \frac{N_c}{C_F} \, \frac{\delta^{ab}}{2 \, N_c}\,.
\eq
For illustration, let us consider the diagram (gttG1) of Fig.~\ref{Fig:3NonAbelianTwo}, illustrated in Fig.~\ref{Fig:gttG1}. It reads for the vector DA,
\beqa
\label{SgttG1V}
\Phi^V_{gttG1}&\!\!\!=&\!\!-e_q \, \frac{1}4 \frac{2}s \frac{(-i) N_c}{C_F} \,  g^2 \, m_\rho \, f_\rho \, \frac{\delta^{ab}}{2 \, N_c}\frac{1}{2s}\!\int\limits^1_0 \! dy_1 \, dy_2 \,  B(y_1, \, y_2)\nonumber\\
&\times& \!\!\int \!\frac{d \kappa}{2 \pi}Tr [\slashchar{e}_\gamma \, (\slashchar{\ell}_1-\slashchar{q}) \, \gamma_\nu \,  \slashchar{p}_1 ]\, d^{\nu \rho}(-q+\ell_1+\ell_2) \nonumber \\
&\times& \frac{V_{\rho \lambda \alpha}(q-\ell_1-\ell_2,\,k_1,\,-k_2-\ell_g)\, d^{\alpha \beta}(k_2+\ell_g)}
{[(\ell_1-q)^2+i \eta][(-q+\ell_1+\ell_2)^2+i \eta] [ (k_2+\ell_g)^2+i \eta]}\nonumber \\
&\times &
V_{\beta \tau \delta}(k_2+\ell_g,\, -k_2,\, -\ell_g)\,  p_2^\lambda \,p_2^\tau \,e^{*\delta}_T\,.
\eqa
It equals, when closing the $\kappa$ contour below on the single pole coming from the third propagator,
\bea
\label{SgttG1VRes}
\Phi^V_{gttG1}&=&\frac{e_q\, g^2}2  \, m_\rho \, f_\rho \, \frac{\delta^{ab}}{2 \, N_c} \frac{N_c}{C_F} \frac{1}{Q^2}
\int\limits^1_0 \! dy_1 \, dy_2 \, B(y_1,\,y_2)\\
&\times&\left\{2y_1(\lb_1\cdot\Pu)\,(\lb_1+\lb_2)\cdot\eu+y_g(\eu\cdot\Pu)\left(\lb_1^2+y_1Q^2\right)\right\}\nonumber\\
&\times&\frac{y_1\yb_2}{(\lb_1^2+\mu_1^2)(y_1\yb_1\,\lb_2^2+y_1\yb_2\,\lb_1^2+2y_1\yb_2\,\lb_1\cdot\lb_2+y_1\yb_2y_g\,Q^2)} \,.\nonumber
 \eea
 The expression (\ref{SgttG1VRes}) 
  is used in the description of the dipole configuration $\{q g\}$. The changes of variables defined in the Table \ref{TableQg} lead to the explicit form for $F^{\{q g\}}_{gttG1,V}(\lb_{qg}, \Lb_{qg}, y_1,y_g)\,,$ defined by Eq.~(\ref{def-Fdiag-V}).

The axial DA contribution from the diagram (gttG1) reads
\beqa
\label{SgttG1A}
\Phi^A_{gttG1}&\!\!\!=&\!\!-e_q \, \frac{i}4 \frac{2}s \frac{(-i) N_c}{C_F} \,  g^2  \, m_\rho \, f_\rho\, \frac{\delta^{ab}}{2 \, N_c}\frac{1}{2s}\!\int\limits^1_0 \! dy_1 \, dy_2 \, D(y_1, \, y_2)\nonumber\\
&\times&\!\!\int \!\frac{d \kappa}{2 \pi}Tr [\slashchar{e}_\gamma \, (\slashchar{\ell}_1-\slashchar{q}) \, \gamma_\nu \,  \slashchar{p}_1 \, \gamma_5]\, d^{\nu \rho}(-q+\ell_1+\ell_2) \nonumber \\
&\times& \frac{V_{\rho \lambda \alpha}(q-\ell_1-\ell_2,\,k_1,\,-k_2-\ell_g)\, d^{\alpha \beta}(k_2+\ell_g)}
{[(\ell_1-q)^2+i \eta][(-q+\ell_1+\ell_2)^2+i \eta] [ (k_2+\ell_g)^2+i \eta]}\nonumber \\
&\times &
V_{\beta \tau \sigma}(k_2+\ell_g,\, -k_2,\, -\ell_g)\,  p_2^\lambda \,p_2^\tau  \,\epsilon^\sigma_{\, \, e^*_T p n} \,.
\eqa
It equals, when closing the $\kappa$ contour below on the single pole coming from the third propagator,
\bea
\label{SgttG1ARes}
&&\Phi^A_{gttG1}=-\frac{e_q\, g^2}2  \, m_\rho \, f_\rho \, \frac{\delta^{ab}}{2 \, N_c} \frac{N_c}{C_F} \frac{1}{Q^2}
\int\limits^1_0 \! dy_1 \, dy_2 \, D(y_1,\,y_2)\\
&&\times\left\{2y_1(\lb_1\cdot\eu)(\lb_1+\lb_2)\cdot\Pu-(\eu\cdot\Pu)\left[2y_1\,\lb_1\cdot\lb_2+(y_1+y_2)\lb_1^2+y_1y_gQ^2\right]\right\}\nonumber\\
&&\times\frac{y_1\yb_2}{(\lb_1^2+\mu_1^2)(y_1\yb_1\,\lb_2^2+y_1\yb_2\,\lb_1^2+2y_1\yb_2\,\lb_1\cdot\lb_2+y_1\yb_2y_g\,Q^2)} \,.\nonumber
\eea
Similarly like in the case of the vector contribution, 
the expression (\ref{SgttG1ARes}) 
  is used in the description of the dipole configuration $\{q g\}$. The changes of variables defined in the Table \ref{TableQg} lead  to the explicit form for $F^{\{q g\}}_{\!gttG1A}(\lb_{qg}, \Lb_{qg}, y_1,y_g)\,,$ defined by Eq.~(\ref{def-Fdiag-A}).

\section*{References}

\providecommand{\href}[2]{#2}\begingroup\raggedright\endgroup

\end{document}